\documentclass[nofootinbib,preprint,onecolumn,11pt,prd]{revtex4-1}

\usepackage{color}
\usepackage{amsmath,bm,amsfonts}
\usepackage{graphicx}

\usepackage{tikz}
\usetikzlibrary{intersections,arrows,
  calc,
  decorations,
  decorations.markings,
  decorations.pathmorphing,
  decorations.pathreplacing,
  graphs,
  patterns,
  positioning,
  shapes.geometric,
}

\tikzset{
  do path picture/.style={
    path picture={
      \pgfpointdiff{\pgfpointanchor{path picture bounding box}{south west}}%
      {\pgfpointanchor{path picture bounding box}{north east}}%
      \pgfgetlastxy\x\y
      \tikzset{x=\x/2,y=\y/2}
      #1
    }
  },
  sin wave/.style={do path picture={    
      \draw [line cap=round] (-3/4,0)
      sin (-3/8,1/2) cos (0,0) sin (3/8,-1/2) cos (3/4,0);
    }},
  cross/.style={do path picture={    
      \draw [line cap=round] (-1,-1) -- (1,1) (-1,1) -- (1,-1);
    }},
  plus/.style={do path picture={    
      \draw [line cap=round] (-3/4,0) -- (3/4,0) (0,-3/4) -- (0,3/4);
    }},
  fermion/.style={
    postaction={decorate},
    decoration={markings,mark=at position .5 with {\arrow[scale=0.7]{triangle 45}}}
  },
  anti fermion/.style={
    postaction={decorate},
    decoration={markings,mark=at position .5 with {\arrow[scale=0.7]{triangle 45 reversed}}}
  },
  gluon/.style={
    decorate, draw=black,
    decoration={coil,aspect=0.75,mirror,
      segment length=1.3mm,  pre length=0.5mm}
  },
  every dot@@/.style={
    /tikz/shape=circle,
    /tikz/graphs/as={},
    /tikz/draw,
    /tikz/fill,
    /tikz/inner sep=0pt,
    /tikz/outer sep=0pt,
    /tikz/minimum size=1mm,
  },
  every dot/.style={/tikz/every dot@@/.append style={#1}},
  dot/.style={
    /tikz/every dot@@,
  },
  every crossed dot@@/.style={
    /tikz/fill=none,
    /tikz/shape=circle,
    /tikz/cross,
    /tikz/minimum size=2mm,
  },
  every crossed dot/.style={/tikz/every crossed dot@@/.append style={#1}},
  crossed dot/.style={
    /tikz/every dot@@,
    /tikz/every crossed dot@@,
  },
}

\usepackage{pgfplots,pgfplotstable}
\usepgfplotslibrary{fillbetween}
\usepgfplotslibrary{groupplots}
\pgfplotsset{compat=newest}

\pgfplotsset{every axis/.style={
    width=12cm,
    height=10cm,
    grid=both,
    scaled ticks=false,
    yticklabel style={/pgf/number format/.cd, fixed,precision=5}
  }
}

\definecolor{darkgreen}{rgb}{0,0.4,0} 
\definecolor{darkblue}{rgb}{0,0,0.6} 
\usepackage[colorlinks=true,linkcolor=darkblue,citecolor=darkgreen]{hyperref}

\newcommand{\as}{\alpha_{\mathrm{s}}}

\newcommand{\LA}{\mathrm{A}}
\newcommand{\LB}{\mathrm{B}}
\newcommand{\LF}{\mathrm{F}}

\newcommand{\LH}{\textsc{H}}
\newcommand{\scH}{\textsc{h}}

\newcommand{\scI}{\textsc{i}}
\newcommand{\LJ}{\mathrm{J}}
\newcommand{\LL}{\mathrm{L}}
\newcommand{\LM}{\mathrm{M}}
\newcommand{\LN}{\mathrm{N}}
\newcommand{\LO}{\mathrm{O}}
\newcommand{\LR}{\mathrm{R}}
\newcommand{\scR}{\textsc{r}}

\newcommand{\LT}{\mathrm{T}}
\newcommand{\La}{\mathrm{a}}
\newcommand{\Lb}{\mathrm{b}}
\newcommand{\Lc}{\mathrm{c}}
\newcommand{\Lf}{\mathrm{f}}

\newcommand{\Lg}{\mathrm{g}}

\newcommand{\Ls}{\mathrm{s}}
\newcommand{\LV}{\mathrm{V}}
\newcommand{\scV}{\textsc{v}}

\newcommand{\bare}{\mathrm{bare}}

\newcommand{\MSbar}{\overline{\mathrm{MS}}}
\newcommand{\msbar}{
  {\overline{\phantom{\raisebox{-0.6 pt}{$\scriptstyle{\textsc{ms}}$}}}
    \hskip - 9.5pt \scriptstyle{\textsc{ms}}}}
\newcommand{\msbarsub}{\!\raisebox{-2pt}{$\msbar$}}

\newcommand{\GeV}{\ \mathrm{GeV}}

\newcommand{\mur}{\mu^2}
\newcommand{\mus}{\mu_{\textsc s}^2}

\newcommand{\cA}{\mathcal{A}}

\newcommand{\cC}{\mathcal{C}}
\newcommand{\cD}{\mathcal{D}}
\newcommand{\cF}{\mathcal{F}}

\newcommand{\cH}{\mathcal{H}}

\newcommand{\cK}{\mathcal{K}}

\newcommand{\cN}{\mathcal{N}}
\newcommand{\cO}{\mathcal{O}}
\newcommand{\cP}{\mathcal{P}}

\newcommand{\cS}{\mathcal{S}}

\newcommand{\cU}{\mathcal{U}}
\newcommand{\cV}{\mathcal{V}}

\newcommand{\cX}{\mathcal{X}}
\newcommand{\cZ}{\mathcal{Z}}

\definecolor{red}{rgb}{1,0,0}

\def\mi{{\mathrm i}}

\def\ket#1{\big|{#1}\big\rangle}
\def\bra#1{\big\langle{#1}\big|}
\def\brax#1{\big\langle{#1}}   
\def\<>#1{\big\langle{#1}\big\rangle}
\def\[]#1{\big[{#1}\big]}

\def\sket#1{\big|{#1}\big)}
\def\sbra#1{\big({#1}\big|}
\def\sbrax#1{\big({#1}}        

\newbox\charbox
\newbox\slabox
\def\s#1{{      
        \setbox\charbox=\hbox{$#1$}
        \setbox\slabox=\hbox{$/$}
        \dimen\charbox=\ht\slabox
        \advance\dimen\charbox by -\dp\slabox
        \advance\dimen\charbox by -\ht\charbox
        \advance\dimen\charbox by \dp\charbox
        \divide\dimen\charbox by 2
        \raise-\dimen\charbox\hbox to \wd\charbox{\hss/\hss}
        \llap{$#1$}
}}

\begin{document}

\title{What is a parton shower?}

\author{Zolt\'an Nagy}

\affiliation{
DESY,
Notkestrasse 85,
22607 Hamburg, Germany
}
\email{Zoltan.Nagy@desy.de}

\author{and Davison E.\ Soper}

\affiliation{
Institute of Theoretical Science,
University of Oregon,
Eugene, OR  97403-5203, USA
}

\email{soper@uoregon.edu}

\begin{abstract}

We consider idealized parton shower event generators that treat parton spin and color exactly, leaving aside the choice of practical approximations for spin and color. We investigate how the structure of such a parton shower generator is related to the structure of QCD. We argue that a parton shower with splitting functions proportional to $\as$ can be viewed not just as a model, but as the lowest order approximation to a shower that is defined at any perturbative order. To support this argument, we present a formulation for a parton shower at order $\as^k$ for any $k$. Since some of the input functions needed are specified by their properties but not calculated, this formulation does not provide a useful recipe for an order $\as^k$ parton shower algorithm. However, in this formulation we see how the operators that generate the shower are related to operators that specify the infrared singularities of QCD.

\end{abstract}

\preprint{DESY 17-063}

\maketitle


\section{Introduction}
\label{sec:introduction}

Parton shower event generators for hadron collisions, such as \textsc{Herwig}  \cite{Herwig}, \textsc{Pythia} \cite{Pythia}, and \textsc{Sherpa} \cite{Sherpa}, perform calculations of cross sections according to an approximation to the standard model or its possible extensions. They are essential for the analysis of experiments at the Large Hadron Collider. The main ideas behind these generators were developed in the 1980s \cite{EarlyPythia, Gottschalk, EarlyHerwig}. There has been extensive development of the algorithms since then \cite{dipolesG, lambdaR, dipolesGP, Herwig1992, Pythia1994, SherpaAlpha, SjostrandSkands, NSI, NSRingberg, Geneva}. The successor programs \cite{Herwig, Pythia, Sherpa, DinsdaleTernickWeinzierl, SchumannKrauss, PlatzerGiesekeI, PlatzerGiesekeII, Deductor, HochePrestel, Vincia2016}, are quite sophisticated. Useful reviews of the field can be found in \cite{review2011, SjostrandReview}. One of the available successor programs is our own, \textsc{Deductor} \cite{NSI, NSII, NSspin, NScolor, Deductor, ShowerTime, PartonDistFctns, ColorEffects, NSThreshold, NSThresholdII}. This paper concerns the perturbative part of these parton shower generators, leaving aside models for the underlying event and hadronization. Furthermore, we consider an idealized version of a parton shower generator in which one accounts exactly for spin and color. Approximations for spin and color are a separate issue, which we do not discuss here.

Our aim in this paper is to investigate how the structure of a shower that treats spin and color exactly is related to the structure of QCD. In particular, we ask whether a shower with splitting functions proportional to $\as$ can be the leading order approximation to something that is defined at any order in $\as$. We find an affirmative answer to this question. Specifically, we find that there is a construction for defining a parton shower that generalizes current showers at any order of perturbation theory. We find also that the problem of relating the structure of a parton shower to the structure of QCD is not as straightforward as one might have naively guessed. First, the construction makes use of functions analogous to the Catani-Seymour dipole splitting functions \cite{CataniSeymour} that specify the infrared behavior of QCD, but beyond leading order the shower splitting functions are not related to the functions that specify the infrared behavior of QCD by anything so simple as just changing their sign. Second, the formulas for the shower automatically includes factors that sum threshold logarithms  \cite{Sterman1987, AppellStermanMackenzie, Catani32, CataniWebberMarchesini, Magnea1990, Korchemsky1993, CataniManganoNason32, SudakovFactorization, KidonakisSterman, KidonakisOderdaSterman, KidonakisOderdaStermanColor, LaenenOderdaSterman, CataniManganoNason, LiJointR, KidonakisOwensPhotons, StermanVogelsang2001, KidonakisOwensJets, LSVJointR, KSVJointR, Kidonakis2005, MukherjeeVogelsang, Bolzoni2006, deFlorianVogelsang, RavindranSmithvanNeerven, RavindranSmith, BonviniForteRidolfi, BFRSaddlePt, deFlorianJets, Catani:2014uta, Ahmed:2014uya, BonviniMarzani, Bonvini:2015ira, ManoharSCET, IdilbiJiSCET, BecherNeubert, BecherNeubertPecjak, BecherNeubertXu, BecherSchwartz, Stewart:2009yx, Beneke:2010da, Bauer:2011uc, Wang:2014mqt, Lustermans:2016nvk, Bonvini:2012az, StermanZeng, Bonvinietalcompare}. These factors are not included in current parton shower generators at even leading order, except for \textsc{Deductor} \cite{NSThreshold, NSThresholdII}. Third, the formulas automatically include matching of the parton shower to a perturbative calculation beyond leading order of the hard scattering that starts the shower. This is fairly straightforward 
\cite{MCatNLO, PowHeg1, PowHeg2, MENLOOPS, PowHegBox, FxFx, MINLO, MINLO2, MEPSatNLO, UNLOPS, PlatzerMatching, NNLOPS, MG5aMCNLO, GENEVANNLO, Czakon, KrkNLO2015, KrkNLO2016, MINLO3, HocheMatching1, HocheMatching2, CKKW, CKKWL, MLM} for a leading order shower, but not for a shower beyond leading order.

We believe that is important to understand that a lowest order parton shower generator can represent the lowest order in a systematically improvable approximation. However, the construction in this paper is not a useful recipe for actually creating a parton shower algorithm beyond the leading order: some of the components of the recipe are specified by their properties but not explicitly constructed. A complete construction will require specifying such choices as a shower ordering variable and momentum mappings. Such a specification will require considerable effort, which lies beyond the scope of this paper.

The papers \cite{JadachJHEP, JadachPolonica, SkandsNLO, HocheNLO1, HocheNLO2} present treatments of a parton shower at order $\as^2$. Ref.~\cite{SkandsNLO} attempts to extend the dipole splittings often used in a leading order shower to a higher order analogue for the case of $e^+ e^-$ annihilation. This approach is similar in spirit to what we do in this paper. An alternative approach \cite{JadachJHEP, JadachPolonica, HocheNLO1, HocheNLO2}, concentrates on the next-to-leading order (NLO) Dokshitzer-Gribov-Lipatov-Altarelli-Parisi (DGLAP) kernel for the evolution of parton distribution functions. Although the parton evolution kernels play a role in our formalism, it is not a central role.

\section{Overview}
\label{sec:overview}

The construction of a parton shower at any perturbative order, presented in Sec.~\ref{sec:toshower}, is rather abstract. In this section, we attempt to provide an overview of what the later mathematics is intending to do.

In our view, it is most useful to think of a parton shower algorithm as beginning with the theorem \cite{factorization} that allows us to write a cross section for an infrared safe observable as a convolution of a hard scattering factor with parton distribution functions. Then the parton shower fills in more detail by using the renormalization group. The parton shower develops with decreasing values of a parameter that is a measure of the hardness of interactions.\footnote{\textsc{Herwig} then rearranges the ordering of splittings in its shower so that larger angle splittings come first.} The essential insight is that the scattering process appears differently depending on the hardness scale at which one examines it. At the hardest scale, the scale of the hard interaction, there are just a few partons (typically quarks and gluons). Then, as the hardness scale at which we examine the process decreases, these partons split, making more partons in a parton shower.\footnote{Thus, with respect to initial state partons, the shower evolution starts from the hard interaction and moves backward in time to softer initial state interactions.} At any stage, a certain amount of structure has emerged, while softer structure remains unresolved.

In this paper, we start with the principle that a parton shower should fully reflect the infrared singularity structure of Feynman graphs for QCD and also the role of parton distribution functions in absorbing initial state singularities. Thus, we start with the infrared sensitive operator associated with the parton distributions and with a perturbative operator $\cD(\mu^2)$ that represents the infrared singularities of QCD Feynman diagrams. We connect $\cD(\mu^2)$ to both the shower splitting kernels and to the subtractions \cite{Daleo:2009yj,Boughezal:2010mc, Gehrmann:2011wi, GehrmannDeRidder:2012ja, Currie:2013vh, Somogyi:2006da, Somogyi:2006db, Somogyi:2013yk} needed to calculate a perturbative cross section beyond the leading order (LO). We work at arbitrary perturbative order. That is, we consider a hard scattering cross section calculated, with subtractions, at $\LN^k\LL \LO$ and a parton shower with $\LN^{k-1}\LL \LO$ splitting functions. (This counts a LO shower as having splitting functions proportional to $\as$.)

The construction that we present is based on the operator $\cD(\mu^2)$. This operator is to contain the infrared singularity structure of Feynman graphs for QCD.   There is no unique recipe for constructing the $\as^{n}$ contribution, $\cD^{(n)}(\mu^2)$, to $\cD(\mu^2)$. As described in Sec.~\ref{sec:pertsigma}, one needs to specify a definition of hardness associated with the integrations in graphs, one needs a momentum mapping, and one needs to specify the form of the functions used as one moves away from the strict soft and collinear limits. At first order, we have made these choices, so that $d\cD^{(1)}(\mu^2)/d\log(\mu^2)$ is part of \textsc{Deductor}. At higher orders, we do not attempt to construct the $\cD^{(n)}(\mu^2)$. Rather, we provide formulas for what to do once one has $\cD^{(n)}(\mu^2)$ for $n \le k$.
 
The formalism uses another operator $\cV(\mu^2)$. This operator is obtained from $\cD(\mu^2)$ and factors associated with the parton distributions but it is obtained by integrating over all of the parton splitting variables, so that it is infrared finite. In a standard first order parton shower, the Sudakov exponent is quite directly related to the part of $\cV(\mu^2)$ that comes from one real parton emission. There is some freedom in setting the color and spin structure of $\cV(\mu^2)$. Thus we leave $\cV(\mu^2)$ partly unspecified.

The parton shower defined here needs to respect the structure of quantum field theory. Thus it includes quantum interference and maintains an exact accounting for the quantum spins and colors of the partons in the shower. The formulation is based on what we call the statistical space, introduce in Ref.~\cite{NSI}. It consists of states that describe the momenta, flavors, colors, and spins of any number of partons. The colors and spins are treated as fully quantum mechanical. This means that the statistical states are density matrices in the quantum color and spin space. We review the statistical space in Sec.~\ref{sec:partonsandrho}.

Using the statistical space, we maintain an exact accounting for the quantum spins and colors. It is not known how to make this practical in computer code, particularly for color. Thus one needs separate approximations, such as the LC+ approximation for color that is used in \textsc{Deductor}. We view the issue of approximations for spin and color as separate from the construction of the shower with exact spin and color. We do not discuss spin and color approximations in this paper. 
 
The final result of the construction, presented in Eq.~(\ref{eq:sigmaU8}), is
\begin{equation}
\begin{split}
\label{eq:preview}
\sigma[J] ={}& 
\sbra{1} \cO_J\, 
\cU(\mu_\Lf^2,\mu_\scH^2)\,
\cU_\cV(\mu_\Lf^2,\mu_\scH^2)\,
\cF(\mu_\scH^2)
\sket{\rho_\scH}
+\cO(\as^{k+B+1}) + \cO(\mu_\Lf^2/Q[J]^2)
\;.
\end{split}
\end{equation}
Here $\sket{\rho_\scH}$ is a statistical state representing the hard scattering, calculated at order $\as^{k+B}$, where $\as^{B}$ is the order of the Born hard scattering process. The hard scattering statistical state $\sket{\rho_\scH}$ includes subtractions, as in a normal $\LN^k\LL\LO$ perturbative calculation. Then $\cF(\mu_\scH^2)$ is an operator on the statistical space that multiplies by appropriate parton distribution functions and a parton luminosity factor. 

The next operator in Eq.~(\ref{eq:preview}),
\begin{equation}
\label{eq:UVexponential00}
\cU_\cV(\mu_\Lf^2,\mu_\scH^{2})
=\mathbb{T} \exp\!\left(
\int_{\mu_\Lf^2}^{\mu_\scH^{2}}\!\frac{d\mu^2}{\mu^2}\,\cS_\cV(\mu^2)
\right)
\;,
\end{equation}
is a process independent operator on the statistical space that leaves the number of partons unchanged and provides perturbative corrections needed to keep the measured cross section correct to order $\as^{k+B}$. This factor also sums threshold logarithms associated with the hard scattering statistical state. The threshold logarithms are an essential part of the construction. As we will discuss, they are included in \textsc{Deductor}, but they are not part of other leading order shower generators. We presented an earlier formulation of threshold summation in a leading order shower in Ref.~\cite{NSThreshold}. This formulation turned out to have certain flaws. In a companion paper \cite{NSThresholdII}, we exhibit the practical effects of the threshold summation according to Eq.~(\ref{eq:preview}) (but with $\sket{\rho_\scH}$ evaluated at leading order only).  

The next operator in Eq.~(\ref{eq:preview}),
\begin{equation}
\label{eq:Vexponential00}
\cU(\mu_\Lf^2,\mu_\scH^{2})
=\mathbb{T} \exp\!\left(
\int_{\mu_\Lf^2}^{\mu_\scH^{2}}\!\frac{d\mu^2}{\mu^2}\,\cS(\mu^2)
\right)
\;,
\end{equation}
is a process independent operator that creates more partons in a parton shower. In the first order case, $k =1$, this is a rather standard parton shower if we average over spins and take the leading color approximation. In general, the splitting generator $\cS(\mu^2)$ consists of terms that are of order $\as^n$ with $1 \le n \le k$. The shower starts at a hardness scale $\mu_\scH^2$ appropriate for the hard scattering and ends at smaller hardness scale $\mu_\Lf^2$ that should be large enough so that perturbation theory at this scale can still be trusted. 

The final operator in Eq.~(\ref{eq:preview}), $\cO_J$, specifies the infrared safe measurement that one wants to make on the parton state after the shower. The hardness scale associated with this measurement is $Q[J]^2$. Finally, $\sbra{1}$ is an instruction to integrate over all of the parton variables. 

The cross section $\sigma[J]$ is then correct to order $\as^{k+B}$ and includes a version of the cross section beyond this order, within the approximations of a parton shower. Notice that the property that the cross section including showering, $\sigma[J]$, is correct to order $\as^{k+B}$ means that the shower is matched to an order $\as^{k+B}$ perturbative calculation of $\sigma[J]$. This matching is an intrinsic part of the shower formulation.

We discuss a very general formulation of parton showers. However, we want to keep the notation simple, so, without loss of generality, we use Higgs boson production as an example. We use five flavors of quarks. In practical applications, one uses a variable flavor number scheme in which non-zero values of $m_\Lb$ and $m_\Lc$ appear. However, this creates complications, especially if we want to work at an arbitrary order of perturbation theory. Thus in this paper we set $m_\Lb = m_\Lc = 0$.

We have found it useful to change some of the notation that we used in our previous papers in order to address a much more general problem.  We hope that this does not cause confusion.  We provide a translation in table \ref{tab:OldNew}.

Following this brief overview, we include a brief Sec.~\ref{sec:factorization} on factorization, which plays an important role in the conceptual development. Then we devote Sec.~\ref{sec:partonsandrho} to partons and the spin and color density operator, which we use to define the statistical space. This leads us describe the perturbative cross section and the infrared sensitive operator $\cD(\mu^2)$ in Sec.~\ref{sec:pertsigma}. Then in Sec.~\ref{sec:toshower} we manipulate the perturbative cross section to define the parton shower. Sec.~\ref{sec:summary} presents a summary and outlook.

There are four appendices. Appendix \ref{sec:toymodel} presents a toy model for the operators used in the construction. We hope that this concrete model will prove instructive. Appendix \ref{sec:K} discusses the definition of parton distribution functions needed for a shower. Appendix \ref{sec:dynamicalscales} discusses how certain scale parameters can be chosen dynamically instead of statically, as in the main text. Appendix \ref{sec:renormalization} discusses the relation of $\MSbar$ renormalization to the definition of the parton shower.

\begin{table}
  \caption{\label{tab:OldNew} New and old notation}
    \begin{tabular}{l|l}
      \multicolumn{1}{c|}{\bf New}  & \multicolumn{1}{c}{\bf Old}  \\
      \hline\hline
      $\cS_{\rm pert}^{(1,0)}$                &\ $\cH_I^{\rm pert}$ \\ 
      $\cS_{\rm pert}^{(0,1)}$                &\ $-\cS^{\rm pert}$  \\
      $[\cF\circ\bar{\cS}^{(1,0)}]\,\cF^{-1}\ $ &\ $\cV$            \\ 
      $\cS^{(1)} - \cS_{\mi \pi}^{(0,1)}$     &\ $\cH_I - \cV$    \\ 
      $\cS_\cV^{(1)} + \cS_{\mi \pi}^{(0,1)}$ &\ $\cV - \cS$      \\
    \end{tabular}
\end{table}

\section{Factorization}
\label{sec:factorization}

We consider a hard scattering process in the collisions of two high energy hadrons, A and B. The hadrons carry momenta $P_\LA$ and $P_\LB$. The hadron energies are high enough that we can simplify the equations describing the collision kinematics by treating the colliding hadrons as being massless. Then with a suitable choice of reference frame, the hadron momenta are
\begin{equation}
\begin{split}
P_\LA ={}& 
\left(P_\LA^+,0,\bm 0 \right)
\;,
\\
P_\LB ={}& 
\left(0, P_\LB^-,\bm 0 \right)
\;.
\end{split}
\end{equation}
Here we use momentum components $(p^+,p^-,\bm p_\perp)$ with $p^\pm = (p^0 \pm p^3)/\sqrt{2}$. We then imagine a parton level process in which a parton from hadron A, with flavor $a$ and momentum $p_\La = \eta_\La P_\LA$ collides with a parton from hadron B, with flavor $b$ and momentum $p_\Lb = \eta_\Lb P_\LB$. 

We will be interested in an inclusive cross section to create some hard state, for instance, a Z boson plus possibly jets, or just jets. We will use the  production of a Higgs boson, $A + B \to \LH + {\it QCD\ partons}$ as our principle example. At the Born level, it is produced via the partonic process $\Lg + \Lg \to \LH$. We treat the Higgs boson as being stable and on shell. We denote the momentum of the Higgs boson by 
\begin{equation}
p_\scH = 
(e^{y_\scH} \sqrt{(m_\scH^2 + \bm p_{\scH,\perp}^2)/2},
 e^{- y_\scH} \sqrt{(m_\scH^2 + \bm p_{\scH,\perp}^2)/2},
  \bm p_{\scH,\perp})
\;.
\end{equation}
The collision also produces QCD partons with flavors $f_i$ and momenta $p_i$, with $i = 1,\dots, m$. In this paper, we consider the QCD partons to be massless. Each final state parton has rapidity $y_i$ and transverse momentum $\bm p_{i,\perp}$, so that the components of its momentum are
\begin{equation}
p_i = (e^{y_i} \sqrt{\bm p_{i,\perp}^2/2}, 
e^{-y_i} \sqrt{\bm p_{i,\perp}^2/2}, \bm p_{i,\perp})
\;.
\end{equation}

It is up to us to decide what we want to measure about the final state of our process. We can consider many cases at once by simply saying that we are interested in a cross section $\sigma[J]$ to measure an observable quantity $J$, leaving the definition of $J$ unspecified. We will see in the following subsection how $\sigma[J]$ can be specified for a general observable $J$. Then parton distribution functions relate $\sigma[J]$ to an analogous cross section $\hat \sigma[J]$ for the collision of two partons. In its briefest form, the relation is
\begin{equation}
  \label{eq:factorization0}
  \sigma[J] \approx \sum_{a,b}\int\!d\eta_\La \int\!d\eta_\Lb\
  f_{a/A}(\eta_\La,\mu^2)\, f_{b/B}(\eta_\Lb,\mu^2)\
  \hat \sigma[J]
  \;.
\end{equation}

\subsection{Infrared safety}
\label{sec:IRsafety}

We demand that the observable $J$ be infrared safe. To specify what that means, we write Eq.~(\ref{eq:factorization0}) in more detail:
\begin{equation}
\begin{split}
\label{eq:sigmaJ}
\sigma[J] ={}& 
\int\! dy_\scH\
\frac{d\sigma_0}{dy_\scH}\
J_0(p_\scH)
+\int\! dy_\scH\,dy_1\,\,d\bm p_{1,\perp}\
\frac{d\sigma_1}
{dy_\scH\,dy_1\,d\bm p_{1,\perp}}\
J_1(p_\scH,p_1)
\\&+
\frac{1}{2!}\int\! dy_\scH\,dy_1\,dy_2\,d\bm p_{1,\perp}\,d\bm p_{2,\perp}\
\frac{d\sigma_2}
{dy_\scH\,dy_1\,dy_2\,d\bm p_{1,\perp}\,d\bm p_{2,\perp}}\
J_2(p_\scH,p_1,p_2)
\\&+
\cdots
\;.
\end{split}
\end{equation}
Here we start with the cross section to produce the Higgs boson plus $m$ partons with momenta 
\begin{equation}
\{p\}_m = \{p_\scH,p_1, \dots, p_m\}
\;.
\end{equation}
We multiply the cross section by a function $J_{m}(\{p\}_m)$ that specifies the measurement that we want to make on the final state partons. These functions are taken to be symmetric under interchange of the QCD momentum arguments $\{p_1, \dots, p_m\}$. Accordingly, we divide by the number $m!$ of permutations of the QCD parton labels. We integrate over the momenta of the final state partons. The transverse momentum of the Higgs boson and the needed momentum fractions for the incoming partons are determined by momentum conservation. Finally, we sum over the number $m$ of final state QCD partons.

Infrared safety is a property of the functions $J_m$ that relates each function $J_{m+1}(\!\{p\}_{m+1}\!)$ to the function $J_{m}(\{p\}_m)$ with one fewer parton. There are two requirements needed for $J$ to be infrared safe.

First, consider the limit in which partons $m + 1$ and $m$ become collinear:
\begin{equation}
\begin{split}
\label{eq:collinearlimit}
&p_{m+1} \to z\tilde p_m
\;,
\\
&p_{m} \to{} (1-z)\tilde p_m
\;.
\end{split}
\end{equation}
Here $\tilde p_m$ is a lightlike momentum and $0 < z < 1$.  We can concentrate on just partons with labels $m+1$ and $m$ because the functions $J$ are assumed to be symmetric under interchange of the parton labels. In order for $J$ to be infrared safe, we demand that
\begin{equation}
\label{eq:IRsafe1}
J_{m+1}(\{p_\scH,p_1,\dots,p_{m-1},p_m,p_{m+1}\})
\to J_m(\{p_\scH,p_1,\dots,p_{m-1},\tilde p_m\})
\end{equation}
in the collinear limit (\ref{eq:collinearlimit}). 

Second, consider also the limit in which parton $m+1$ becomes collinear to one of the beams,
\begin{equation}
\label{eq:collinearA}
p_{m+1} \to \xi  P_\LA
\end{equation}
or
\begin{equation}
\label{eq:collinearB}
p_{m+1} \to \xi  P_\LB
\;.
\end{equation}
Here $0 \le \xi$. When $\xi = 0$, parton $m+1$ is simply becoming infinitely soft. In order for $J$ to be infrared safe, we demand that
\begin{equation}
\label{eq:IRsafe2}
J_{m+1}(\{p_\scH,p_1,\dots,p_m,p_{m+1}\})
\to J_m(\{p_\scH,p_1,\dots,p_{m}\})
\end{equation}
in either limit (\ref{eq:collinearA}) or (\ref{eq:collinearB}).

Briefly, then, infrared safety means that the result of the measurement is not sensitive to whether or not one parton splits into two almost collinear partons and it is not sensitive to any partons that have very small momenta transverse to the beam directions.

\subsection{A more quantitative view of infrared safety}

We now discuss infrared safety a little more quantitatively. Consider, as above, two final state partons that are nearly collinear. This is modeled in a parton shower algorithm as a splitting of a parton with momentum $\tilde p_m$ into two partons with momenta $p_m$ and $p_{m+1}$. We can measure how close we are to the collinear limit by calculating\footnote{One could choose other hardness measures $\mu^2$. This one is based on the parton shower formulation in \cite{Deductor, ShowerTime}.}
\begin{equation}
\label{eq:musplit}
\mu_{\rm split}^2 = \frac{\sqrt{Q_\LH^2}}{E_m + E_{m+1}}\ (p_m + p_{m+1})^2
\;.
\end{equation}
Here $Q_\LH$ is a momentum vector that describes the hard scattering, for instance the momentum $p_\LH$ of the produced Higgs boson in our example. We define the parton energies in the rest frame of $Q_\LH$. The limit expressed in Eq.~(\ref{eq:collinearlimit}) is $\mu_{\rm split}^2 \to 0$. 

Alternatively, we can consider a splitting of an initial state parton as modeled in a parton shower algorithm. It suffices to consider the splitting of an initial state parton in hadron A. The initial state parton with momentum $\tilde p_\La$ becomes a new initial state parton with momentum $p_\La$ and a new final state parton with momentum $p_{m+1}$.\footnote{This is in the ``backwards evolution'' picture. Going forward in time, the parton with momentum $p_\La$ splits.}  We can measure how close we are to the collinear limit by calculating 
\begin{equation}
\mu_{\rm split}^2 = -\frac{\sqrt{Q_\LH^2}}{E_\La - E_{m+1}}\ (p_\La - p_{m+1})^2
\;.
\end{equation}
Again, the limit expressed in Eq.~(\ref{eq:collinearA}) is $\mu_{\rm split}^2 \to 0$.

We now suppose that we measure the cross section $\sigma[J]$ corresponding to an infrared safe measurement function $J$, using Eq.~(\ref{eq:sigmaJ}).
When $\mu_{\rm split}^2 = 0$ in either example, application of Eq.~(\ref{eq:sigmaJ}) gives a cross section that we can call $\sigma_0[J]$. In applying Eq.~(\ref{eq:sigmaJ}), we can use the term with $m + 1$ final state partons, with parton $m+1$ exactly collinear with either parton $m$ or parton ``a.'' Equivalently, we can use the $m$ parton state before the splitting. Because of Eq.~(\ref{eq:IRsafe1}) or Eq.~(\ref{eq:IRsafe2}), the result is exactly the same. Now, when $\mu_{\rm split}^2$ is small but not zero, we get a slightly different result, $\sigma[J]$. Let $\delta\sigma[J] = \sigma_0[J] - \sigma[J]$. We assume that $J_{m+1}(\{p\}_{m+1})$ is a smooth function of the parton momenta, at least near the soft or collinear limits. Then we will have $\delta\sigma[J] \to 0$ as $\mu_{\rm split}^2 \to 0$. A typical case is $\delta\sigma[J] \propto \mu_{\rm split}^2$ as $\mu_{\rm split}^2 \to 0$. Then we can define a scale $Q^2[J]$ that is characteristic of the observable by 
\begin{equation}
\label{eq:IRsensitivity}
\left\langle
\frac{\delta\sigma[J]}{\mu_{\rm split}^2}
\right\rangle
= \frac{\sigma[J]}{Q^2[J]}
\;.
\end{equation}
The ratio in Eq.~(\ref{eq:IRsensitivity}) can be sensitive to the parton configuration, so we average over all configurations with the same $\mu_{\rm split}^2$. The scale $Q^2[J]$ measures how sensitive the cross section is to parton splittings.

An example may be helpful. Let $\sigma[J]$ measure the one-jet-inclusive cross section for jet transverse momentum $P_\LJ$ at small rapidity, using the anti-$k_\LT$ jet algorithm with radius parameter $R$. Consider a very narrow jet in which one parton splits into two, one of which is soft. Then a simple estimate using the definitions above is 
\begin{equation}
\label{eq:jetQ^2J}
Q^2[J] \approx \frac{\sqrt{Q_\LH^2}}{P_\LJ}\,
\frac{(R P_\LJ)^2}{N}
\;.
\end{equation}
Here $N$ measures how fast the jet cross section falls with increasing $P_\LJ$:
\begin{equation}
N = \left|\frac{P_\LJ}{\sigma(P_\LJ)}\frac{d\sigma(P_\LJ)}{dP_\LJ}\right|
\;.
\end{equation}
This is a fairly large number, so that the effective scale $Q^2[J]$ for the jet measurement is smaller than $P_\LJ^2$. Additionally, $R$ is typically chosen to be less than one. The smaller $R$ is, the smaller $Q^2[J]$ is.

We see that we can understand infrared safety in terms of measurements that we might make in a parton shower simulation of a high energy scattering event. As the shower progresses, there are splittings. The corresponding values of $\mu_{\rm split}^2$ get smaller and smaller. If we measure $\sigma[J]$ at each stage of the shower, we will see that $\sigma[J]$ changes as the shower develops, but the changes get smaller and smaller according to
\begin{equation}
\label{eq:IRerror}
\delta\sigma[J] \sim \frac{\mu_{\rm split}^2}{Q^2[J]}\ \sigma[J]
\;.
\end{equation}

This leads us to a limit on the possible accuracy of a perturbative approach to calculating $\sigma[J]$. When the parton shower has evolved to a 1 GeV scale, then we have reached the limits of perturbation theory. The fractional uncertainty associated with one more splitting is then
\begin{equation}
\delta\sigma[J] \sim \frac{(1 \GeV)^2}{Q^2[J]}\ \sigma[J]
\;.
\end{equation}
The accuracy of the perturbative calculation can never be better than this.

\subsection{Factorization}

We can now state how one calculates the cross section for whatever observable $J$ we want -- as long as $J$ is infrared safe. The formula we use was stated in Eq.~(\ref{eq:factorization0}) and we restate it here in a slightly more detailed form \cite{factorization}:
\begin{equation}
\label{eq:factorization}
\sigma[J] = \sum_{a,b}\int\!d\eta_\La \int\!d\eta_\Lb\
f_{a/A}(\eta_\La,\mu^2)\, f_{b/B}(\eta_\Lb,\mu^2)\
\hat \sigma_{a,b}[\eta_\La,\eta_\Lb,\mu^2;J]
+ {\cal O}\!\left( \left[m/Q\right]^n\right)
\;.
\end{equation}
The intuitive basis for this is very simple. The factor $f_{a/A}(\eta_\La,\mu^2)\, d\eta_\La$ represents the probability to find a parton of flavor $a$ in a hadron of flavor $A$. For the other hadron, the corresponding probability is  $f_{b/B}(\eta_\Lb,\mu^2)\, d\eta_\Lb$. Then $\hat \sigma[J]$ is the cross section to obtain the observable $J$ from the scattering of these partons, as given in Eq.~(\ref{eq:sigmaJ}). Naturally, this parton level cross section depends on the parton variables $a,b,\eta_\La,\eta_\Lb$. Here the differential cross sections to produce $m$ final state partons contain delta functions that relate the momentum fractions $\eta_\La$ and $\eta_\Lb$ to the final state parton momenta. The parton distributions depend on a scale $\mu$. This is often called the factorization scale $\mu_\LF$ and distinguished from the argument of $\as$ and other running couplings, which is called the renormalization scale $\mu_\LR$. In order to keep our notation simple, we set $\mu_\LF = \mu_\LR = \mu$.

The cross section $\hat \sigma[J]$ has a perturbative expansion in powers of $\as(\mu^2)$. That is
\begin{equation}
\begin{split}
\hat \sigma_{a,b}[\eta_\La,\eta_\Lb,\mu^2;J]
={}&
\hat \sigma_{a,b}^{(0)}[\eta_\La,\eta_\Lb,\mu^2;J]
+
\left[\frac{\as(\mu^2)}{2\pi}\right]
\hat \sigma_{a,b}^{(1)}[\eta_\La,\eta_\Lb,\mu^2;J]
\\&
+
\left[\frac{\as(\mu^2)}{2\pi}\right]^2
\hat \sigma_{a,b}^{(2)}[\eta_\La,\eta_\Lb,\mu^2;J]
+
\cdots
\;.
\end{split}
\end{equation}
Here we do not display the factors of $\as$ or $\alpha_{\rm ew}$ that appear in the Born level cross section $\hat\sigma^{(0)}$. Perturbative calculations can be at lowest order (LO), corresponding to one term in the expansion, next-to-lowest order (NLO) with two terms, sometimes NNLO, and, in general, $\LN^k \LL\LO$.

One useful property is that the dependence of the calculated cross section on $\mu^2$ diminishes as we go to higher orders. Indeed, the cross section in nature, $\sigma[J]$, does not depend on $\mu^2$. Thus if we calculate to order $\as^{k}$, the derivative of the calculated cross section with respect to $\mu^2$ will be of order $\as^{k+1}$.

There is an error term in Eq.~(\ref{eq:factorization}). No matter how many terms are included in $\hat \sigma$, there are contributions that are left out. These terms are suppressed by a power of $m \sim 1 \GeV$ divided by a large scale parameter $Q$ that characterizes the hard scattering process to be measured.  These contributions arise from the approximations needed to derive Eq.~(\ref{eq:factorization}). For instance if a loop momentum $l$ flows through the wave function of quarks in a proton, we have to neglect $l$ compared to the hard momenta, say the transverse momentum of an observed jet. Not much is known about the general form of the power corrections for hadron-hadron collisions. In the rest of this paper, we will assume that the power $n$ in Eq.~(\ref{eq:factorization}) is $n = 2$ and we use $\sqrt{Q^2[J]}$ from Eq.~(\ref{eq:IRerror}) for $Q$. However, even if we lack a good estimate of the power corrections, it is important that they are there. If $Q$ is of order 100 GeV, then the power corrections are completely negligible. However, if $Q$ is of order 5 GeV, then we ought not to claim 1\% accuracy in the calculation of $\sigma[J]$, no matter how many orders of perturbation theory we use.

\section{Partons and the density operator}
\label{sec:partonsandrho}

We will describe perturbative calculations of cross sections, how these are connected to the parton shower description of these same cross sections, and how this is connected to factorization. We begin in this section with definitions that we need to describe the evolution of a parton shower. We follow the framework of Ref.~\cite{NSI}. 

\subsection{Amplitudes and the density operator in spin and color} 
\label{sec:densityoperator}

In a perturbative calculation of a cross section, one constructs an amplitude $\ket{M(\{p,f\}_m)}$. This amplitude depends on the momenta and flavors of two initial state partons, whatever outgoing electroweak partons there are, and $m$ outgoing QCD partons. For our example of Higgs boson production, the momentum and flavor observables are
\begin{equation}
\{p,f\}_m = \{\eta_\La,a,\eta_\Lb,b, p_\scH, p_1,f_1,\cdots,p_m,f_m\}
\;.
\end{equation}
The partons carry spin and color, so the amplitude is a vector in the partonic spin and color space for $m$ final state QCD partons plus the two incoming partons, as indicted by the representation of the amplitude as a ket vector  $\ket{M(\{p,f\}_m)}$. We use basis vectors $\ket{\{s\}_m}$ for the partonic spin space and basis vectors $\ket{\{c\}_m}$ for the partonic color space.\footnote{The spin basis vectors can be chosen in a very simple way, but it is not so trivial to choose useful basis vectors for the color space. The choice that we make for \textsc{Deductor} is specified in Ref.~\cite{NSI}. With this choice, the color basis vectors are not exactly normalized and the basis vectors for different colors are not exactly orthogonal.} To describe the evolution of a parton shower, it is useful to use quantum statistical mechanics, keeping the full quantum nature of the colors and spins. Thus we use the density operator in the color and spin space. The density operator is a linear combination of basis operators $\ket{\{s,c\}_m}\bra{\{s',c'\}_m}$. Thus to describe the system at a certain stage of evolution, we use a function $\rho$ of the number $m$ of final state QCD partons and of the momenta and flavors $\{p,f\}_m$, where the value of $\rho$ is a color-spin density operator. That is
\begin{equation}
\rho(\{p,f\}_m) = \sum_{\{s,s',c,c'\}_m}
\rho(\{p,f,s,s',c,c'\}_m)
\ket{\{s,c\}_m}\bra{\{s',c'\}_m}
\;.
\end{equation}
The interpretation of this is that the differential probability $dP$ for there to be $m$ final state QCD partons with momenta and flavors $\{p,f\}_m$, times the expectation value of an operator $O$ on the color and spin space, is
\begin{equation}
\label{eq:probability}
dP \times \langle O\rangle = [d\{p\}_m] \sum_{\{s,s',c,c'\}_m}
\rho(\{p,f,s,s',c,c'\}_m)\,
\bra{\{s',c'\}_m} O\ket{\{s,c\}_m}
\;,
\end{equation}
where
\begin{equation}
\label{eq:pfmeasure}
\begin{split}
\big[d\{p\}_{m}\big] \equiv{}& 
\frac{d^d p_\scH}{(2\pi)^{d}}\,
2\pi\delta_{+}(p_\scH^{2} - m^2_\scH)
\prod_{i=1}^m
\left\{\frac{d^dp_i}{(2\pi)^{d}}\,
2\pi\delta_{+}(p_{i}^{2})\right\}
d\eta_{\La}\,
d\eta_{\Lb}
\\& \times
(2\pi)^{d}
\delta\bigg(p_\La + p_\Lb - p_\scH -\sum_{i=1}^{m}p_{i}\bigg)\
\;.
\end{split}
\end{equation}
Here we use dimensional regularization with $d = 4 - 2\epsilon$.

The set of all such functions $\rho$ constitutes a vector space, which we call the statistical space. We represent the vector $\rho$ as a ket vector, $\sket{\rho}$. The rounded end of the ket is meant to distinguish a vector in the statistical space from a vector in the quantum spin$\otimes$color space.

Notice that we use the symbol $\rho$ for four different but related concepts. First,   for each choice of $m$ and $\{p,f,s,s',c,c'\}_m$, $\rho(\{p,f,s,s',c,c'\}_m)$ is a complex number. Second, for each choice of $m$ and $\{p,f\}_m$, $\rho(\{p,f\}_m)$ is a linear operator on the quantum spin$\otimes$color space for the $m + 2$ partons. Third, $\rho$ is a linear map from $m$ and $\{p,f\}_m$ to the space of operators on the quantum spin$\otimes$color space. Fourth, $\sket{\rho}$ is this linear map considered as an element of a vector space, the statistical space. This may seem complicated, but in the end we use almost entirely the statistical vectors $\sket{\rho}$. This then gives us what we think is a compact and powerful notation.

We can define basis vectors $\sket{\{p,f,s,s',c,c'\}_m}$ in the statistical space in such a way that 
\begin{equation}
\sbrax{\{p,f,s,s',c,c'\}_m}\sket{\rho} = \rho(\{p,f,s,s',c,c'\}_m)
\;.
\end{equation}
The completeness relation for the basis vectors is
\begin{equation}
1 = \sum_m \frac{1}{m!} \int\![d\{p\}_m] \sum_{\{f\}_m}\sum_{\{s,s',c,c'\}_m}
\sket{\{p,f,s,s',c,c'\}_m}\sbra{\{p,f,s,s',c,c'\}_m}
\;.
\end{equation}

\subsection{Making an inclusive measurement}

There is a special vector $\sbra{1}$ defined by
\begin{equation}
\sbrax{1}\sket{\{p,f,s,s',c,c'\}_m} = \brax{\{s',c'\}_m}\ket{\{s,c\}_m}
\;.
\end{equation}
With this definition, 
\begin{equation}
\sbrax{1}\sket{\rho} = \sum_m \frac{1}{m!} 
\int [d\{p\}_m] \sum_{\{f\}_m}\sum_{\{s,s',c,c'\}_m}
\rho(\{p,f,s,s',c,c'\}_m)\,
\brax{\{s',c'\}_m}\ket{\{s,c\}_m}
\end{equation}
is the total probability associated with the statistical state $\sket{\rho}$ as defined in Eq.~(\ref{eq:probability}) with $O = 1$. 

With this notation, we begin with perturbatively calculated amplitudes $\ket{M(\{p,f\}_m)}$ for just a few partons. Thus we begin with a perturbatively calculated vector $\sket{\rho}$ in the statistical space. Then we use perturbative operations  that are represented as linear operators on the statistical space. Similarly, the measurement $J$ in Sec.~\ref{sec:IRsafety} is represented as a linear operator on the statistical space. Finally, multiplication by $\sbra{1}$ allows us to obtain the expectation value of the measurement operator.

\subsection{Scales}
\label{sec:scales}

The formalism also uses a reference vector $Q_\LH$ and several scales: a renormalization scale $\mu_\LR^2$, a factorization scale $\mu_\LF^2$, and an ultraviolet cutoff scale $\mu_\Ls^2$. For simplicity, all of these scales are set to a single scale $\mu^2$. The vector $Q_\LH$ is used to set the value $\mu_\scH^2$ of the common scale associated with the hard state $\sket{\rho_\scH}$, defined later in Eq.~(\ref{eq:rhoH}): $\mu_\scH^2 = Q_\LH^2$. A parton shower needs a measure of hardness of parton splittings. We also use $Q_\LH$ as a vector to help define one possible measure of hardness, $\Lambda^2$ defined in Eq.~(\ref{eq:Lambdadef}). For this purpose, $Q_\LH$ should be roughly in the direction of $p_\La + p_\Lb$ in an imagined initial hard scattering. The simplest way to define $Q_\LH$ is to use the hard core part of the intended measurement. For instance, if we are looking for the cross section to produce a Higgs boson with rapidity near zero, we can take $Q_\LH^2 = m_\scH^2$ with zero rapidity and zero transverse part for $Q_\LH$. We will mention another, more dynamic, way to define $Q_\LH$ later, in Appendix \ref{sec:dynamicalscales}.

\subsection{Multiplying by parton distribution functions}

In order to turn matrix elements into cross sections, we divide by a parton luminosity factor\footnote{We have noticed that this factor was too small by a factor 2 in Eq.\ (3.15) of \cite{NSI}.} $n_\Lc(a) n_\Ls(a)\, n_\Lc(b) n_\Ls(b)\ 4 p_\La\cdot p_\Lb$. Here the color counting factor is $n_\Lc(f) = 3$ for a quark flavor $f$ and $n_\Lc(f) = 8$ for $f = \Lg$. The spin counting factor is $n_\Ls(f) = 2$ for a quark flavor $f$ and $n_\Ls(f) = 2(1-\epsilon)$ for $f = \Lg$ when we work in $4-2\epsilon$ dimensions in order to regularize infrared divergences. Then we need to multiply by a parton distribution factor 
\begin{equation}
F_{a,b}(\eta_\La,\eta_\Lb,\mu^2) = f_{a/A}(\eta_\La,\mu^2)\,f_{b/B}(\eta_\Lb,\mu^2)
\;.
\end{equation}
The parton distribution functions here could be the five flavor $\MSbar$ parton distribution functions, or they could have a different definition. We combine the two parton distribution functions as one operator $\cF(\mu^2)$ that acts on the statistical space:
\begin{equation}
\label{eq:cFdef}
\cF(\mu^2)\sket{\{p,f,s,s',c,c'\}_m}
= \frac{F_{a,b}(\eta_\La,\eta_\Lb,\mu^2)}
{n_\Lc(a) n_\Ls(a)\, n_\Lc(b) n_\Ls(b)\ 4 p_\La\cdot p_\Lb}\,
\sket{\{p,f,s,s',c,c'\}_m}
\;.
\end{equation}
We sometimes need a more general parton factor in which the parton distributions are convolved with a function that is also a matrix in the parton flavors. For instance, the evolution equation for the product of parton distribution functions is
\begin{equation}
\mu^2 \frac{d}{d\mu^2}F_{a,b}(\eta_\La,\eta_\Lb,\mu^2)
= F'_{a,b}(\eta_\La,\eta_\Lb,\mu^2)
\;,
\end{equation}
where the prime denotes differentiation and $F'$ is given by the convolution product of $F$ with an evolution kernel $P$,
\begin{equation}
\label{eq:FoP}
F' = F\circ P
\;.
\end{equation}
The precise definition is\footnote{Based on the order of the flavor indices, it would be more conventional to write this as $P\circ F$, but we believe that the notation in Eq.~(\ref{eq:FoP}) better expresses the physics in the context of this paper.}
\begin{equation}
\label{eq:FPconvolution}
F'_{a,b}(\eta_\La,\eta_\Lb,\mu^2) = \sum_{a',b'} \int_0^1\!\frac{dz_\La}{z_\La}
\int_0^1\!\frac{dz_\Lb}{z_\Lb}\ 
F_{a',b'}(\eta_\La/z_\La,\eta_\Lb/z_\Lb,\mu^2)\;
P_{a,a',b,b'}(z_\La,z_\Lb,\mu^2)
\;.
\end{equation}
The evolution kernel for the product of parton distributions is the sum of parton evolution kernels for each of the two parton distribution functions:
\begin{equation}
\begin{split}
P_{a,a',b,b'}(z_\La,z_\Lb,\mu^2) ={}& 
P_{a,a'}(z_\La,\mu^2)\,\delta_{b,b'}\,\delta(z_\Lb - 1)
+ \delta_{a,a'}\,\delta(z_\La - 1)\,P_{b,b'}(z_\Lb,\mu^2)
\;.
\end{split}
\end{equation}
This gives us an operator $\cF'(\mu^2)$ defined by
\begin{equation}
\cF'(\mu^2)\sket{\{p,f,s,s',c,c'\}_m}
= \frac{F'_{a,b}(\eta_\La,\eta_\Lb,\mu^2)}
{n_\Lc(a) n_\Ls(a)\, n_\Lc(b) n_\Ls(b)\ 4 p_\La\cdot p_\Lb}\,
\sket{\{p,f,s,s',c,c'\}_m}
\;.
\end{equation}
In place of $\cF'(\mu^2)$, we use a notation that directly displays that $F'_{a,b}(\eta_\La,\eta_\Lb,\mu^2)$ is constructed according to Eq.~(\ref{eq:FPconvolution}),
\begin{equation}
\cF'(\mu^2) = [\cF(\mu^2)\circ \cP(\mu^2)]
\;.
\end{equation}
The circle indicates the convolution and the square brackets $[\cdots]$ indicate what is included in the convolution.

\section{The perturbative cross section}
\label{sec:pertsigma}

We now consider the cross section for some hard process. We can use any hard process that can lead to an infrared safe cross section, but the details of the notation depend on what hard process we consider. In order to keep the notation simple, we consider a specific process, the cross section make a Higgs boson plus QCD partons, calculated at $\LN^k\LL\LO$. We let $J$ represent an infrared safe measurement of interest as in Sec.~\ref{sec:IRsafety}. The hardest scale associated with this measurement is $\mu_\scH^2$. The scale $Q^2[J]$ introduced in Eq.~(\ref{eq:IRsensitivity}) could be much smaller, as long as it is large compared to $1 \GeV^2$.  In that case, the sort of fixed order calculation discussed in this section is not so useful because there are large logarithms, $\log(\mu_\scH^2/Q^2[J])$. Thus in this section, it may be helpful to imagine that $Q^2[J]$ is not very different from $\mu_\scH^2$.

\subsection{The Born cross section}

Let us begin at the Born level. We call the statistical state corresponding to the Born level matrix element $\sket{\rho^{(0,0)}(\mu^2)}$. The measurement $J$ can be represented as an operator $\cO_J$. To make a cross section, we need a luminosity factor and parton distribution functions. We define an operator $\cF_{\msbarsub}(\mu^2)$ as in Eq.~(\ref{eq:cFdef}) by 
\begin{equation}
\cF_{\msbarsub}(\mu^2)\sket{\{p,f,s,s',c,c'\}_m}
= \frac{f_{a/A}^{\msbar}(\eta_\La,\mu^2)\,f_{b/B}^{\msbar}(\eta_\Lb,\mu^2)}
{n_\Lc(a) n_\Ls(a)\, n_\Lc(b) n_\Ls(b)\ 4 p_\La\cdot p_\Lb}\,
\sket{\{p,f,s,s',c,c'\}_m}
\;.
\end{equation}
We use $\MSbar$ parton distributions for five flavors.  Then the cross section for measurement operator $\cO_J$ is
\begin{equation}
\sigma[J] = \sbra{1}\cF_{\msbarsub}(\mu^2)\cO_J\sket{\rho^{(0,0)}(\mu^2)}
\;.
\end{equation}

\subsection{The higher order cross section}
\label{sec:higherorderrho}

Now we consider higher orders of perturbation theory. We define $\as$ and the parton distribution functions that we start with using $\MSbar$ renormalization, as reviewed in Appendix \ref{sec:renormalization}. Including terms up to order $\as^k$, the $\LN^k\LL\LO$ cross section can be written as
\begin{equation}
  \label{eq:NkLOsigma}
  \sigma[J] = 
  \sbra{1}\! \left[\cF_{\msbarsub}(\mu^2)\circ
    \cZ_F(\mu^2)
  \right]  \cO_J \sket{\rho(\mu^2)}
  +\cO(\as^{k+1})
  \;.
\end{equation}
Here we convolve $\cF_{\msbarsub}(\mu^2)$ with the renormalization factor $\cZ_F(\mu^2)$ for five flavor $\MSbar$ parton distribution functions.\footnote{This is for massless partons. Conceptually, $\cZ_F(\mu^2)$ should be understood as the inverse, in the sense of convolutions, of the product of two parton-in-a-parton distribution functions with on-shell massless incoming parton states. Then this factor removes infrared poles from the cross section. At the level of bare operators, $f^{\rm bare}_{a/b}(\xi) = \delta_{a b}\delta(1-\xi)$. Convolving the renormalized parton-in-a-parton distribution functions with $\cZ_F$ gives the bare distribution functions, leading to Eq.~(\ref{eq:NkLOsigma}). See Appendix \ref{sec:renormalization}.} This factor is just 1 at order $\as^0$. Its higher order contributions are defined by working in $4 - 2\epsilon$ dimensions and consist of the $1/\epsilon^n$ pole terms needed to remove the ultraviolet poles from the renormalized operator that defines parton distributions. Eq.~(\ref{eq:NkLOsigma}) is a perturbative formula. We are to expand all of the factors up to the desired order $\as^k$ and neglect the remainder, indicated by the error estimate $\cO(\as^{k+1})$.

The statistical state $\sket{\rho(\mu^2)}$ has a perturbative expansion
\begin{equation}
\label{eq:rhoexpansion}
\sket{\rho(\mu^2)} = \sket{\rho^{(0)}(\mu^2)} 
+ \sum_{n = 1}^k \left[\frac{\as(\mu^2)}{2\pi}\right]^n
\sket{\rho^{(n)}(\mu^2)}
+ \cO(\as^{k+1})
\;.
\end{equation}
The order $n$ contribution to the statistical state is a sum of terms,\footnote{It may be helpful to note that we define $\sket{\rho^{(n)}(\mu^2)}$ using on-shell matrix elements $\ket{M(\{p,f\}_m)}$ and their complex conjugates, including the factors needed to make $\ket{M(\{p,f\}_m)}$ into an S-matrix element. Then the right hand side of Eq.~(\ref{eq:rhoexpansion}) is invariant under changes of the renormalization scale $\mu^2$ up to order $\as^{k+1}$.}
\begin{equation}
\label{eq:rhoexpansionparts}
\sket{\rho^{(n)}(\mu^2)} = \sum_{n_\scR = 0}^n \sum_{n_\scV = 0}^n 
\theta(n_\scR + n_\scV = n)\,\sket{\rho^{(n_\scR,n_\scV)}(\mu^2)}
\;.
\end{equation}
In $\sket{\rho^{(n_\scR,n_\scV)}(\mu^2)}$, there are $n_\scR$ final state partons and $n_\scV$ virtual loops. This is for Higgs boson production as the Born process. If we had chosen two jet production as the Born level hard process, then  there would be $2 + n_\scR$ partons in the final state.

The $n_\scV$ virtual loops in $\sket{\rho^{(n_\scR,n_\scV)}(\mu^2)}$ can each produce $1/\epsilon$ and $1/\epsilon^2$ poles. The $n_\scR$ partons in the final state can give soft and collinear singularities. The statistical state vector is singular when any of these partons become soft or collinear with the beam directions or collinear with each other. In the case of a Born process that has final state QCD partons, there are also singularities when any of the $n_\scR$ additional partons becomes collinear with starting final state partons. In these singular regions, the infrared safe measurement operator $\cO_J$ sees the partons that are collinear to a given direction as equivalent to a single parton and it does not see partons that are soft or collinear to the beam directions at all. Thus, it is as if we had a completely inclusive measurement as defined by left multiplying by $\sbra{1}$. Then we again get $1/\epsilon$ and $1/\epsilon^2$ poles. Many of the poles cancel between real and virtual graphs. There are, however, some poles associated with momenta that are collinear with the initial state parton momenta. These cancel the poles in $\cZ_F(\mu^2)$. We are left with a finite result.


\subsection[Introduction of the infrared sensitive operator $\cD(\mu^2)$]
{Introduction of the infrared sensitive operator \boldmath{$\cD(\mu^2)$}}
\label{sec:cD}

The formula (\ref{eq:NkLOsigma}) is not completely practical because each term $\sket {\rho^{(n_\scR,n_\scV)} (\mu^2)}$ generates infrared singularities or poles. Only at the end of a calculation, which includes some complicated integrations that include the measurement function, do the poles produced by the infrared singularities cancel. To make this more practical, we define a certain operator $\cD(\mu^2)$ and insert a factor $\cD(\mu^2)\cD^{-1}(\mu^2)$ into Eq.~(\ref{eq:NkLOsigma}), giving 
\begin{equation}
  \label{eq:NkLOsigmaDDinv}
  \sigma[J] = 
  \sbra{1}\!
  \left[\cF_{\msbarsub}(\mu^2)\circ\cZ_F(\mu^2)\right]
  \cD(\mu^2)\,\cD^{-1}(\mu^2)\, \cO_J \sket{\rho(\mu^2)}
  +\cO(\as^{k+1})
  \;.
\end{equation}
The operator $\cD(\mu^2)$ depends on the dimensional regularization parameter $\epsilon$, but we do not display this dependence explicitly. It depends on a second scale $\mu_\Ls^2$ along with $\mu^2$, but we set $\mu^2 = \mu_\Ls^2$. 

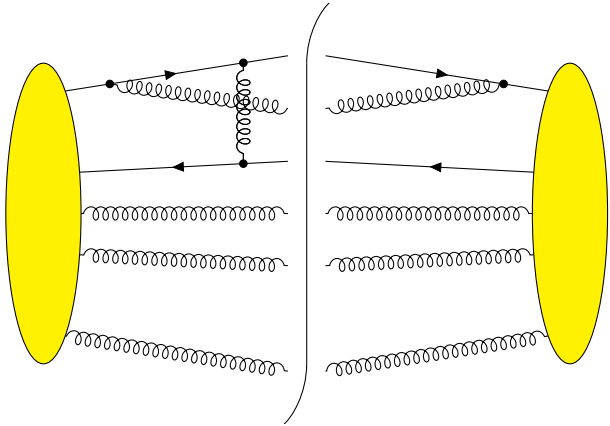
\begin{figure}
  \centering
  \def\finalstatecut {
    \draw (0.3,3.5) .. controls (0.2, 3.4) and (0,3) .. (0,2.5);
    \draw (0.0,2.5) -- (0,-2.5);
    \draw (-0.3,-3.5) .. controls (-0.2, -3.4) and (0,-3) .. (0,-2.5);
  }  

  \def\aampl#1 {
    \begin{scope}[#1]
      \draw [fill=yellow, name path=blob](0,0) ellipse (0.5 and 2); 
      
      \coordinate(o) at(-10,0);
      \coordinate(o1) at(5,0);
      \coordinate(o2) at(0,0);
      
      \path [gray!30!white, name path=p] (3.25,3)--(3.25,-3);

      \coordinate[rotate around={9:(o)}](a0) at(0,0);
      \coordinate[rotate around={3:(o)}](b0) at(0,0);
      \coordinate[rotate around={0:(o)}](c0) at(0,0);
      \coordinate[rotate around={-3:(o)}](d0) at(0,0);
      \coordinate[rotate around={-9:(o)}](e0) at(0,0);
      
      \coordinate[rotate around={9:(o)}](a00) at(o1);
      \coordinate[rotate around={3:(o)}](b00) at(o1);
      \coordinate[rotate around={0:(o)}](c00) at(o1);
      \coordinate[rotate around={-3:(o)}](d00) at(o1);
      \coordinate[rotate around={-9:(o)}](e00) at(o1);

      \path [gray!30!white, name path=g1] (a0)--(a00);
      \path [gray!30!white, name path=g2] (b0)--(b00);
      \path [gray!30!white, name path=g3] (c0)--(c00);
      \path [gray!30!white, name path=g4] (d0)--(d00);
      \path [gray!30!white, name path=g5] (e0)--(e00);

      \path[name intersections={of=g1 and blob, by={bi}}];
      \path[name intersections={of=g1 and p, by={b1i}}];

      \coordinate(a) at(bi);
      \coordinate(a1) at(b1i);
          
      \path[name intersections={of=g2 and blob, by={bi}}];
      \path[name intersections={of=g2 and p, by={b1i}}];

      \coordinate(b) at(bi);
      \coordinate(b1) at(b1i);

      \coordinate [dot](r) at ($(a)!0.2!(a1)$){};
      \coordinate [](r1) at ($(a1)!0.5!(b1)$);

      \path[name intersections={of=g3 and blob, by={bi}}];
      \path[name intersections={of=g3 and p, by={b1i}}];

      \coordinate(c) at(bi);
      \coordinate(c1) at(b1i);
      
      \path[name intersections={of=g4 and blob, by={bi}}];
      \path[name intersections={of=g4 and p, by={b1i}}];

      \coordinate(d) at(bi);
      \coordinate(d1) at(b1i);
      
      \path[name intersections={of=g5 and blob, by={bi}}];
      \path[name intersections={of=g5 and p, by={b1i}}];

      \coordinate(e) at(bi);
      \coordinate(e1) at(b1i);
    \end{scope}
  }

  \begin{tikzpicture}[scale=1, transform shape]
    \aampl{}
    \coordinate[dot](l) at ($(a)!0.8!(a1)$){};	
    \coordinate(ll) at($(l) - (0,5)$);
    \coordinate[dot](l1) at(intersection of l--ll and b--b1);
    
    \draw [fermion] (a) -- (a1);
    \draw [anti fermion] (b) -- (b1);
    \draw [gluon] (l) -- (l1);
    \draw [gluon]  (r) -- (r1);
    \draw [gluon]  (c1) -- (c);
    \draw [gluon]  (d1) -- (d);
    \draw [gluon]  (e1) -- (e);

    \aampl{xscale=-1, xshift=-7cm}
    \draw [anti fermion] (a) -- (a1);
    \draw [fermion] (b) -- (b1);
    \draw [gluon] (l) -- (l1);
    \draw [gluon]  (r1) -- (r);
    \draw [gluon]  (c) -- (c1);
    \draw [gluon]  (d) -- (d1);
    \draw [gluon]  (e) -- (e1);	
     
    \begin{scope}[yscale = 0.8, xshift=3.5cm]
      \finalstatecut
    \end{scope}
\end{tikzpicture}
\caption{An infrared singular diagram. Each yellow blob represents a graph, possibly with loops, in which everything is harder than the scale $\mu_{\rm hard}^2$.}
\label{fig:D0}
\end{figure}

The idea behind $\cD(\mu^2)$ is that a contribution to $\sket{\rho(\mu^2)}$ has poles from virtual loops and has singularities when some of its external lines become collinear or soft. This is illustrated in Fig.~\ref{fig:D0}. It is simplest to think of the graphs depicted as being in a physical gauge.  There are two hard subgraphs, represented as yellow blobs, one for the amplitude and one for the conjugate amplitude. The subgraphs can be tree graphs or can contain virtual loops. We suppose that everything inside the hard subgraphs is harder than a reference scale $\mu_{\rm hard}^2$. That is, all of the internal propagators are far off shell. Two initial state lines and $m$ final state parton lines emerge from the hard subgraph. Here $m = 3$. Then there are additional interactions. Some number $n_\scR$ of additional partons are emitted and $n_\scV$ parton lines are exchanged. Here $n_\scR = 1$ and $n_\scV = 1$. The external parton momenta are labeled $\{\hat p\}_{m + n_\scR}$. There is an infrared divergence when the the virtual gluon becomes soft and there are singularities when the real gluon momentum becomes soft or collinear to the antiquark line while the virtual gluon momentum is becoming soft.

\begin{figure}
  \centering
\def\finalstatecut {
  \draw (0.3,3.5) .. controls (0.2, 3.4) and (0,3) .. (0,2.5);
  \draw (0.0,2.5) -- (0,-2.5);
  \draw (-0.3,-3.5) .. controls (-0.2, -3.4) and (0,-3) .. (0,-2.5);
}  

\def\aampl#1 {
  \begin{scope}[#1]
    \draw [fill=yellow, name path=blob](0,0) ellipse (0.5 and 2); 
    
    \coordinate(o) at(-10,0);
    \coordinate(o1) at(5,0);
    \coordinate(o2) at(0,0);
    
    \coordinate(p0) at($(3.5,3)-(2.2,0)$);
    \coordinate(p1) at($(3.5,-3)-(2.2,0)$);
    \coordinate(p2) at($(3.5,3)-(2.45,0)$);
    \coordinate(p3) at($(3.5,-3)-(2.45,0)$);
    
    \path [red!30!white, name path=pr,draw] (p0)--(p1);
    \path [red!30!white, name path=pr1] (p2)--(p3);
    \path [red!30!white, name path=p] (3.25,3)--(3.25,-3);

    \coordinate[rotate around={9:(o)}](a0) at(0,0);
    \coordinate[rotate around={3:(o)}](b0) at(0,0);
    \coordinate[rotate around={0:(o)}](c0) at(0,0);
    \coordinate[rotate around={-3:(o)}](d0) at(0,0);
    \coordinate[rotate around={-9:(o)}](e0) at(0,0);
    
    \coordinate[rotate around={9:(o)}](a00) at(o1);
    \coordinate[rotate around={3:(o)}](b00) at(o1);
    \coordinate[rotate around={0:(o)}](c00) at(o1);
    \coordinate[rotate around={-3:(o)}](d00) at(o1);
    \coordinate[rotate around={-9:(o)}](e00) at(o1);

    \path [gray!30!white, name path=g1] (a0)--(a00);
    \path [gray!30!white, name path=g2] (b0)--(b00);
    \path [gray!30!white, name path=g3] (c0)--(c00);
    \path [gray!30!white, name path=g4] (d0)--(d00);
    \path [gray!30!white, name path=g5] (e0)--(e00);

    \path[name intersections={of=g1 and blob, by={bi}}];
    \path[name intersections={of=g1 and p, by={b1i}}];
    \path[name intersections={of=g1 and pr, by={b2i}}];
    \path[name intersections={of=g1 and pr1, by={b3i}}];

    \coordinate [](a) at(bi) ;
    \coordinate [](a1) at(b1i);
    \coordinate [crossed dot](a2) at(b2i){};
    \coordinate [](a3) at(b3i);

    \path[name intersections={of=g2 and blob, by={bi}}];
    \path[name intersections={of=g2 and p, by={b1i}}];
    \path[name intersections={of=g2 and pr, by={b2i}}];
    \path[name intersections={of=g2 and pr1, by={b3i}}];

    \coordinate [](b) at(bi);
    \coordinate [](b1) at(b1i);
    \coordinate [crossed dot](b2) at(b2i){};
    \coordinate [](b3) at(b3i);

    \coordinate [dot](r) at ($(a2)!0.2!(a1)$){};
    \coordinate [](r1) at ($(a1)!0.5!(b1)$);

    \path[name intersections={of=g3 and blob, by={bi}}];
    \path[name intersections={of=g3 and p, by={b1i}}];
    \path[name intersections={of=g3 and pr, by={b2i}}];
    \path[name intersections={of=g3 and pr1, by={b3i}}];

    \coordinate [](c) at(bi);
    \coordinate [](c1) at(b1i);
    \coordinate [crossed dot](c2) at(b2i);
    \coordinate [](c3) at(b3i);

    \path[name intersections={of=g4 and blob, by={bi}}];
    \path[name intersections={of=g4 and p, by={b1i}}];
    \path[name intersections={of=g4 and pr, by={b2i}}];
    \path[name intersections={of=g4 and pr1, by={b3i}}];
    
    \coordinate [](d) at(bi);
    \coordinate [](d1) at(b1i);
    \coordinate [crossed dot](d2) at(b2i);
    \coordinate [](d3) at(b3i);

    \path[name intersections={of=g5 and blob, by={bi}}];
    \path[name intersections={of=g5 and p, by={b1i}}];
    \path[name intersections={of=g5 and pr, by={b2i}}];
    \path[name intersections={of=g5 and pr1, by={b3i}}];

    \coordinate [](e) at(bi);
    \coordinate [](e1) at(b1i);
    \coordinate [crossed dot](e2) at(b2i);
    \coordinate [](e3) at(b3i);

  \end{scope}
}

  \begin{tikzpicture}[scale=1, transform shape]
    \aampl{}
    
    \coordinate [dot](l) at ($(a)!0.8!(a1)$);	
    \coordinate(ll) at($(l) - (0,5)$);
    \coordinate[dot](l1) at(intersection of l--ll and b--b1);
    
    \draw[fermion] (a) -- (a3);
    \draw[fermion] (a2) -- (a1);
    \draw[anti fermion] (b) -- (b3);
    \draw[anti fermion] (b2) -- (b1);
    \draw[gluon] (l) -- (l1);
    
    \draw[gluon] (r) --(r1);
    \draw[gluon] (c3) -- (c);
    \draw[gluon] (c1) -- (c2);
    \draw[gluon] (d3) -- (d);
    \draw[gluon] (d1) -- (d2);
    \draw[gluon] (e3) -- (e);
    \draw[gluon] (e1) -- (e2);

    \coordinate(w0) at(p1);
     
    \aampl{xscale=-1, xshift=-7cm}
    
    \draw[anti fermion] (a) -- (a3);
    \draw[anti fermion] (a2) -- (a1);
    \draw[fermion](b) -- (b3);
    \draw[ fermion](b2) -- (b1);
    
    \draw[gluon] (r1) --(r);
    \draw[gluon] (c) -- (c3);
    \draw[gluon] (c2) -- (c1);
    \draw[gluon] (d) -- (d3);
    \draw[gluon] (d2) -- (d1);
    \draw[gluon] (e) -- (e3);
    \draw[gluon] (e2) -- (e1);

    \coordinate(w1) at(p1);

    \draw [decoration={brace}, decorate] (w1) -- (w0)
    node [pos=0.5, below] {${\cal D}(\mu^2)$};
    
    \begin{scope}[yscale = 0.8, xshift=3.5cm]
      \finalstatecut
    \end{scope}
\end{tikzpicture}
\caption{The diagram in Fig.~\ref{fig:D0} after separating the hard diagram from $\cD(\mu^2)$.}
\label{fig:D1}
\end{figure}
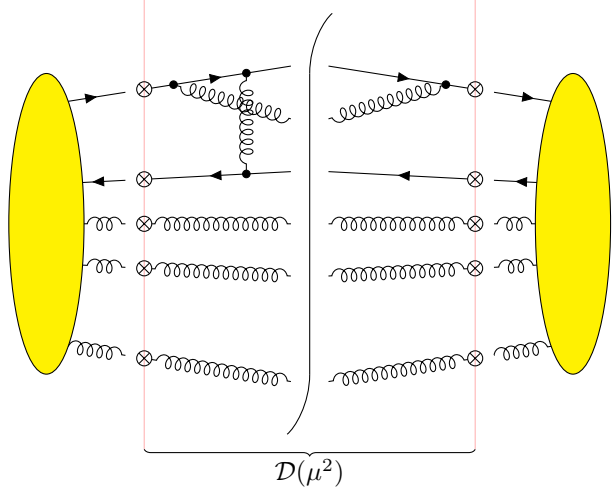

We want to capture the structure of these infrared singularities, as illustrated in Fig.~\ref{fig:D1}. We note that near the singularities, the partons emerging from the hard subgraphs are nearly on shell and their momenta lie in the directions of the external parton momenta. The momenta carried by lines internal to the hard subgraphs are almost unchanged. Therefore, we can approximate the graph by letting the momenta $\{p\}_{m}$ of the partons emerging from the hard subgraphs be exactly on shell. Their momenta are given as functions of the momenta $\{\hat p\}_{m + n_\scR}$ of the external partons. Here, we need to define a momentum mapping $\{\hat p\}_{m + n_\scR} \to \{p\}_{m}$. Then we can approximate the original graph by a hard part $\sbrax{\{p,f,s,s',c,c'\}_m}\sket{\rho_{\rm hard}(\mu^2)}$ and a singular factor. We call the singular factor $\sbra{\{\hat p,\hat f,\hat s,\hat s',\hat c,\hat c'\}_{m+n_\scR}}\cD(\mu^2)\sket{\{p,f,s,s',c,c'\}_m}$. The singular factor is derived just from the singular part of the graph. It is independent of what is in the hard part.

We thus assert that any such set of poles and singularities can be organized into a hard subgraph, $\sket{\rho_{\rm hard}(\mu^2)}$, convolved with a singular factor: 
\begin{equation}
  \begin{split}
    \big(\{\hat p,\hat f,\hat s,\hat s',\hat c,{}&\hat c'\}_{m+n_\scR}\sket{\rho(\mu^2)}
    \\
    \sim{}&
    \frac{1}{m!} \int\![d\{p\}_m] \sum_{\{f\}_m}\sum_{\{s,s',c,c'\}_m}
    \\&\qquad\times
    \sbra{\{\hat p,\hat f,\hat s,\hat s',\hat c,\hat c'\}_{m+n_\scR}}\cD(\mu^2)\sket{\{p,f,s,s',c,c'\}_m}
    \\&\qquad\times
    \sbrax{\{p,f,s,s',c,c'\}_m}\sket{\rho_{\rm hard}(\mu^2)}
    \;.
\end{split}
\end{equation}
The division between singular and hard factors depends on the singularity to be examined. In the hard factor, the external momenta $\{p\}_m$ and any internal loop momenta are to be hard at some scale that we can call $\mu^2_{\rm hard}$. This means that they are not closely collinear to each other or soft at scales softer than $\mu^2_{\rm hard}$.

The singular factor is typically represented as separate factors labeled ${\it soft}$ and ${\it jet}_i$, where ${\it soft}$ can include Glauber exchanges and two of the jets are in the beam directions. However, we do not need to separate the singular factor into separate subfactors. An early and instructive analysis of the singularities of QCD was given by Libby and Sterman \cite{LibbySterman}. An extensive modern analysis can be found in Collins \cite{JCCbook}. For one real gluon emission, an example of $\cD(\mu^2)$ can be defined from the Catani-Seymour dipole splitting functions \cite{CataniSeymour}. The operators $\bm{Sp}$ of Catani, de Florian, and Rodrigo \cite{CataniFactViolated} are also closely related to $\cD(\mu^2)$ for certain cases. At one loop, our version of $d \cD(\mu^2)/d \log\mu^2$ is implemented in \textsc{Deductor}.

The operator $\cD(\mu^2)$ has a perturbative expansion
\begin{equation}
\label{eq:Dexpansion}
\cD(\mu^2) = 1 
+ \sum_{n = 1}^k \left[\frac{\as(\mu^2)}{2\pi}\right]^n
\cD^{(n)}(\mu^2)
+ \cO(\as^{k+1})
\;.
\end{equation}
The order $n$ contribution, $\cD^{(n)}(\mu^2)$, is a sum of infrared sensitive operators,
\begin{equation}
\label{eq:Dexpansionparts}
\cD^{(n)}(\mu^2) = \sum_{n_\scR = 0}^n \sum_{n_\scV = 0}^n 
\theta(n_\scR + n_\scV = n)\,\cD^{(n_\scR,n_\scV)}(\mu^2)
\;.
\end{equation}
Acting on a state $\sket{\{p,f,s,s',c,c'\}_m}$ with $m$ final state QCD partons, $\cD^{(n_\scR,n_\scV)}(\mu^2)$ produces a state with $m + n_\scR$ final state QCD partons with momenta $\{\hat p\}_{m + n_\scR}$. There are integrations over the loop momenta $\{\ell_1, \dots, \ell_{n_\scV}\}$ of $n_\scV$ virtual loops. 

The operator $\cD(\mu^2)$ depends on a hardness scale $\mu^2_\Ls$ that defines an infrared sensitive region $R(\mu^2_\Ls)$ in the space of the momenta $\{\hat p\}_{m + n_\scR}$ and $\{\ell_1, \dots, \ell_{n_\scV}\}$. We can think of $\mu^2_\Ls$ as being comparable to the scale $\mu^2_{\rm hard}$ of $\sket{\rho_{\rm hard}(\mu^2)}$. The infrared sensitive region $R(\mu^2_\Ls)$ surrounds the leading singularity, at which each of the momenta $\{\hat p\}_{m + n_\scR}$ and $\{\ell_1, \dots, \ell_{n_\scV}\}$ is soft or collinear to one of the input momenta $\{p\}_m$. In the case of the output momenta $\{\hat p\}_{m + n_\scR}$, this means that, at the leading singularity, these momenta form $m$ infinitely narrow jets with momenta $\{p\}_{m}$. Of course, if we look just a little bit away from the limit of infinitely narrow jets, we see that the jets can have subjets. The singularity structure of the subjets, including both real and virtual momenta, is included in $\cD(\mu^2)$.

We need the infrared sensitive region $R(\mu^2_\Ls)$ because,
when we form $\cD(\mu^2)$ by making approximations that apply near the leading singularity, we necessarily simplify the behavior away from this singularity. We introduce cuts such that $\cD^{(n_\scR,n_\scV)}(\mu^2)$ gets contributions only from inside $R(\mu^2_\Ls)$. The leading singularity is inside the region for any $\mu^2_\Ls$, but for larger $\mu^2_\Ls$, the region is larger.  Of course, there is more than one way to introduce $\mu_\Ls^2$.

For simplicity, we set the renormalization and factorization scale $\mu^2$ equal to $\mu_\Ls^2$.

We will say more about the $\cD^{(n)}(\mu^2)$ later, although we do not construct them. For now we simply assume that they are available and investigate how they can be used to construct subtractions for a fixed order calculation and splitting functions for a parton shower. See Appendix \ref{sec:toymodel} for an example of the operators $\cD^{(n)}(\mu^2)$ in a toy model.

\subsection{Subtractions for the perturbative cross section}

Given $\cD(\mu^2)$, we can construct 
\begin{equation}
\label{eq:Dinvexpansion}
\cD^{-1}(\mu^2) = 1 
- \sum_{n = 1}^k \left[\frac{\as(\mu^2)}{2\pi}\right]^n
\widetilde\cD^{(n)}(\mu^2)
+ \cO(\as^{k+1})
\;.
\end{equation}
The perturbative coefficients $\widetilde\cD^{(n)}(\mu^2)$ are defined by
\begin{equation}
\label{eq:DinvD}
\cD^{-1}(\mu^2)\cD(\mu^2)=1
\;.
\end{equation}
This gives, for instance,
\begin{equation}
\begin{split}
\widetilde\cD^{(1)}(\mu^2) ={}& \cD^{(1)}(\mu^2)
\;,
\\
\widetilde\cD^{(2)}(\mu^2) ={}&  \cD^{(2)}(\mu^2) - 
\cD^{(1)}(\mu^2)\,\cD^{(1)}(\mu^2)
\;,
\end{split}
\end{equation}
or for the higher orders
\begin{equation}
  \widetilde\cD^{(n)}(\mu^2) = 
  \cD^{(n)}(\mu^2) 
  - \sum_{k=1}^{n-1}\cD^{(k)}(\mu^2)\,\widetilde\cD^{(n-k)}(\mu^2)\;.
\end{equation}
The order $n$ contribution, $\widetilde\cD^{(n)}(\mu^2)$, is a sum of operators,
\begin{equation}
\label{eq:tildeDexpansionparts}
\widetilde\cD^{(n)}(\mu^2) = \sum_{n_\scR = 0}^n \sum_{n_\scV = 0}^n 
\theta(n_\scR + n_\scV = n)\,\widetilde\cD^{(n_\scR,n_\scV)}(\mu^2)
\;.
\end{equation}
Acting on a state $\sket{\{p,f,s,s',c,c'\}_m}$ with $m$ partons, $\widetilde\cD^{(n_\scR,n_\scV)}(\mu^2)$ produces a state with $m + n_\scR$ partons while integrating over $n_\scV$ virtual loops. One constructs $\widetilde\cD^{(n_\scR,n_\scV)}(\mu^2)$ using Eq.~(\ref{eq:DinvD}).

The operator $\cD^{-1}(\mu^2)$ is very useful. In Eq.~(\ref{eq:NkLOsigmaDDinv}), we have the factor $\cD^{-1}(\mu^2) \cO_J \sket{\rho(\mu^2)}$. We are to expand this product in powers of $\as$, keeping the terms up to order $\as^k$. The factor $\cO_J \sket{\rho(\mu^2)}$ has infrared singularities and poles, but the operator $\cD^{-1}(\mu^2)$ removes them.  The simple argument is that $\cD(\mu^2)\sket{\{p,f,s,s',c,c'\}_m}$ contains the infrared singularities and poles produced by QCD from a hard parton state $\sket{\{p,f,s,s',c,c'\}_m}$ but, according to Eq.~(\ref{eq:DinvD}), when we apply $\cD^{-1}(\mu^2)$ to this state, the poles and singular terms are cancelled. That is, $\cD^{-1}(\mu^2)$ provides the subtraction terms that we need to remove the singularities and poles from an $\LN^k\LL\LO$ perturbative calculation.

To understand what the operator $\cD^{-1}(\mu^2)$ does, it is helpful to examine the familiar case of an NLO calculation. At this order, Eq.~(\ref{eq:NkLOsigmaDDinv}) becomes
\begin{equation}
  \begin{split}
    \label{eq:sigmaNLO1}
    \sigma[J] ={}& 
    \sbra{1}\! \left[\cF_{\msbarsub}(\mu^2)\circ\cZ_F(\mu^2)\right]
    \cD(\mu^2)
    \\&\quad\times
    \bigg\{
    \cO_J\sket{\rho^{(0,0)}(\mu^2)}
    \\&\qquad\quad
    + \frac{\as(\mu^2)}{2\pi}
    \left[\cO_J\sket{\rho^{(0,1)}(\mu^2)}
      - \cD^{(0,1)}(\mu^2)\cO_J\sket{\rho^{(0,0)}(\mu^2)}
    \right]
    \\&\qquad\quad
    + \frac{\as(\mu^2)}{2\pi}
    \left[\cO_J\sket{\rho^{(1,0)}(\mu^2)}
      - \cD^{(1,0)}(\mu^2)\cO_J\sket{\rho^{(0,0)}(\mu^2)}
    \right]
    \bigg\}
    \\&+\cO(\as^{2})
    \;.
  \end{split}
\end{equation}
We have defined $\cD^{(0,1)}(\mu^2)$ so that it leaves the number of partons unchanged and so that it has infrared poles. Furthermore, the poles in $\cD^{(0,1)}(\mu^2) \sket{\rho^{(0,0)}}$ should directly cancel those of $\sket{\rho^{(0,1)}(\mu^2)}$. We have defined $\cD^{(1,0)}(\mu^2)$ so that, acting on the state $\sket{\rho^{(0,0)}(\mu^2)}$, it adds one parton and so that when this parton is soft or nearly collinear with one of the existing partons (in our example, the initial state partons) $\cD^{(1,0)}(\mu^2)\cO_J \sket{\rho^{(0,0)}(\mu^2)}$ approaches $\cO_J\sket{\rho^{(1,0)}(\mu^2)}$. In a standard application, one performs the integrations over the momentum of the emitted parton numerically. The integrand in the subtraction cancels the integrand in $\sket{\rho^{(1,0)}(\mu^2)}$ in the infrared region, so that one obtains a convergent integration. Having subtracted the operators $\cD^{(1,0)}$ and $\cD^{(0,1)}$, we add them back as part of $\cD$. Now, in a standard application, all of the integrations corresponding to the first line of Eq.~(\ref{eq:sigmaNLO1}) are performed analytically. All of the $1/\epsilon^2$ and $1/\epsilon$ poles cancel and we are left with a completely finite order $\as$ contribution to the cross section. Note that the contribution from the first line beyond just the parton distribution functions is infrared finite, but it is not zero. It forms a significant part of the NLO calculation. In Eq.~(\ref{eq:sigmaNLO1}), we have a product of one infrared finite object times another, each expanded to order $\as$. In the customary NLO calculation, one expands the product to order $\as$ and drops the $\as^2$ term, but one could keep the $\as^2$ term if desired.

We note that if we had wanted to use $\cD(\mu^2)$ for the single purpose of defining subtractions for the hard scattering, $\sket{\rho(\mu^2)}$, we could have used a fixed scale. We need an adjustable scale $\mu_\Ls^2$ to use $\cD(\mu^2)$ in a shower algorithm because the hardness scale of the shower changes as the shower progresses.

\subsection[Properties of the infrared sensitive operator]
{Properties of the infrared sensitive operator}

We can now say a little more about the infrared sensitive operator $\cD(\mu^2)$, without giving a detailed specification. This operator is decomposed into operators $\cD^{(n_\scR,n_\scV)}(\mu^2)$ according to Eqs.~(\ref{eq:Dexpansion}) and (\ref{eq:Dexpansionparts}). Acting on a state $\sket{\{p,f,s,s',c,c'\}_m}$ with $m$ partons, $\cD^{(n_\scR,n_\scV)}(\mu^2)$ produces a state with $m + n_\scR$ partons while adding $n_\scV$ virtual loops. The resulting state can be expanded in basis states for $m + n_\scR$ partons, $\sket{\{\hat p,\hat f,\hat s,\hat s',\hat c,\hat c'\}_{m + n_\scR}}$. There is then an invertible mapping between the new momenta $\{\hat p\}_{m + n_\scR}$ and the starting momenta $\{p\}_{m}$ together with a set of splitting variables $\zeta_p({n_\scR})$. One can choose what this mapping is.

To make this a little more concrete, consider $\cD^{(1,0)}(\mu^2)$ with one parton emitted. There is one term in the emission probability for each final state parton $l \in \{1,\dots, m\}$ and one for each initial state parton $l \in \{\La,\Lb\}$. We think of $l$ as the emitting parton and let the emission probability be singular when $\hat p_{m+1}$ becomes collinear with $\hat p_l$. The emission probabilities are also singular in the limit in which $\hat p_{m+1}$ becomes soft, $\hat p_{m+1} \to 0$. \textsc{Deductor} uses something similar to the Catani-Seymour \cite{CataniSeymour} dipole splitting functions to construct $\cD^{(1,0)}(\mu^2)$.  The splitting variables $\zeta_p$ are taken to be an azimuthal angle $\phi$, a momentum fraction $z$, and a measure of the hardness of the splitting.  In \textsc{Deductor}, the hardness variable is the virtuality of the splitting divided by the energy of the mother parton,
\begin{equation}
\label{eq:Lambdadef}
\Lambda^2 \equiv \frac{2 \hat p_l \cdot \hat p_{m+1}}{2 p_{l}\cdot Q_\LH}\,Q_\LH^2
\;,
\end{equation}
where the vector $Q_\LH$ is defined globally as described in Sec.~\ref{sec:scales}. Alternatively, it can be defined dynamically as described in Appendix \ref{sec:dynamicalscales}.

There is freedom to choose the functional form of $\cD^{(1,0)}(\mu^2)$ away from the limits of soft and collinear emissions. 

There is also freedom to choose the momentum mapping. The simplest case is a splitting of a parton $l$ into two partons $l$ and $m+1$. Then we cannot have $p_l$ be the same as $\hat p_l + \hat p_{m+1}$ with $p_l^2 = \hat p_l^2 = \hat p_{m+1}^2 = 0$, so the momentum mapping has to take a some momentum from the other partons and supply it to $\hat p_l + \hat p_{m+1}$. In \textsc{Deductor}, we use a global mapping, taking a small amount of momentum from each of the other partons. A second possibility  beyond a simple splitting is interference between emission of a gluon $m+1$ from parton $l_\LL$ in the ket state and emission from another parton $l_\LR$ in the bra state. In this case, $\cD^{(1,0)}(\mu^2)$ in \textsc{Deductor} is a linear combination of contributions that use the momentum mappings for a splitting of parton $l_\LL$ and for a splitting of parton $l_\LR$. The coefficients in the linear combination are a ``dipole partitioning'' function $A'$ that is specified in \textsc{Deductor}.

The operator $\cD^{(1,0)}(\mu^2)$, acting on a state $\sket{\{p,f,s,s',c,c'\}_m}$, produces a state with one more parton, parton $m+1$. It is crucial that there be an ultraviolet cutoff for $\hat p_{m+1}$. The cutoff is specified by a parameter that we call $\mu_\Ls^2$. In \textsc{Deductor}, we use $\Lambda^2$ as given in Eq.~(\ref{eq:Lambdadef}) to define the cutoff. In $\cD^{(1,0)}(\mu^2)$, we require
\begin{equation}
\Lambda^2 < \mu_\Ls^2
\;.
\end{equation}

A similar cutoff applies inside the integration for a virtual loop in $\cD^{(0,1)}(\mu^2)$. Defining this cutoff is more involved than we can review here. The calculations are described in Refs.~\cite{NSThreshold, NSThresholdII}.

\section{From the perturbative cross section to a parton shower}
\label{sec:toshower}

In this section, we begin with Eq.~(\ref{eq:NkLOsigmaDDinv}) for the perturbative cross section. We set the scale to $\mu_\scH^2$, which we take to be equal to $Q_\LH^2$. Here $Q_\LH^2$ is defined in Sec.~\ref{sec:scales} to be a fixed vector, although we can use a dynamical definition as described in Appendix \ref{sec:dynamicalscales}. We now seek a more powerful formulation that will enable us to use more general measurement operators $\cO_J$ for which a perturbative expansion of the cross section might contain large logarithms of the generic form $\as^n \log^{2n}(k^2/Q^2)$. Often, a parton shower can approximately sum such logarithms.

As just stated, we set the scale $\mu^2$ in Eq.~(\ref{eq:NkLOsigmaDDinv}) to $\mu_\scH^2$. In our example in which the Born process is Higgs boson production, we might choose $\mu_\scH^2 = m_\scH^2$. This affects the scale at which $\as$ and the parton distribution functions are evaluated. It also affects the upper cutoff on the scale of emissions in the subtraction terms in $\cD^{-1}(\mu_\scH^2)$ and in the parton shower that we will discuss below. However, in the multi-parton matrix elements in $\sket{\rho(\mu_\scH^2)}$, the partons can have any momenta. Parton emissions with small transverse momenta need subtractions, but parton emissions with very large transverse momenta do not need subtractions. If we work at order $\as^k$, then we can have up to $k$ high transverse momentum jets in addition to the Higgs boson in $\sket{\rho(\mu_\scH^2)}$.\footnote{If we are limited to a $k=1$ shower, then Higgs plus one jet is LO in $\sket{\rho(\mu_\scH^2)}$. Then one might want define a lower cutoff on the $P_\LT$ of the jet and use a calculation with $p + p \to H + J$ as the Born process, calculated at NLO. In that case, one has two calculations and one may want to define a procedure to merge them. If a $k = 2$ shower is available, then $\sket{\rho(\mu_\scH^2)}$ can include inclusive Higgs production at NNLO and Higgs plus one jet at NLO. Then merging different calculations is less needed.}

\subsection{Moving the measurement operator}

The first step towards a more general formulation is to interchange the order of the measurement operator $\cO_J$ and the operators $\cD^{-1}(\mu_\scH^2)$ and $\cD(\mu_\scH^2)$. This doesn't change the result, since $\cD\,\cD^{-1} = 1$ and $\cO_J$ commutes with 1. With the $\cO_J$ moved, we have
\begin{equation}
  \label{eq:sigmaNLO3}
  \sigma[J] =
  \sbra{1}\!
  \left[\cF_{\msbarsub}(\mu_\scH^2)\circ\cZ_F(\mu_\scH^2)\right]
  \cO_J\,\cD(\mu_\scH^2)
  \sket{\rho_\scH}
  +\cO(\as^{k+1})
  \;.
\end{equation}
Here we have denoted
\begin{equation}
  \label{eq:rhoH}
  \sket{\rho_\scH}
  = 
  \cD^{-1}(\mu_\scH^2)\sket{\rho(\mu_\scH^2)}
  \;,
\end{equation}
where the product is expanded to $\LN^k \LL \LO$ using Eqs.~(\ref{eq:rhoexpansion}), (\ref{eq:rhoexpansionparts}), (\ref{eq:Dinvexpansion}), and (\ref{eq:tildeDexpansionparts}):
\begin{equation}
  \sket{\rho_\scH} = \sum_{n=0}^k\left[\frac{\as(\mu_\scH^2)}{2\pi}\right]^n
  \sum_{n_\scR = 0}^n \sum_{n_\scV = 0}^n 
  \theta(n_\scR + n_\scV = n)\, \sket{\rho_\scH^{(n_\scR, n_\scV)}}
  +\cO(\as^{k+1})
  \;,
\end{equation}
where
\begin{equation}
\begin{split}
    \sket{\rho_\scH^{(n_\scR, n_\scV)}}
    =
    \sket{\rho^{(n_\scR, n_\scV)}(\mu_\scH^2)}
    - \mathop{\sum_{r=0}^{n_\scR}\sum_{l=0}^{n_\scV}}_{r+l > 0}
    \widetilde\cD^{(r,l)}(\mu_\scH^2)\,
    \sket{\rho^{(n_\scR-r, n_\scV-l)}(\mu_\scH^2)}
    \;.
\end{split}
\end{equation}
These quantities $\sket{\rho_\scH^{(n_\scR, n_\scV)}}$ are finite without dimensional regularization.

\subsection{Introducing shower oriented parton distribution functions}
\label{sec:showerpdfs}

We can do a little more by introducing an operator $\cF(\mu_\scH^2)$ that multiplies by parton distribution functions $f_{a/A}(\eta_\La,\mu^2)\,f_{b/B}(\eta_\Lb,\mu^2)$ and a luminosity factor. However, these are not the five-flavor $\MSbar$ parton distribution functions that we used in $\cF_{\msbarsub}(\mu^2)$. Rather, they are adapted to the choice of the definition for $\cD(\mu^2)$ that we use. 

The shower oriented parton operator $\cF(\mu^2)$ is related to $\cF_{\msbarsub}(\mu^2)$ by factor $\cK(\mu^2)$,
\begin{equation}
\label{eq:F5toF}
\cF_{\msbarsub}(\mu^2) = [\cF(\mu^2)\circ \cK(\mu^2)]
\;.
\end{equation}
This is a rather compact notation, so it is worthwhile to write it in more detail. The left hand side is defined by
\begin{equation}
\label{eq:cFdefMSbar}
\cF_{\msbarsub}(\mu^2)\sket{\{p,f,s,s',c,c'\}_m}
= \frac{f_{a/A}^\msbar(\eta_\La,\mu^2)\,f_{b/B}^\msbar(\eta_\Lb,\mu^2)}
{n_\Lc(a) n_\Ls(a)\, n_\Lc(b) n_\Ls(b)\ 4 p_\La\cdot p_\Lb}\,
\sket{\{p,f,s,s',c,c'\}_m}
\;.
\end{equation}
The right hand side is
\begin{equation}
  \begin{split}
    [\cF(\mu^2)\circ{}& \cK(\mu^2)]\sket{\{p,f,s,s',c,c'\}_m}
    \\
    ={}& \sum_{a',b'} \int_0^1\!\frac{dz_\La}{z_\La}
    \int_0^1\!\frac{dz_\Lb}{z_\Lb}\ 
    \frac{f_{a'/A}(\eta_\La/z_\La,\mu^2)\,f_{b'/B}(\eta_\Lb/z_\Lb,\mu^2)}
    {n_\Lc(a) n_\Ls(a)\, n_\Lc(b) n_\Ls(b)\ 4 p_\La\cdot p_\Lb}
    \\
    &\qquad\times
    K^{(\La)}_{a,a'}(z_\La,\mu^2,\{p,f\}_m)\,
    K^{(\Lb)}_{b,b'}(z_\Lb,\mu^2,\{p,f\}_m)\,
    \sket{\{p,f,s,s',c,c'\}_m}
    \;.
  \end{split}
\end{equation}
Here we take the kernel $K$ to be a product, so that each of the two parton distributions is transformed separately. We allow each kernel to depend on the momentum and flavor variables of the parton state to which $[\cF(\mu^2)\circ \cK(\mu^2)]$ is applied. The kernels each have a perturbative expansion beginning with
\begin{equation}
K^{(\La)}_{a,a'}(z,\mu^2,\{p,f\}_m) = 
\delta_{a,a'} \delta(1-z)
+ \frac{\as(\mu^2)}{2\pi}\,K^{(\La,1)}_{a,a'}(z,\mu^2,\{p,f\}_m)
+ \cO(\as^2)
\;.
\end{equation}
The choice of $\cK(\mu^2)$ defines the shower-oriented parton distribution functions. The evolution of these parton distribution functions needs to be matched to the parton splitting functions introduced in the following sections. In particular, the choice of $\cK(\mu^2)$ is largely determined by the definition of the cutoff $\mu_\Ls^2$ that we use  for the shower. We provide an example in Appendix \ref{sec:K}.

The parton operators $\cF_{\msbarsub}(\mu^2)$ and $\cF(\mu^2)$ obey evolution equations 
\begin{equation}
\begin{split}
\label{eq:Fevolution}
\mu^2 \frac{d}{d\mu^2}\,\cF(\mu^2)
={}& [\cF(\mu^2)\circ \cP(\mu^2)]
\;,
\\
\mu^2 \frac{d}{d\mu^2}\,\cF_{\msbarsub}(\mu^2)
={}& [\cF_{\msbarsub}(\mu^2)\circ \cP_{\msbarsub}(\mu^2)]
\;.
\end{split}
\end{equation}
Using Eq.~(\ref{eq:F5toF}), we see that the evolution kernels are related by
\begin{equation}
\cP(\mu^2) = [\cK(\mu^2)\circ \cP_{\msbarsub}(\mu^2)\circ \cK^{-1}(\mu^2)]
- \left[\left(\mu^2 \frac{d}{d\mu^2}\,\cK(\mu^2)\right)
\circ \cK^{-1}(\mu^2)\right]
\;.
\end{equation}

With the transformation from $\cF_{\msbarsub}(\mu^2)$ to $\cF(\mu^2)$, we write
\begin{equation}
\label{eq:F5ZFtoFKZF}
[\cF_{\msbarsub}(\mu^2)\circ \cZ_F(\mu^2)] = 
[\cF(\mu^2)\circ \cK(\mu^2)\circ \cZ_F(\mu^2)]
\;.
\end{equation}
Thus our cross section is
\begin{equation}
  \label{eq:sigmaNLO4}
  \sigma[J] =
  \sbra{1}\!
  \left[\cF(\mu_\scH^2)\circ \cK(\mu_\scH^2)\circ \cZ_F(\mu_\scH^2)\right]
  \cO_J\,\cD(\mu_\scH^2)
  \sket{\rho_\scH}
  +\cO(\as^{k+1})
  \;.
\end{equation}
We will introduce $\cF(\mu^2)$ into another place in the formalism shortly.

\subsection{Changing the scale of the subtraction operators}

Next, we would like to change the scale of the operators in Eq.~(\ref{eq:sigmaNLO4}) from a large scale $\mu_\scH^2$ to something smaller. We let $\mu_\scI^2$ be an ``intermediate'' size scale that is much smaller than the scale $Q^2[J]$ associated with the operator $\cO_J$ but is nevertheless large compared to $1 \GeV^2$ and is certainly large enough to allow the use of perturbation theory in $\as(\mu_\scI^2)$. We can change the parton factor to be evaluated at scale $\mu_\scI^2$ because this factor is a renormalization group invariant:
\begin{equation}
\left[\cF(\mu_\scH^2)\circ \cK(\mu_\scH^2)\circ
\cZ_F(\mu_\scH^2)\right]
=
\left[\cF(\mu_\scI^2)\circ \cK(\mu_\scI^2)\circ
\cZ_F(\mu_\scI^2)\right]
\;.
\end{equation}

The operator $\cD(\mu^2)$ is not invariant under changes of scale. However, we can write
\begin{equation}
\cD(\mu_1^2) = \cD(\mu_2^2)\,
\cU_{\rm pert}(\mu_2^2,\mu_1^2)
\;,
\end{equation}
where 
\begin{equation}
\label{eq:Upert}
\cU_{\rm pert}(\mu_2^2,\mu_1^2)
=\cD^{-1}(\mu_2^2)\,\cD(\mu_1^2)\,
\;.
\end{equation}
Here we note that $\cD(\mu_1^2)$ generates $1/\epsilon$ poles and infrared singularities, but $\cD^{-1}(\mu_2^2)$ provides the proper subtractions to remove the poles and infrared singularities when we expand the product of operators to a fixed order of perturbation theory. Thus we can evaluate $\cU_{\rm pert}(\mu_2^2,\mu_1^2)$ in four dimensions instead of $4 - 2 \epsilon$ dimensions. The perturbative evolution operator $\cU_{\rm pert}(\mu^2,\mu^{\prime\,2})$ obeys the differential equation
\begin{equation}
\label{eq:Upertevolution}
\mu^2 \frac{d}{d \mu^2}\,\cU_{\rm pert}(\mu^2,\mu^{\prime\,2})
= - S_{\rm pert}(\mu^2)\, \cU_{\rm pert}(\mu^2,\mu^{\prime\,2})
\;,
\end{equation}
where
\begin{equation}
\label{eq:Spert}
S_{\rm pert}(\mu^2) =  \cD^{-1}(\mu^2)\, 
\mu^2\frac{d}{d \mu^2} \cD(\mu^2)
\;.
\end{equation}
Since $\cU_{\rm pert}(\mu_2^2,\mu_1^2)$ is infrared finite, so is $S_{\rm pert}(\mu^2)$. We can write the solution of Eq.~(\ref{eq:Upertevolution}) as
\begin{equation}
\cU_{\rm pert}(\mu^2,\mu^{\prime\,2})
= \mathbb{T} \exp\!\left(
\int_{\mu^2}^{\mu^{\prime\,2}}\!\frac{d\mu^2}{\mu^2}\,S_{\rm pert}(\mu^2)
\right)
\;,
\end{equation}
where $\mathbb{T}$ indicates $\mu^2$ ordering of the exponential with smaller $\mu^2$ to the left. Working to order $\as$ with use of Eqs.~(\ref{eq:Dexpansion}) and (\ref{eq:Dexpansionparts}), we have
\begin{equation}
\label{eq:Spertexpansion}
S_{\rm pert}(\mu^2)
= \frac{\as(\mu^2)}{2\pi}\,\cS^{(1,0)}_{\rm pert}(\mu^2)
+ \frac{\as(\mu^2)}{2\pi}\,\cS^{(0,1)}_{\rm pert}(\mu^2)
+\cO(\as^2)
\;,
\end{equation}
where $\as(\mu^2)$ is the running coupling in the four dimensional theory and 
\begin{equation}
\begin{split}
\label{eq:Spertcoefficients}
\cS^{(1,0)}_{\rm pert}(\mu^2) ={}& 
\mu^2\frac{d}{d \mu^2}\,\cD^{(1,0)}(\mu^2)
\;,
\\
\cS^{(0,1)}_{\rm pert}(\mu^2) ={}& 
\mu^2\frac{d}{d \mu^2}\,\cD^{(0,1)}(\mu^2)
\;.
\end{split}
\end{equation}
Recall that $\mu^2 = \mu_\Ls^2$. Thus $\cS^{(1,0)}_{\rm pert}(\mu^2)$ is the derivative of (approximated) real emission graphs with respect to the ultraviolet cutoff that we impose. Similarly, $\cS^{(0,1)}_{\rm pert}(\mu^2)$ is the derivative of approximated one loop virtual graphs with respect to the ultraviolet cutoff. The subscript ``pert'' emphasizes that only perturbative Feynman diagrams are used to obtain $\cS^{(1,0)}_{\rm pert}(\mu^2)$ and $\cS^{(0,1)}_{\rm pert}(\mu^2)$.

With these changes, we have
\begin{equation}
  \begin{split}
    \label{eq:sigmaU1}
    \sigma[J] ={}& 
    \sbra{1}\! \left[\cF(\mu_\scI^2)\circ \cK(\mu_\scI^2)\circ
      \cZ_F(\mu_\scI^2)
    \right]
    \cO_J\,\cD(\mu_\scI^2)\, 
    \cU_{\rm pert}(\mu_\scI^2,\mu_\scH^2)
    \sket{\rho_\scH}
    +\cO(\as^{k+1})
    \;.
  \end{split}
\end{equation}

Now we note that very soft or collinear splittings at scales much smaller than $Q[J]^2$ are not resolved by the measurement operator $\cO_J$. The operator $\cD(\mu_\scI^2)$ generates splittings at scales $\mu_\scI^2$ and smaller. Since we have chosen $\mu_\scI^2 \ll Q[J]^2$, the operator $\cO_J$ commutes with $\cD(\mu_\scI^2)$ to a good approximation, with an error of order $\mu_\scI^2/Q[J]^2$. Thus Eq.~(\ref{eq:sigmaNLO4}) can be written as
\begin{equation}
  \begin{split}
    \label{eq:sigmaU2}
    \sigma[J] ={}& 
    \sbra{1}\!\left[\cF(\mu_\scI^2)\circ \cK(\mu_\scI^2)\circ
      \cZ_F(\mu_\scI^2)
    \right]
    \cD(\mu_\scI^2)\, \cO_J\,
    \cU_{\rm pert}(\mu_\scI^2,\mu_\scH^2)
    \sket{\rho_\scH}
    \\&
    +\cO(\as^{k+1}) +\cO(\mu_\scI^2/Q[J]^2)
    \;.
  \end{split}
\end{equation}

\subsection[The inclusive infrared finite operator $\cV(\mu^2)$]
{The inclusive infrared finite operator \boldmath{$\cV(\mu^2)$}
\label{sec:cVdef}}

We now introduce an operator $\cX(\mu^2)$ defined by 
\begin{equation}
\label{eq:X}
\cX(\mu^2) = 
\left[\cF(\mu^2)\circ \cK(\mu^2)\circ
\cZ_F(\mu^2)
\right]
\cD(\mu^2)\,
\cF^{-1}(\mu^2)
\;.
\end{equation}
Using $\cX(\mu^2)$, Eq.~(\ref{eq:sigmaU2}) is more compact:
\begin{equation}
\begin{split}
\label{eq:sigmaU3}
\sigma[J] ={}& 
\sbra{1} \cX(\mu_\scI^2)\,\cF(\mu_\scI^2) \cO_J\,
\cU_{\rm pert}(\mu_\scI^2,\mu_\scH^2)
\sket{\rho_\scH}
+\cO(\as^{k+1}) +\cO(\mu_\scI^2/Q[J]^2)
\;.
\end{split}
\end{equation}
The operator $\cX(\mu^2)$ involves parton distribution functions and purely perturbative operators. If we evaluate the perturbative operators at order zero, we get simply $\cF(\mu^2)\cF^{-1}(\mu^2)$. Thus
\begin{equation}
\cX(\mu^2) = 1 + \cO(\as)
\;.
\end{equation}
The operator $\cX(\mu^2)$, when expanded in powers of $\as$, creates partons, up to $k$ partons at order $\as^k$. 

The operator $\cX(\mu^2)$ is infrared sensitive. When we apply $\cX(\mu^2)$ to a state $\sket{\{p,f,s,s',c,c'\}_m}$, we get a state $\cX(\mu^2)\sket{\{p,f,s,s',c,c'\}_m}$ containing poles $1/\epsilon$ and singularities when the partons that $\cX(\mu^2)$ creates become soft or collinear with other partons or with each other. In that sense, $\cX(\mu^2)$ is like $\cD(\mu^2)$. However, $\cX(\mu^2)$ contains the parton factor $\left[\cF(\mu^2)\circ \cK(\mu^2)\circ \cZ_F(\mu^2)\right]$. This factor gives $\cX(\mu^2)$ a property not shared by $\cD(\mu^2)$. If we integrate over the momenta of the partons created by $\cX(\mu^2)$ and sum over their colors and flavors by forming the inclusive sum $\sbra{1} \cX(\mu^2) \sket{\{p,f,s,s',c,c'\}_m}$, then the singularities cancel and we obtain a finite result.

In fact, we need to ensure that $\sbra{1} \cX(\mu^2) \sket{\{p,f,s,s',c,c'\}_m}$ is not only finite after dimensional regularization is removed but that it vanishes in the limit $\mu^2 \to 0$. For example, if this quantity arises from an integration
\begin{equation}
\sbra{1} \cX(\mu^2) \sket{\{p,f,s,s',c,c'\}_m} = \int_0^{\mu^2}\!\frac{dk^2}{k^2}\
G(k^2)
\;,
\end{equation}
then, with subtractions included, $G(k^2)$ needs to be a smooth function that is well enough behaved for $k^2 \to 0$ that the integral is convergent. This property is needed later in Eq.~(\ref{eq:cVmuf}).

Suppose for a moment that we worked in a modified theory, denoted by subscripts M,  in which partons carried only momenta and flavors, but not color and spin. Then from the inclusive sum $\sbra{1} \cX_\LM(\mu^2) \sket{\{p,f\}_m}$ we could define another operator $\cV_\LM(\mu^2)$ that leaves the number of partons, their momenta and flavors unchanged:
\begin{equation}
\label{eq:Vproperty}
\cV_\LM(\mu^2)\sket{\{p,f\}_m} = \lambda(\{p,f\}_m) \sket{\{p,f\}_m}
\;.
\end{equation}
Then $\lambda(\{p,f\}_m,\mu^2) = \sbra{1} \cV_\LM(\mu^2)\sket{\{p,f\}_m}$. We define $\cV_\LM(\mu^2)$ by Eq.~(\ref{eq:Vproperty}) and
\begin{equation}
\label{eq:XtoV0}
\sbra{1} \cV_\LM(\mu^2)\sket{\{p,f\}_m} =
\sbra{1} \cX_\LM(\mu^2)\sket{\{p,f\}_m}
\;.
\end{equation}

Now return to QCD. With spin and color, we can define an operator $\cV(\mu^2)$ that satisfies $\sbra{1} \cV(\mu^2) = \sbra{1} \cX(\mu^2)$. However, its structure is more complex. The operator $\cX(\mu^2)$ can be expanded in powers of $\as$:
\begin{equation}
\label{eq:Xexpansion}
\cX(\mu^2) = 1 
+ \sum_{n = 1}^k \left[\frac{\as(\mu^2)}{2\pi}\right]^n
\cX^{(n)}(\mu^2)
+ \cO(\as^{k+1})
\;.
\end{equation}
The order $n$ contribution, $\cX^{(n)}(\mu^2)$, is a sum of infrared sensitive operators,
\begin{equation}
\label{eq:Xexpansionparts}
\cX^{(n)}(\mu^2) = \sum_{n_\scR = 0}^n \sum_{n_\scV = 0}^n 
\theta(n_\scR + n_\scV = n)\,\cX^{(n_\scR,n_\scV)}(\mu^2)
\;.
\end{equation}
Acting on a state $\sket{\{p,f,s,s',c,c'\}_m}$ with $m$ final state partons, $\cX^{(n_\scR,n_\scV)}(\mu^2)$ produces a state with $m + n_\scR$ final state partons with momenta and flavors $\{\hat p, \hat f\}_{m + n_\scR}$. There are integrations over the loop momenta $\{\ell_1, \dots, \ell_{n_\scV}\}$ of $n_\scV$ virtual loops. 

We need to understand the color and spin structure of $\cX^{(n_\scR,n_\scV)}(\mu^2)$. Suppose that we have constructed a basis of operators that act on the quantum spin$\otimes$color space and create a quantum spin$\otimes$color state for $n_\scR$ more partons. We label the basis operators by an index $i$. A convenient choice would be 
\begin{equation}
i = \{m,\{\hat s,\hat c\}_{m+n_\scR},\{s,c\}_{m}\}\;.
\end{equation}
Then we could let
\begin{equation}
\sigma_i^{(n_\scR)}\ket{\{s',c'\}_{m'}} =
\begin{cases}
\ket{\{\hat s,\hat c\}_{m+n_\scR}} & 
m = m'\ {\rm \&}\ \{s,c\}_{m} = \{s',c'\}_{m}\\
0 & {\rm otherwise}
\end{cases}
\;.
\end{equation}
Using these basis operators, we can expand $\cX^{(n_\scR,n_\scV)}(\mu^2)$ as
\begin{equation}
\cX^{(n_\scR,n_\scV)}(\mu^2) = \sum_{i,j} 
\cX^{(n_\scR,n_\scV)}_{i,j}(\mu^2)\ 
\sigma_i^{(n_\scR)}\otimes \sigma_j^{(n_\scR)\dagger}
\;.
\end{equation}
Here $\cX^{(n_\scR,n_\scV)}_{i,j}(\mu^2)$ is still an operator on the momentum and flavor part of the statistical space, which has basis vectors $\sket{\{p,f\}_m}$. In the case $n_\scR = 0$, this operator adds no partons and leaves the parton momenta and flavors $\{p,f\}_m$ unchanged.

Now we wish to define another operator $\cV(\mu^2)$ with an expansion 
\begin{equation}
\label{eq:Vexpansion}
\cV(\mu^2) = 1 
+ \sum_{n = 1}^k \left[\frac{\as(\mu^2)}{2\pi}\right]^n
\cV^{(n)}(\mu^2)
+ \cO(\as^{k+1})
\;.
\end{equation}
The order $n$ contribution is to add no partons and leave the parton momenta and flavors $\{p,f\}_m$ unchanged, but it still can be a non-trivial operator on the spin$\otimes$color space
\begin{equation}
\cV^{(n)}(\mu^2) = \sum_{i,j} 
\cV^{(n)}_{i,j}(\mu^2)\ 
\sigma_i^{(0)}\otimes \sigma_j^{(0)\dagger}
\;.
\end{equation}
Here $\cV^{(n)}_{i,j}(\mu^2)$ is still an operator on the momentum and flavor part of the statistical space. We want $\cV(\mu^2)$ to be related to $\cX(\mu^2)$ by
\begin{equation}
\label{eq:XtoV}
\sbra{1} \cV(\mu^2) = \sbra{1} \cX(\mu^2)
\;.
\end{equation}
Thus we want
\begin{equation}
  \begin{split}
    \sbra{1}
    \sum_{i,j} 
    \cV^{(n)}_{i,j}{}&(\mu^2)\ 
    \sigma_i^{(0)}\otimes \sigma_j^{(0)\dagger}
    \sket{\{p,f,s,s',c,c'\}_m}
    \\ &=
    \sum_{n_\scR = 0}^n
    \sum_{i,j} 
    \sbra{1}
    \cX^{(n_\scR,n - n_\scR)}_{i,j}(\mu^2)\ 
    \sigma_i^{(n_\scR)}\otimes \sigma_j^{(n_\scR)\dagger}
    \sket{\{p,f,s,s',c,c'\}_m}
    \;.
  \end{split}
\end{equation}
This is the same as
\begin{equation}
  \begin{split}
    \sum_{i,j} &
    \bra{\{s',c'\}_m}\sigma_j^{(0)\dagger}\sigma_i^{(0)}\ket{\{s,c\}_m}
    \sbra{1}
    \cV^{(n)}_{i,j}(\mu^2)\ 
    \sket{\{p,f\}_m}
    \\ &=
    \sum_{n_\scR = 0}^n
    \sum_{i,j}
    \bra{\{s',c'\}_m}\sigma_j^{(n_\scR)\dagger}\sigma_i^{(n_\scR)}\ket{\{s,c\}_m}
    \sbra{1}
    \cX^{(n_\scR,n - n_\scR)}_{i,j}(\mu^2)\ 
    \sket{\{p,f\}_m}
    \;.
  \end{split}
\end{equation}
This needs to work for any choice of $m$-parton spin$\otimes$color states $\bra{\{s',c'\}_m}$ and $\ket{\{s,c\}_m}$, so we need an identity of spin$\otimes$color operators, 
\begin{equation}
  \begin{split}
    \label{eq:Vcolorspinidentity}
    \sum_{i,j}
    \sigma_j^{(0)\dagger}\sigma_i^{(0)}
    \sbra{1}&
    \cV^{(n)}_{i,j}(\mu^2)\ 
    \sket{\{p,f\}_m}
    \\ ={}&
    \sum_{n_\scR = 0}^n
    \sum_{i,j}
    \sigma_j^{(n_\scR)\dagger}\sigma_i^{(n_\scR)}
    \sbra{1}
    \cX^{(n_\scR,n - n_\scR)}_{i,j}(\mu^2)\ 
    \sket{\{p,f\}_m}
    \;.
  \end{split}
\end{equation}
The right hand side of Eq.~(\ref{eq:Vcolorspinidentity}) is an operator on the spin$\otimes$color space for $m$ final state partons. On the left hand side, the operators $\sigma_i^{(0)}$ form a basis for this space of operators, as do the operators $\sigma_j^{(0)\dagger}$, so the operators $\sigma_j^{(0)\dagger}\sigma_i^{(0)}$ span this space and are, in fact, over-complete. That is, one can always find coefficients $\sbra{1} \cV^{(n)}_{i,j}(\mu^2)\ \sket{\{p,f\}_m}$ so that we match the operator on the right hand side. However, the choice is not unique. At order $n = 1$, we have made a simple choice in \textsc{Deductor}. It is beyond our scope here to investigate what choices might be best at NLO, $n = 2$.

Since $\sbra{1} \cX(\mu^2)$ is infrared finite, Eq.~(\ref{eq:XtoV}) tells us that $\cV(\mu^2)$ is infrared finite.

\subsection{A more sophisticated shower evolution operator}

Using Eq.~(\ref{eq:XtoV}), Eq.~(\ref{eq:sigmaU3}) becomes
\begin{equation}
\begin{split}
\label{eq:sigmaU4}
\sigma[J] ={}& 
\sbra{1} \cV(\mu_\scI^2)\,\cF(\mu_\scI^2) \cO_J\,
\cU_{\rm pert}(\mu_\scI^2,\mu_\scH^2)
\sket{\rho_\scH}
+\cO(\as^{k+1}) +\cO(\mu_\scI^2/Q[J]^2)
\;.
\end{split}
\end{equation}
Since $\cV(\mu_\scI^2)\,\cF(\mu_\scI^2)$ does not change the number, momenta, or flavors of partons, it commutes with $\cO_J$. Thus
\begin{equation}
\begin{split}
\label{eq:sigmaU5}
\sigma[J] ={}& 
\sbra{1} \cO_J\, \cV(\mu_\scI^2)\,\cF(\mu_\scI^2)\,
\cU_{\rm pert}(\mu_\scI^2,\mu_\scH^2)
\sket{\rho_\scH}
+\cO(\as^{k+1}) +\cO(\mu_\scI^2/Q[J]^2)
\;.
\end{split}
\end{equation}

Now we can define the shower evolution operator that we need,
\begin{equation}
\label{eq:Udef}
\cU(\mu_2^2,\mu_1^2) = 
\cV(\mu_2^2)\,\cF(\mu_2^2)\,
\cU_{\rm pert}(\mu_2^2,\mu_1^2)
\cF^{-1}(\mu_1^2)\,\cV^{-1}(\mu_1^2)
\,.
\end{equation}
With this definition, the cross section is
\begin{equation}
\begin{split}
\label{eq:sigmaU6}
\sigma[J] ={}& 
\sbra{1} \cO_J\, 
\cU(\mu_\scI^2,\mu_\scH^2)\,
\cV(\mu_\scH^2)\,\cF(\mu_\scH^2)
\sket{\rho_\scH}
+\cO(\as^{k+1}) +\cO(\mu_\scI^2/Q[J]^2)
\;.
\end{split}
\end{equation}
This moves $\cF$ next to $\sket{\rho_\scH}$ so that at the hard interaction we have the proper factors to make a cross section. It also moves $\cV$ next to $\sket{\rho_\scH}$. We will see later what the consequences of this are.

In Eq.~(\ref{eq:sigmaU6}), we use a scale $\mu_\scI^2$ that was left undefined except that it should be small compared to $Q^2[J]$ (which was the scale of $\cO_J$) and should be large enough to allow the use of perturbation theory with coupling $\as(\mu_\scI^2)$. Our cross section is independent of the value of $\mu_\scI^2$. Let us now fix on a standard choice near the lower end of this range. We take $\mu_\scI^2 \to \mu_\Lf^2$, where $\mu_\Lf^2$ is on the order of $1 \GeV^2$. Then
\begin{equation}
\begin{split}
\label{eq:sigmaU7}
\sigma[J] ={}& 
\sbra{1} \cO_J\, 
\cU(\mu_\Lf^2,\mu_\scH^2)\,
\cV(\mu_\scH^2)\,\cF(\mu_\scH^2)
\sket{\rho_\scH}
+\cO(\as^{k+1}) +\cO(\mu_\Lf^2/Q[J]^2)
\;.
\end{split}
\end{equation}

We can write $\cU(\mu_2^2,\mu_1^2)$ in a simpler form. We note that, using Eqs.~(\ref{eq:Upert}) and (\ref{eq:X}),
\begin{equation}
\begin{split}
\cU(\mu_2^2,\mu_1^2) ={}& 
\cV(\mu_2^2)\,\cF(\mu_2^2)\,
\cD^{-1}(\mu_2^2)\,\cD(\mu_1^2)\,
\cF^{-1}(\mu_1^2)\,\cV^{-1}(\mu_1^2)
\\
={}& \cV(\mu_2^2)\,\cX^{-1}(\mu_2^2)\,
\left[\cF(\mu_2^2)\circ \cK(\mu_2^2)\circ
\cZ_F(\mu_2^2)
\right]
\\&\times
\left[\cF(\mu_1^2)\circ \cK(\mu_1^2)\circ
\cZ_F(\mu_1^2)
\right]^{-1}
\cX(\mu_1^2)\,
\cV^{-1}(\mu_1)
\;.
\end{split}
\end{equation}
Since the operator $\left[\cF(\mu^2)\circ \cK(\mu^2)\circ \cZ_F(\mu^2) \right]$ is independent of scale, this is
\begin{equation}
\begin{split}
\label{eq:Udef2}
\cU(\mu_2^2,\mu_1^2) ={}& 
\cV(\mu_2^2)\,\cX^{-1}(\mu_2^2)\,
\cX(\mu_1^2)\,
\cV^{-1}(\mu_1)
\;.
\end{split}
\end{equation}

\subsection[Probability preservation in $\cU(\mu_2^2,\mu_1^2)$]
{\label{sec:probpreserv}Probability preservation in \boldmath{$\cU(\mu_2^2,\mu_1^2)$}}

The operator $\cU(\mu_2^2,\mu_1^2)$ has an important property, which we now derive. 
From Eq.~(\ref{eq:Udef2}), we have
\begin{equation}
\sbra{1}\,\cU(\mu_2^2,\mu_1^2)
= 
\sbra{1}\,\cV(\mu_2^2)\,\cX^{-1}(\mu_2^2)\,
\cX(\mu_1^2)\,\cV^{-1}(\mu_1^2)
\;.
\end{equation}
Then using Eq.~(\ref{eq:XtoV}) twice, we have
\begin{equation}
\label{eq:probabilitypreserving}
\sbra{1}\,\cU(\mu_2^2,\mu_1^2) = \sbra{1}
\;.
\end{equation}

Multiplying any statistical state $\sket{\rho}$ by $\sbra{1}$ gives the total probability associated with that state. Thus Eq.~(\ref{eq:probabilitypreserving}) says that shower evolution as represented in $\cU(\mu_2^2,\mu_1^2)$ is probability preserving. Current parton shower algorithms are typically constructed to have this property. Here probability preservation is a derived property. 

\subsection{Factorization}

Let $Q^2[J]$ be the smallest scale associated with the measurement operator $\cO_J$ in Eq.~(\ref{eq:sigmaU7}), as discussed in Sec.~\ref{sec:IRsafety}. If $Q^2[J]$ us close to the scale $\mu_\scH^2$ of the hard process with which we start the shower, then measuring $\sigma[J]$ does not make use of the full power of a parton shower. Suppose now that $Q^2[J] \gg 1 \GeV^2$ but that $Q^2[J] \ll \mu_\scH^2$. Then perturbation theory for $\sigma[J]$ is still applicable, but there may be large logarithms, $\log(\mu_\scH^2/Q^2[J])$ in the perturbative expansion of $\sigma[J]$. In many cases, a parton shower is useful for summing such logarithms.  It takes a dedicated analysis to show that a given parton shower algorithm does sum the logarithms associated with a given operator $\cO_J$, but there is at least a chance that if we use a parton shower we will do better than if we simply use fixed order perturbation theory. Thus we consider this sort of measurement operator and examine how factorization works when $Q^2[J] \ll \mu_\scH^2$.

We argued in Sec.~\ref{sec:IRsafety} that shower splittings at scale $\mu^2$ can change the measurement by a fraction $\mu^2/Q^2[J]$. We can neglect these modifications as long as $\mu^2$ is small enough, say
\begin{equation}
\mu^2 < \epsilon_\Ls Q^2[J]
\;.
\end{equation}
We want $\epsilon_\Ls$ to be small enough that we can regard fractional errors of order $\epsilon_\Ls$ as negligible. However, we may want $\log(1/\epsilon_\Ls)$ not to be large.

We can use our knowledge of the scale of $\cO_J$ by writing
\begin{equation}
\cU(\mu_\Lf^2,\mu^2) = 
\cU(\mu_\Lf^2,\epsilon_\Ls\, Q^2[J])\,
\cU(\epsilon_\Ls\, Q^2[J],\mu_\scH^2)
\;.
\end{equation}
Then writing 
\begin{equation}
\label{eq:factorizationOJ}
\cO_J\,\cU(\mu_\Lf^2,\epsilon_\Ls\, Q^2[J])
\approx
\cU(\mu_\Lf^2,\epsilon_\Ls\, Q^2[J])\,\cO_J
\end{equation}
results in a negligible error. With this substitution, Eq.~(\ref{eq:sigmaU7}) becomes
\begin{equation}
\begin{split}
\label{eq:sigmaU7mod}
\sigma[J] ={}& 
\sbra{1}  
\cU(\mu_\Lf^2,\epsilon_\Ls\, Q^2[J])\,
\cO_J\,
\cU(\epsilon_\Ls\, Q^2[J],\mu_\scH^2)\,
\cV(\mu_\scH^2)\,\cF(\mu_\scH^2)
\sket{\rho_\scH}
+\cO(\as^{k+1})+\cO(\epsilon_\Ls)
\;.
\end{split}
\end{equation}
Now, factorization for the cross section measured by $\cO_J$ requires that splittings at scales smaller than $\epsilon_\Ls\, Q^2[J]$ not affect the cross section. Thus we need
\begin{equation}
\begin{split}
\label{eq:sigmaU9}
\sigma[J] ={}& 
\sbra{1}  
\cO_J\,
\cU(\epsilon_\Ls\, Q^2[J],\mu_\scH^2)\,
\cV(\mu_\scH^2)\,\cF(\mu_\scH^2)
\sket{\rho_\scH}
+\cO(\as^{k+1})+\cO(\epsilon_\Ls)
\;.
\end{split}
\end{equation}
This follows by using Eq.~(\ref{eq:probabilitypreserving}) to obtain $\sbra{1}  \cU(\mu_\Lf^2,\epsilon_\Ls\, Q^2[J]) = \sbra{1}$.

We should emphasize that in order to measure the cross section corresponding to the infrared safe operator $\cO_J$ with scale $Q^2[J]$, it is not necessary to cut off the shower at scale $\epsilon_\Ls\, Q^2[J]$ as in Eq.~(\ref{eq:sigmaU9}). Rather, one simply runs the shower down to $\mu_\Lf^2$ and measures $\cO_J$ on the final state produced by the full shower, as in Eq.~(\ref{eq:sigmaU7}). When we do that, we are setting $\epsilon_\Ls = \mu_\Lf^2/Q^2[J]$, so the error estimate $\cO(\epsilon_\Ls)$ becomes $\cO(\mu_\Lf^2/Q^2[J])$.

\subsection{The shower evolution equation}

Using its definition Eq.~(\ref{eq:Udef}), we see that the shower evolution operator $\cU(\mu^2,\mu^{\prime\,2})$ obeys an evolution equation of the form
\begin{equation}
\label{eq:Uevolution}
\mu^2\frac{d}{d\mu^2}\,\cU(\mu^2,\mu^{\prime\,2})
= -\cS(\mu^2)\,\cU(\mu^2,\mu^{\prime\,2})
\;.
\end{equation}
Thus
\begin{equation}
\label{eq:Vexponential}
\cU(\mu^2,\mu^{\prime\,2})
=\mathbb{T} \exp\!\left(
\int_{\mu^2}^{\mu^{\prime\,2}}\!\frac{d\mu^2}{\mu^2}\,\cS(\mu^2)
\right)
\;.
\end{equation}
Since, according to Eq.~(\ref{eq:probabilitypreserving}), $\sbra{1} \cU(\mu^2,\mu^{\prime\,2}) = \sbra{1}$, we have
\begin{equation}
\sbra{1}\cS(\mu^2) = 0
\;.
\end{equation}

Using Eqs.~(\ref{eq:Udef}) and (\ref{eq:Upertevolution}), we see that the shower generator $\cS$ in Eq.~(\ref{eq:Uevolution}) is
\begin{equation}
\begin{split}
\label{eq:Sdef}
\cS(\mu^2) ={}& 
\cV(\mu^2)\,\cF(\mu^2)\,
\cS_{\rm pert}(\mu^2)\,
\cF^{-1}(\mu^{2})\,
\cV^{-1}(\mu^{2})
\\&
- \left(\mu^2\frac{d}{d\mu^2}\,\cV(\mu^2)\,\cF(\mu^2)
\right)
\cF^{-1}(\mu^{2})\,
\cV^{-1}(\mu^{2})
\\ ={}&
\cV(\mu^2)\,\cF(\mu^2)\,
\cS_{\rm pert}(\mu^2)\,
\cF^{-1}(\mu^{2})\,
\cV^{-1}(\mu^{2})
\\&
-
\cV(\mu^{2})\left(\mu^2\frac{d}{d\mu^2}\cF(\mu^2)
\right)
\cF^{-1}(\mu^2)
\cV^{-1}(\mu^{2})
- \left(\mu^2\frac{d}{d\mu^2}\,\cV(\mu^2)\right)
\cV^{-1}(\mu^2)
\;.
\end{split}
\end{equation}
It is convenient to define
\begin{equation}
\label{eq:SVdef}
\cS_\cV(\mu^2) =
\cV^{-1}(\mu^2)\,
\mu^2\frac{d}{d\mu^2}\,\cV(\mu^2)
\;.
\end{equation}
Also, we can use Eq.~(\ref{eq:Fevolution}) for the evolution of $\cF(\mu^2)$ and we can note that since $\cV(\mu^2)$ does not change the number of partons or their momenta or flavors, $\cV(\mu^2)$ commutes with $\cF(\mu^2)$. Then
\begin{equation}
\begin{split}
\label{eq:Sdef3}
\cS(\mu^2) ={}& \cV(\mu^2)\,\cF(\mu^2)\,
\cS_{\rm pert}(\mu^2)\,
\cF^{-1}(\mu^{2})\,
\cV^{-1}(\mu^{2})
\\&
-
[\cF(\mu^2)\circ \cP(\mu^2)]\cF^{-1}(\mu^2)
- \cV(\mu^2)\,\cS_\cV(\mu^2)\,\cV^{-1}(\mu^2)
\;.
\end{split}
\end{equation}
The operator $\cV(\mu^2)$ here has a perturbative expansion beginning with
\begin{equation}
\cV(\mu^2) = 1 + \frac{\as(\mu^2)}{2\pi}\,
\cV^{(1)}(\mu^2)
+ \cdots
\;.
\end{equation}
Then also
\begin{equation}
\cS_\cV(\mu^2) = \frac{\as(\mu^2)}{2\pi}\,
\cS_\cV^{(1)}(\mu^2)
+ \cdots
\;.
\end{equation}
A sensible procedure for determining $\cS(\mu^2)$ is to expand it perturbatively to whatever order is known, {\em e.g.} order $\as^k$,
\begin{equation}
\cS(\mu^2) = \frac{\as(\mu^2)}{2\pi}\,
\cS^{(1)}(\mu^2)
+ \cdots
\;.
\end{equation}

It is of interest to see how this works out at order $\as$. Since $\cS_{\rm pert}(\mu^2)$ and $\cS_\cV$ are already order $\as$, we can simply replace $\cV(\mu^2)$ by 1 in the first and third terms of Eq.~(\ref{eq:Sdef3}). Then we have
\begin{equation}
\begin{split}
\label{eq:S1storder1}
\cS^{(1)}(\mu^2) ={}& \cF(\mu^2)\,
\cS_{\rm pert}^{(1)}(\mu^2)\,
\cF^{-1}(\mu^{2})\,
-
[\cF(\mu^2)\circ \cP^{(1)}(\mu^2)]
\cF^{-1}(\mu^2)
- \cS_\cV^{(1)}(\mu^2)
\;.
\end{split}
\end{equation}
Let us multiply by $\sbra{1}$ and use $\sbra{1}\cS^{(1)}(\mu^2) = 0$. In Ref.~\cite{NSThreshold}, we found that $\sbra{1}\cS_{\rm pert}^{(1,0)}(\mu^2)$ has a simple form,
\begin{equation}
\sbra{1}\cF(\mu^2)\,
\cS_{\rm pert}^{(1,0)}(\mu^2)\,
\cF^{-1}(\mu^2) =
\sbra{1}
[\cF(\mu^2)\circ\bar{\cS}^{(1,0)}(\mu^2)]\,\cF^{-1}(\mu^2)
\;,
\end{equation}
where $\bar{\cS}_{\rm pert}^{(1,0)}(\mu^2)$ leaves the number of partons, their momenta and their flavors unchanged but has a non-trivial color structure.\footnote{The operator $[\cF(\mu^2)\circ\bar{\cS}^{(1,0)}(\mu^2)]\,\cF^{-1}(\mu^2)$ was called $\cV$ in Ref.~\cite{NSThreshold}, but here we are letting $\cV$ denote a different operator.} We also divide $\cS_{\rm pert}^{(0,1)}(\mu^2)$ into two pieces
\begin{equation}
\cS_{\rm pert}^{(0,1)}(\mu^2) = 
\cS_{\mi \pi}^{(0,1)}(\mu^2) + \cS_{\rm Re}^{(0,1)}(\mu^2)
\;.
\end{equation}
Here $\cS_{\mi \pi}^{(0,1)}(\mu^2)$ is the contribution from the virtual graphs that is proportional to $\mi\pi$, while $\cS_{\rm Re}^{(0,1)}(\mu^2)$ is the remaining part. We note that \cite{NSThreshold}
\begin{equation}
\sbra{1}\cS_{\mi \pi}^{(0,1)}(\mu^2) = 0
\;.
\end{equation}

This gives us
\begin{equation}
  \begin{split}
    \sbra{1}\cS_\cV^{(1)}(\mu^2) 
    =
    \sbra{1}\Big\{&
    [\cF(\mu^2)\circ\bar{\cS}^{(1,0)}(\mu^2)]\,\cF^{-1}(\mu^2) 
    + \cS_{\rm Re}^{(0,1)}(\mu^2)
    \\
    &-[\cF(\mu^2)\circ \cP^{(1)}(\mu^2)]
    \cF^{-1}(\mu^2)
    \Big\}
    \;.
  \end{split}
\end{equation}
All of the operators here leave the parton state unchanged except for being operators on the color and spin space. We define $\cS_\cV^{(1)}(\mu^2)$ to have the color and spin structure of the right hand side of the equation, so that
\begin{equation}
\begin{split}
\label{eq:SV1storder}
\cS_\cV^{(1)}(\mu^2) =
[\cF(\mu^2)\circ\bar{\cS}^{(1,0)}(\mu^2)]\,\cF^{-1}(\mu^2) 
+ {\cS}_{\rm Re}^{(0,1)}(\mu^2)
-[\cF(\mu^2)\circ \cP^{(1)}(\mu^2)]
\cF^{-1}(\mu^2)
\;.
\end{split}
\end{equation}
This result also gives us\footnote{In Ref.~\cite{NSThreshold}, we neglected the $\mi\pi$ term and we averaged over spin. Then the right hand side of Eq.~(\ref{eq:S1storder2}) was called $\cH_I - \cV$.}
\begin{equation}
\begin{split}
\label{eq:S1storder2}
\cS^{(1)}(\mu^2) ={}& \cF(\mu^2)\,
\cS_{\rm pert}^{(1,0)}(\mu^2)\,
\cF^{-1}(\mu^{2})
- [\cF(\mu^2)\circ\bar{\cS}^{(1,0)}(\mu^2)]\,\cF^{-1}(\mu^2)
+\cS_{\mi \pi}^{(0,1)}(\mu^2)
\;.
\end{split}
\end{equation}
Compare this to the more general Eq.~(\ref{eq:Ssimplified}) in Appendix \ref{sec:renormalization}.

To use $\cS^{(1)}(\mu^2)$, one solves Eq.~(\ref{eq:Uevolution}) in the form
\begin{equation}
  \begin{split}
    \label{eq:practicalevolution}
    \cU(\mu_2^2, \mu_1^2) ={}& \cN(\mu_2^2, \mu_1^2)
    \\&
    + \int_{\mu_2^2}^{\mu_1^2}\!\frac{d\mu^2}{\mu^2}\
    \cU(\mu_2^2, \mu^2)\,
    \cF(\mu^2)\,
    \frac{\as(\mu^2)}{2\pi}
    \cS_{\rm pert}^{(1,0)}(\mu^2)\,
    \cF^{-1}(\mu^{2})\,
    \cN(\mu^2, \mu_1^2)
    \;,
  \end{split}
\end{equation}
where
\begin{equation}
  \begin{split}
    \label{eq:nosplitting}
    \cN(\mu_2^2, &\mu_1^2)
    \\={}& 
    \mathbb{T} \exp\!\left(
      \int_{\mu_2^2}^{\mu_1^2}\!\frac{d\mu^2}{\mu^2}\,
      \frac{\as(\mu^2)}{2\pi}
      \left\{- [\cF(\mu^2)\circ\bar{\cS}^{(1,0)}(\mu^2)]\,\cF^{-1}(\mu^2)
        +\cS_{\mi \pi}^{(0,1)}(\mu^2)
      \right\}
    \right)
    \;.
  \end{split}
\end{equation}
Here the Sudakov factor $\cN$ is the exponential of the part of $\cS$ that does not change the number of partons or their momenta or flavors. Normally its spin and color structure is simplified and the $\mi\pi$ contribution is not included. The splitting operator $\cF(\mu^2)\,\cS_{\rm pert}^{(1,0)}(\mu^2)\, \cF^{-1}(\mu^{2})$ adds one parton. Its spin and color structure is also normally simplified. Then Eq.~(\ref{eq:practicalevolution}) is implemented by solving it iteratively, so that there are some number of splittings interleaved with Sudakov factors.

\subsection[The structure of $\cV(\mu^2)$]
{The structure of \boldmath{$\cV(\mu^2)$}}

In Eq.~(\ref{eq:sigmaU7}), there is a factor $\cV(\mu_\scH^2)$ that multiplies $\cF(\mu_\scH^2)$ and $\sket{\rho_\scH}$. We can write this operator as
\begin{equation}
\cV(\mu_\scH^2) = \cV(\mu_\Lf^2)\,
\cU_\cV(\mu_\Lf^2,\mu_\scH^2)
\;,
\end{equation}
where
\begin{equation}
\cU_\cV(\mu_2^2,\mu_1^2)
=
\cV^{-1}(\mu_2^2)\,\cV(\mu_1^2)
\;.
\end{equation}
The operator $\cU_\cV(\mu^2,\mu^{\prime\, 2})$ obeys the evolution equation
\begin{equation}
\mu^2 \frac{d}{d\mu^2}\,\cU_\cV(\mu^2,\mu^{\prime\, 2})
=
-\cS_\cV(\mu^2)\,
\cU_\cV(\mu^2,\mu^{\prime 2})
\;,
\end{equation}
where $\cS_\cV(\mu^2)$ was defined in Eq.~(\ref{eq:SVdef}). Thus
\begin{equation}
\label{eq:UVexponential}
\cU_\cV(\mu_\Lf^2,\mu_\scH^{2})
=\mathbb{T} \exp\!\left(
\int_{\mu_\Lf^2}^{\mu_\scH^{2}}\!\frac{d\mu^2}{\mu^2}\,\cS_\cV(\mu^2)
\right)
\;.
\end{equation}
To use Eq.~(\ref{eq:UVexponential}), we apply $\cU_\cV(\mu_\Lf^2,\mu_\scH^{2})$ to a statistical state $\sket{\{p,f,s,s',c,c'\}_m}$ that contributes to $\sket{\rho_\scH}$. We expand $\cS_\cV(\mu^2)$ in powers of $\as(\mu^2)$ with factors involving the running parton distributions $\cF(\mu^2)$. The available scales other than $\mu^2$ come from $\sket{\rho_\scH}$, so the relevant matrix elements involve $\mu^2/\mu_\scH^2$. Since $\cV(\mu^2)$ is an infrared finite operator, the perturbative coefficients in $\cS_\cV(\mu^2)$ will then be proportional to $\mu^2/\mu_\scH^2$, possibly times logarithms of $\mu^2/\mu_\scH^2$. Thus the low $\mu^2$ end of the integration is power suppressed. However, it is important that there is a lower bound $\mu_\Lf^2$ on the integration. That is because the running coupling $\as(\mu^2)$ and the running parton distributions $\cF(\mu^2)$ are not well defined for very small $\mu^2$.

We are left with a factor 
\begin{equation}
\cV(\mu_\Lf^2) = 1 + \cO(\as(\mu_\Lf^2))\;.
\end{equation}
We can expand this perturbatively in powers of $\as(\mu_\Lf^2)$, using parton distributions $\cF(\mu_\Lf^2)$. The coefficients of $\as^n(\mu_\Lf^2)$ for $n \ge 1$ are then proportional to $\mu_\Lf^2/\mu_\scH^{2}$, possibly times logarithms of $\mu_\Lf^2/\mu_\scH^{2}$. Since $\mu_\Lf^2 \ll \mu_\scH^{2}$, we can safely neglect all of the higher order terms and simply replace
\begin{equation}
\label{eq:cVmuf}
\cV(\mu_\Lf^2) \to 1
\;.
\end{equation}
Thus we make the replacement $\cV(\mu_\scH^2) \to \cU_\cV(\mu_\Lf^2,\mu_\scH^2)$ in Eq.~(\ref{eq:sigmaU7}), giving us
\begin{equation}
\begin{split}
\label{eq:sigmaU8}
\sigma[J] ={}& 
\sbra{1} \cO_J\, 
\cU(\mu_\Lf^2,\mu_\scH^2)\,
\cU_\cV(\mu_\Lf^2,\mu_\scH^2)\,
\cF(\mu_\scH^2)
\sket{\rho_\scH}
+\cO(\as^{k+1}) + \cO(\mu_\Lf^2/Q[J]^2)
\;.
\end{split}
\end{equation}
The factor $\cU_\cV(\mu_\Lf^2,\mu_\scH^2)$ does two things. First, it provides perturbative corrections to the hard scattering state $\sket{\rho_\scH}$, which we need in order to calculate the cross section correct to $\LN^k\LL\LO$. For this purpose, it would suffice to expand the exponential in $\cU_\cV (\mu_\Lf^2, \mu_\scH^2)$ to the desired perturbative order. The second function of $\cU_\cV(\mu_\Lf^2,\mu_\scH^2)$ is to sum threshold logarithms. For this purpose, it is important that $\cU_\cV(\mu_\Lf^2,\mu_\scH^2)$ is an exponential.

To understand the relation of the operator $\cU_\cV(\mu_\Lf^2,\mu_\scH^{2})$ to threshold logarithms, it is instructive to look at it at order $\as$ with $\Lambda$ ordering for the shower. It is structurally the same as the operator introduced in Ref.~\cite{NSThreshold}, which concerned the summation of threshold logarithms.\footnote{In Ref.~\cite{NSThreshold}, an approximate version of $\cU_\cV(\mu_2^2,\mu_1^2)$ was used to sum threshold logarithms, but it appeared between each pair of parton splittings at scales $\mu_1^2$ and $\mu_2^2$ and also between the initial hard interaction and the first splitting. This caused problems, which were alleviated by inserting an artificial cut in $\cS_\cV(\mu^2)$.}  The analysis in Ref.~\cite{NSThreshold} simply averaged over spins, so we leave out spin here. The operator $\cS_\cV(\mu^2)$ in Ref.~\cite{NSThreshold} contains several terms. Rather than listing them all, we simply recall the most important terms:
\begin{equation}
  \begin{split}
    \label{eq:LOThreshold}
    \cS_\cV(\mu^2)\sket{\{p,f,c,c'\}_{m}}
    ={}& 
    \frac{\as}{2\pi}\,
    \int_{1/(1+\mu^2/\mu_\scH^2)}^1\!dz\,
    \Bigg[
    \left(1 -
      \frac{
        f_{a/A}(\eta_{\La}/z,\mu^2)}
      {f_{a/A}(\eta_{\La},\mu^2)}
    \right)
    \frac{2C_a}{1-z}\,
    \\ & \qquad 
    + \left(1 -
      \frac{
        f_{b/B}(\eta_{\Lb}/z,\mu^2)}
      {f_{b/B}(\eta_{\Lb},\mu^2)}
    \right)
    \frac{2C_b}{1-z}\Bigg]\,
    [1\otimes 1]\
    \sket{\{p,f,c,c'\}_{m}}
    \\&
    + \cdots
    \;,
  \end{split}
\end{equation}
where $C_a = C_\LF$ when $a$ is a quark flavor and $C_\Lg = C_\LA$.
The operator $[1\otimes 1]$ is the unit operator in the color space. Note that when $\mu^2 \ll \mu_\scH^2$, the range of the $z$ integration shrinks to  just a tiny region near $z = 1$. Thus, this contribution might seem unimportant. However, in Eq.~(\ref{eq:LOThreshold}) there is a factor $1/(1-z)$. This multiplies a factor involving the parton distribution functions. The result is that $\cU_\cV(\mu_\Lf^2,\mu_\scH^{2})$ is substantially different from 1 when the parton distribution functions are falling quickly as the momentum fraction grows. This gives a ``threshold logarithm'' effect that we can sum in a leading approximation by using $\cU_\cV(\mu_\Lf^2,\mu_\scH^{2})$.

The form of $\cS_\cV(\mu^2)$ depends on the definition of the infrared sensitive operator $\cD(\mu^2)$. In particular, shower evolution uses parton distributions $\cF(\mu^2)$ that are related to the $\MSbar$ parton distributions according to Eq.~(\ref{eq:F5toF}). This relation is discussed in Appendix \ref{sec:K}. Briefly, $\MSbar$ parton distributions are defined by removing ultraviolet divergences using $\MSbar$ renormalization, while the ultraviolet region of shower splittings is removed with a cutoff at scale $\mu_\Ls^2$ as defined in $\cD(\mu^2)$. See Sec.~\ref{sec:cD}. The evolution kernel $\cP(\mu^2)$ for the parton distributions reflects the definition of $\cF(\mu^2)$. This evolution kernel appears in $\cS_\cV(\mu^2)$, Eqs.~(\ref{eq:SV1storder}) and (\ref{eq:SVdefencore}). Thus the choice of shower oriented parton distribution functions affects how the summation of threshold logarithms appears in the overall result (\ref{eq:sigmaU8}) for the cross section: part of the summation of threshold logarithms appears as $\cU_\cV( \mu_\Lf^2, \mu_\scH^2)$ and part appears as the redefinition of the parton distributions, $\cF(\mu^2)$ in Eq.~(\ref{eq:sigmaU8}). This is described in Sec.~(9.4) of Ref.~\cite{NSThreshold}. If one uses $k_\LT$ ordering instead of $\Lambda$ ordering for the shower, then the shower oriented parton distributions are very close to the $\MSbar$ parton distributions and all of the leading threshold logarithms appear in $\cU_\cV( \mu_\Lf^2, \mu_\scH^2)$. On the other hand, in the factorization scheme in ref.~\cite{KrkNLO2015, KrkNLO2016}, a different definition of factorization results in the leading threshold corrections being absorbed into the parton distribution functions.

\section{Summary and outlook}
\label{sec:summary}

We began with an expression (\ref{eq:NkLOsigma}) for the cross section $\sigma[J]$ for an infrared safe measurement $J$ calculated at $\LN^k\LL\LO$. The pieces in this expression are infrared divergent in four dimensions, so that they are defined by working in $4 - 2\epsilon$ dimensions. Integrating over the phase space that is unresolved by the measurement leads to some cancellations of poles $1/\epsilon$. Other poles cancel after factorization of initial state infrared sensitivity into parton distribution functions. This leaves a result that is finite in four dimensions, even though it consists of pieces that are divergent in four dimensions. We introduced an infrared sensitive operator $\cD(\mu^2)$ and an inclusive infrared finite operator $\cV(\mu^2)$ to help organize the cancellations.

After some analysis, we have represented the cross section as Eq.~(\ref{eq:sigmaU8}). Here the separate factors are all finite in four dimensions. If we expand this expression to order $\as^k$, we have the same cross section that we started with except for a power suppressed contribution that we have dropped.

In Eq.~(\ref{eq:sigmaU8}), we have a hard scattering state $\sket{\rho_\scH}$ and a factor $\cF(\mu_\scH^2)$ that supplies parton distribution functions and a parton luminosity factor so that if we trace over colors and spins, we have a differential cross section in the space of parton number, flavors, and momenta. Then we have a factor $\cU_\cV(\mu_\Lf^2,\mu_\scH^2)$ that supplies a summation of threshold logarithms associated with the hard state and also part of the $\LN^k\LL\LO$ perturbative corrections to the hard scattering cross section. Next, we have a complete parton shower generated by $\cU(\mu_\Lf^2,\mu_\scH^2)$. The parton shower operator preserves inclusive probabilities: $\sbra{1}\cU(\mu_2^2,\mu_1^2) = \sbra{1}$. We have ended the shower at a scale $\mu_\Lf^2$. After that, the factor $\sbra{1} \cO_J$ represents the measurement that we want to make. We suppose that this is an infrared safe measurement that is not sensitive to soft or collinear parton splittings at the scale $\mu_\Lf^2$ or below. That means that the error in the calculation, estimated by $\cO(\mu_\Lf^2/Q[J]^2)$, is small. With such an infrared safe measurement, the result of the measurement is not sensitive to hadronization. If we wanted to use a measurement operator that is sensitive to hadronization, then we would need to include a model of hadronization before the measurement operator. It is then less certain what a good error estimate is.

There is some temptation to imagine Eq.~(\ref{eq:sigmaU8}) as being simpler than it is. In our Higgs boson example, if we expand $\sbra{1} \cU_\cV(\mu_\Lf^2, \mu_\scH^2)\,\cF(\mu_\scH^2) \sket{\rho_\scH}$ to order $\as^k$, it is the $\LN^k\LL\LO$ inclusive cross section to make a Higgs boson. The operator $\cU(\mu_\Lf^2,\mu_\scH^2)$ generates a probability preserving parton shower. Thus it might seem that one takes the hard scattering cross section and then distributes the probability across different final states according to what the shower generates. However, $\cU_\cV(\mu_\Lf^2, \mu_\scH^2)\,\cF(\mu_\scH^2) \sket{\rho_\scH}$ is not a cross section. It is a statistical state, representing different numbers of final state partons, which come with their own quantum color and spin states. The shower operator acts separately on each component of this statistical state. Then if we measure $\sigma[J]$ for an observable that is more complicated than just the inclusive measurement of a Higgs boson, the separate contributions are not sensible by themselves, but they sum to give $\sigma[J]$ correct to order $\as^k$ with only a power suppressed correction.

We have spoken of getting $\sigma[J]$ correct to order $\as^k$, but, of course, that is not the point of a parton shower. In applying Eq.~(\ref{eq:sigmaU8}), one would evaluate the splitting function $\cS(\mu^2)$ in the exponent of $\cU(\mu_\Lf^2,\mu_\scH^2)$ to order $\as^k$, then retain $\cU(\mu_\Lf^2,\mu_\scH^2)$ as an exponential. When the desired measurement operator $\cO_J$ contains widely different scales, the cross section will contain large logarithms. Then $\cU(\mu_\Lf^2,\mu_\scH^2)$ has the potential to sum these logarithms. After all, it exponentiates the soft and collinear singularities of QCD at order $\as^k$. Unfortunately, one needs to study the structure of $\cU(\mu_\Lf^2,\mu_\scH^2)$ as it relates to the structure of $\cO_J$ in order to check how well the shower does in summing the logarithms.

One can wonder whether the formalism of this paper is of any use for just a LO shower. We suggest that it is. If one averages over spins and makes the leading color approximation, the shower operator $\cU(\mu_\Lf^2,\mu_\scH^2)$ generates a rather conventional probability preserving dipole shower. With $\Lambda^2$ as the ordering variable, it is the leading color version of the shower in \textsc{Deductor}. One can choose other ordering variables. The operator $\cU_\cV(\mu_\Lf^2,\mu_\scH^2)$ generates threshold corrections, as described in Ref.~\cite{NSThreshold}. These corrections are numerically important in some cases and could be included in standard parton shower programs. 

In fact, Eq.~(\ref{eq:sigmaU8}) has been useful in improving \textsc{Deductor}. While working on our paper \cite{NSThreshold} on threshold corrections, we did not have Eq.~(\ref{eq:sigmaU8}). The result was a structure that had certain undesirable features that needed to be controlled by means of an {\em ad hoc} cutoff. The current more general formulation in Eq.~(\ref{eq:sigmaU8}) removes this problem, although it does not much change the numerical results. We present phenomenological results from the new version of \textsc{Deductor} in a separate paper \cite{NSThresholdII}.

The formalism is based on an operator $\cD(\mu^2)$ that encodes the infrared structure of QCD starting with a state with any number $m$ of final state partons. If we know $\cD(\mu^2)$ up to order $\as^k$, then we generate, in an automatic way, the subtraction terms for an $\LN^k\LL\LO$ perturbative calculation. See the example of the toy model in Appendix \ref{sec:toymodel}. This appears to us to be simpler than constructing the subtraction terms directly \cite{Daleo:2009yj,Boughezal:2010mc, Gehrmann:2011wi, GehrmannDeRidder:2012ja, Currie:2013vh, Somogyi:2006da, Somogyi:2006db, Somogyi:2013yk}. From $\cD(\mu^2)$ at order $\as^k$, we also generate, in an automatic way, the shower splitting kernels at order $\as^k$.

The perturbative contributions to $\cD(\mu^2)$ are not simple in full QCD. Furthermore, their form depends on choices like the momentum mapping scheme and the choice of a hardness ordering variable. At order $\as^1$, we have made the required choices, made suitable approximations, and calculated the corresponding splitting functions $\cS(\mu^2)$ in Refs.~\cite{NSThreshold, NSThresholdII}.  Similarly, from $\cD(\mu^2)$ we generate the inclusive infrared finite operator $\cV(\mu^2)$. In general, there are some choices that one can make in defining $\cV(\mu^2)$. At order $\as^1$, we have made the required choices, made suitable approximations, and calculated $\cS_\cV(\mu^2)$ in Ref.~\cite{NSThreshold, NSThresholdII}.

In the first order \textsc{Deductor} shower, we need only $d \cD^{(1)}(\mu^2)/d\log(\mu^2)$, which is finite in four dimensions. For subtractions and matching to the shower, one needs the full integrated operator $\cD^{(1)}(\mu^2)$. Techniques for this are described in Ref.~\cite{Czakon} and the earlier papers \cite{Robens1, Robens2, Czakon1}.

We leave it to future work to make suitable choices for a momentum mapping scheme, a hardness ordering variable, and definitions away from the strict soft and collinear limits so as to construct order $\as^2$ contributions to $\cD(\mu^2)$. With a choice for color structure, we could then also construct $\cV(\mu^2)$. We thus hope that the formalism presented in this paper might prove useful in developing a parton shower with order $\as^2$ splitting functions.

We also hope that the formalism presented in this paper might provide support for the view that a parton shower is similar to a more straightforward perturbative calculation at $\LN^k\LL\LO$. In this view, the parton shower is an approximate way to calculate cross sections, but the approximation is systematically improvable by working at higher perturbative order. In a practical program, there may be further approximations with respect to color and spin. These need a separate justification.

\acknowledgments{ 
This work was supported in part by the United States Department of Energy under grant DE-SC0011640. This project was begun while the authors were at the Munich Institute for Astro- and Particle Physics program ``Challenges, Innovations and Developments in Precision Calculations for the LHC.''   It was continued while the authors were at the Kavli Institute for Theoretical Physics at the University of California, Santa Barbara, program ``LHC Run II and the Precision Frontier,'' which was supported by the U.\ S.\ National Science Foundation under Grant No.\ NSF PHY11-25915. DS thanks the Erwin Schr\"odinger Institute of the University of Vienna for its hospitality at its program ``Challenges and Concepts for Field Theory and Applications in the Era of the LHC Run-2.'' We thank the MIAPP, the KITP, and the ESI for providing stimulating research environments. We thank Tim Cohen, Marat Freytsis, and Jonathan Gaunt for helpful discussions.}

\appendix 

\section{A toy model for parton shower operators}
\label{sec:toymodel}

The construction in this paper relies on an operator $\cD$ that contains the infrared singularities of statistical states $|\rho)$ that represent the color and spin density matrix elements of QCD. The operators $\cD$ can be used to construct both the infrared subtractions needed for a perturbative calculation of the cross section at $\LN^k\LL\LO$ and also the splitting kernels needed to construct a parton shower at the corresponding order, $\LN^{k-1}\LL\LO$. This construction has been quite abstract, especially since we lack an example of $\cD$ at $\LN^2\LL\LO$, which would correspond to an $\LN\LL\LO$ shower.

In this appendix, we illustrate some of the ideas of the paper using a toy model that provides a concrete example of $|\rho)$ and $\cD$ at $\LN^2\LL\LO$. In this example, we construct the splitting kernel $\cS_{\rm pert}$ at $\LN\LL\LO$. The toy model is very simple. There are no parton distributions. The coupling $\as$ does not run. The momenta are one dimensional. There is no spin. There is quantum color, but the color structure is vastly simplified compared to what one has in real QCD.

\subsection{Statistical states in the toy model}

We use momentum states $\{p_1,p_2,\cdots,p_m\}$ for $m$ partons, with each $p_i$ being a real number in the range $0 < p_i < \infty$. We use ``color'' represented by basis states labeled by an index pair $(c_\scR,c_\scV)$ with $c_\scR \in \{0,1,\dots, m\}$, $c_\scV \in \{0,1,\dots\}$. The statistical states have the form
\begin{equation}
\sket{\{p\}_m,(c_\scR,c_\scV)}=
\sket{\{p_1,p_2,\cdots,p_m\},(c_\scR,c_\scV)}
\;.
\end{equation}
These are defined to be invariant under permutations of the $p_i$. The Born level cross section is $\sket{\{\}_0,(0,0)}$. 

We will make use of color operators $\cC_\LR$ and $\cC_\LV$ that act on the space of statistical states. These have the form
\begin{equation}
\begin{split}
\cC_\LR ={}& 1 + \cA^\dagger_\LR
\;,
\\
\cC_\LV ={}& 1 + \cA^\dagger_\LV
\;,
\end{split}
\end{equation}
where $\cA^\dagger_\LR$ and $\cA^\dagger_\LR$ are raising operators:
\begin{equation}
\begin{split}
\cA^\dagger_\LR \sket{\{p\}_m,(c_\scR,c_\scV)} ={}& 
\sket{\{p\}_m,(c_\scR + 1,c_\scV)}
\;,
\\
\cA^\dagger_\LV \sket{\{p\}_m,(c_\scR,c_\scV)} ={}& 
\sket{\{p\}_m,(c_\scR,c_\scV+1)}
\;.
\end{split}
\end{equation}
The inclusive probabilities corresponding to our statistical states are defined by
\begin{equation}
\sbrax{1}\sket{\{p\}_m,(c_\scR,c_\scV)} = 
\left(\frac{1}{N_\Lc^2}\right)^{\! c_\scR + c_\scV}
\;.
\end{equation}
Here $N_\Lc = 3$ represents the number of colors. Thus keeping only $c_\scR = c_\scV = 0$ is analogous to the leading color approximation.

We include an operator representing an observable. The observable depends only on the parton momenta, not on their color state:
\begin{equation}
\cO_J\sket{\{p\}_m,(c_\scR,c_\scV)}
=
\left( 1 + \frac{p_1^2 + p_2^2 + \cdots p_m^2}{Q^2}\right)
\sket{\{p\}_m,(c_\scR,c_\scV)}
\;.
\end{equation}

\subsection{Perturbative hard scattering states}

We take the perturbative hard scattering states in the toy model to have the form
\begin{equation}
  \sket{\rho} = \sket{\rho^{(0,0)}}
  + \as \left(\sket{\rho^{(1,0)}}+\sket{\rho^{(0,1)}}\right)
  + \as^2 \left(\sket{\rho^{(2,0)}}+\sket{\rho^{(1,1)}}+\sket{\rho^{(0,2)}}\right)
  + \cO(\as^3)
  \;.
\end{equation}
Here $|\rho^{(n_\scR,n_\scV)})$ represents a perturbative contribution with $n_\scR$ real partons emitted and $n_\scV$ virtual loops. The contributions are defined to be
\begin{equation}
  \begin{split}
    \sket{\rho^{(0,0)}} ={}& \sket{\{ \}_0,(0,0)}
    \;,
    \\
    \sket{\rho^{(1,0)}} ={}& \int_0^{Q^2}\!\frac{dp_1^2}{p_1^2}\,
    \left(\frac{p_1^2}{Q^2}\right)^{\!\!\epsilon}\,
    \frac{Q^2}{p_1^2 + Q^2}\,
    \cC_\LR
    \sket{\{p\}_1,(0,0)}
    \;,
    \\
    \sket{\rho^{(0,1)}} ={}& -\int_0^{\infty}\!\frac{dk_1^2}{k_1^2}\,
    \left(\frac{k_1^2}{Q^2}\right)^{\!\!\epsilon}\,
    \frac{Q^2}{k_1^2 + Q^2}\,
    \cC_\LV
    \sket{\{ \}_0,(0,0)}
    \;,
    \\
    \sket{\rho^{(2,0)}} ={}& 
    \int_0^{Q^2}\!\frac{dp_2^2}{p_2^2}\,
    \left(\frac{p_2^2}{Q^2}\right)^{\!\!\epsilon}\,
    \int_0^{Q^2}\!\frac{dp_1^2}{p_1^2 + 2 p_2^2}\,
    \left(\frac{p_1^2}{Q^2}\right)^{\!\!\epsilon}\,
    \frac{Q^2}{p_1^2 + p_2^2 + Q^2}\,
    \cC_\LR^2
    \sket{\{p\}_2,(0,0)}
    \;,
    \\
    \sket{\rho^{(1,1)}} ={}& 
    -\bigg\{\int_0^{Q^2}\!\frac{dp_1^2}{p_1^2}\,
    \left(\frac{p_1^2}{Q^2}\right)^{\!\!\epsilon}\,
    \int_0^{\infty}\!\frac{dk_1^2}{k_1^2 + 2p_1^2}\,
    \left(\frac{k_1^2}{Q^2}\right)^{\!\!\epsilon}\,
    \frac{Q^2}{k_1^2 + p_1^2 + Q^2}\,
    \\& \qquad+
    \int_0^{\infty}\!\frac{dk_1^2}{k_1^2}\,
    \left(\frac{k_1^2}{Q^2}\right)^{\!\!\epsilon}\,
    \int_0^{Q^2}\!\frac{dp_1^2}{p_1^2 + 2k_1^2}\,
    \left(\frac{p_1^2}{Q^2}\right)^{\!\!\epsilon}\,
    \frac{Q^2}{p_1^2 + k_1^2 + Q^2}
    \bigg\}
    \\&\times
    \cC_\LR\cC_\LV
    \sket{\{p\}_1,(0,0)}
    \;,
    \\
    \sket{\rho^{(0,2)}} ={}& 
    \int_0^{\infty}\!\frac{dk_2^2}{k_2^2}\,
    \left(\frac{k_2^2}{Q^2}\right)^{\!\!\epsilon}\,
    \int_0^{\infty}\!\frac{dk_1^2}{k_1^2 + 2k_2^2}\,
    \left(\frac{k_1^2}{Q^2}\right)^{\!\!\epsilon}\,
    \frac{Q^2}{k_1^2 + k_2^2 + Q^2}\,
    \cC_\LV^2
    \sket{\{\}_0,(0,0)}
    \;.
  \end{split}
\end{equation}
The integrals here are regularized in the infrared by factors $(p^2/Q^2)^\epsilon$. The individual contributions $|\rho^{(n_\scR,n_\scV)})$ contain singularities when parton momenta become small and $1/\epsilon$ poles that arise from integration from virtual parton momenta. Furthermore, the contributions with real parton emissions have different color states than the corresponding contributions with virtual loops. However, when we calculate the cross section $(1|\cO_J |\rho)$, we can use the fact that
\begin{equation}
\sbra{1}\cC_\LR = \sbra{1}\cC_\LV
\;,
\end{equation}
so that the color contributions from real emissions and virtual loops match. Then, in fact, there are $\mathit{real} - \mathit{virtual}$ cancellations, with the result that $(1|\cO_J |\rho)$ is infrared finite.

\subsection{The infrared sensitive operator}

The infrared structure of this is fairly simple and can be represented using the infrared sensitive operator $\cD(\mu_\Ls^2)$ with
\begin{equation}
\begin{split}
\cD(\mu_\Ls^2) = 1
&+ \as \left(\cD^{(1,0)}(\mu_\Ls^2)+\cD^{(0,1)}(\mu_\Ls^2)\right)
\\&
+ \as^2 \left(\cD^{(2,0)}(\mu_\Ls^2)+\cD^{(1,1)}(\mu_\Ls^2)
+\cD^{(0,2)}(\mu_\Ls^2)\right)
+ \cO(\as^3)
\;,
\end{split}
\end{equation}
with
\begin{equation}
\begin{split}
\label{eq:cDmodel}
\cD^{(1,0)}(\mu_\Ls^2)\sket{\{p\}_m,(c_\scR,c_\scV)} 
={}& \int_0^{\mu_\Ls^2}\!\frac{dp_{m+1}^2}{p_{m+1}^2}\,
\left(\frac{p_{m+1}^2}{Q^2}\right)^{\!\!\epsilon}\,
\cC_\LR
\sket{\{p\}_{m+1},(c_\scR,c_\scV)}
\;,
\\
\cD^{(0,1)}(\mu_\Ls^2)\sket{\{p\}_m,(c_\scR,c_\scV)} 
={}& -\int_0^{\mu_\Ls^2}\!\frac{dk_1^2}{k_1^2}\,
\left(\frac{k_1^2}{Q^2}\right)^{\!\!\epsilon}\,
\cC_\LV
\sket{\{p\}_m,(c_\scR,c_\scV)}
\;,
\\
\cD^{(2,0)}(\mu_\Ls^2)\sket{\{p\}_m,(c_\scR,c_\scV)} ={}& 
\int_0^{\mu_\Ls^2}\!\frac{dp_{m+2}^2}{p_{m+2}^2}\,
\left(\frac{p_{m+2}^2}{Q^2}\right)^{\!\!\epsilon}\,
\int_0^{\mu_\Ls^2}\!\frac{dp_{m+1}^2}{p_{m+1}^2 + 2p_{m+2}^2}\,
\left(\frac{p_{m+1}^2}{Q^2}\right)^{\!\!\epsilon}
\\& \times
\cC_\LR^2
\sket{\{p\}_{m+2},(c_\scR,c_\scV)}
\;,
\\
\cD^{(1,1)}(\mu_\Ls^2)\sket{\{p\}_m,(c_\scR,c_\scV)} ={}& 
-\bigg\{\int_0^{\mu_\Ls^2}\!\frac{dp_{m+1}^2}{p_{m+1}^2}\,
\left(\frac{p_{m+1}^2}{Q^2}\right)^{\!\!\epsilon}\,
\int_0^{\mu_\Ls^2}\!\frac{dk_1^2}{k_1^2 + 2p_{m+1}^2}\,
\left(\frac{k_1^2}{Q^2}\right)^{\!\!\epsilon}\,
\\&\qquad +
\int_0^{\mu_\Ls^2}\!\frac{dk_1^2}{k_1^2}\,
\left(\frac{k_1^2}{Q^2}\right)^{\!\!\epsilon}\,
\int_0^{\mu_\Ls^2}\!\frac{dp_{m+1}^2}{p_{m+1}^2 + 2k_1^2}\,
\left(\frac{p_{m+1}^2}{Q^2}\right)^{\!\!\epsilon}
\bigg\}
\\& \times
\cC_\LR\cC_\LV
\sket{\{p\}_{m+1},(c_\scR,c_\scV)}
\;,
\\
\cD^{(0,2)}(\mu_\Ls^2)\sket{\{p\}_m,(c_\scR,c_\scV)} ={}& 
\int_0^{\mu_\Ls^2}\!\frac{dk_2^2}{k_2^2}\,
\left(\frac{k_2^2}{Q^2}\right)^{\!\!\epsilon}\,
\int_0^{\mu_\Ls^2}\!\frac{dk_1^2}{k_1^2 + 2k_2^2}\,
\left(\frac{k_1^2}{Q^2}\right)^{\!\!\epsilon}
\\&\times
\cC_\LV^2
\sket{\{p\}_m,(c_\scR,c_\scV)}
\;.
\end{split}
\end{equation}

From $\cD(\mu_\Ls^2)$ we can construct $\cD^{-1}(\mu_\Ls^2)$,
\begin{equation}
  \begin{split}
    \cD^{-1}(\mu_\Ls^2) = 1
    & - \as \left(\cD^{(1,0)}(\mu_\Ls^2)+\cD^{(0,1)}(\mu_\Ls^2)\right)
    \\&
    - \as^2\,\Big(\cD^{(2,0)}(\mu_\Ls^2)+\cD^{(1,1)}(\mu_\Ls^2)
    +\cD^{(0,2)}(\mu_\Ls^2)
    \\&\quad\qquad
    - \cD^{(1,0)}(\mu_\Ls^2)\cD^{(1,0)}(\mu_\Ls^2)
    - \cD^{(1,0)}(\mu_\Ls^2)\,\cD^{(0,1)}(\mu_\Ls^2)
    \\&\quad\qquad
    - \cD^{(0,1)}(\mu_\Ls^2)\,\cD^{(1,0)}(\mu_\Ls^2)
    - \cD^{(0,1)}(\mu_\Ls^2)\,\cD^{(0,1)}(\mu_\Ls^2)
    \Big)
    \\&+ \cO(\as^3)
    \;.
  \end{split}
\end{equation}

\subsection{Subtractions for the hard scattering states}

We can now construct the subtracted statistical state including the measurement operator,
\begin{equation}
\sket{\hat \rho} = \cD^{-1}(\mu_\Ls^2)\cO_J\sket{\rho}
\;.
\end{equation}
This has the expansion
\begin{equation}
\sket{\hat\rho} = \sket{\rho^{(0,0)}}
+ \as \left(\sket{\hat\rho^{(1,0)}}+\sket{\hat\rho^{(0,1)}}\right)
+ \as^2 \left(\sket{\hat\rho^{(2,0)}}+\sket{\hat\rho^{(1,1)}}
+\sket{\hat\rho^{(0,2)}}\right)
+ \cO(\as^3)
\;.
\end{equation}

At first order in $\as$, there are two terms. The first is
\begin{equation}
\sket{\hat\rho^{(1,0)}}
=
\cO_J\sket{\rho^{(1,0)}}
-\cD^{(1,0)}(\mu_\Ls^2)\cO_J\sket{\rho^{(0,0)}}
\;.
\end{equation}
This is
\begin{equation}
\begin{split}
\sket{\hat\rho^{(1,0)}}
=
\int_0^{\infty}&\frac{dp_1^2}{p_1^2}\,
\left(\frac{p_1^2}{Q^2}\right)^{\!\!\epsilon}\,
\left\{
\theta(p_1^2 < Q^2)\,
\frac{Q^2}{p_1^2 + Q^2}
\left(
1 + \frac{p_1^2}{Q^2}
\right)
- \theta(p_1^2 < \mu_\Ls^2)
\right\}
\\&\times
\cC_\LR
\sket{\{p\}_1,(0,0)}
\;.
\end{split}
\end{equation}
The first term is singular when $p_1^2 \to 0$, but the subtraction from $\cD^{(1,0)}(\mu_\Ls^2)$ eliminates the singularity. Then $(1|\hat\rho^{(1,0)})$ is finite at $\epsilon = 0$. The coefficient of $\as$ that corresponds to virtual graphs is
\begin{equation}
\sket{\hat\rho^{(0,1)}}
=
\cO_J\sket{\rho^{(0,1)}}
-\cD^{(0,1)}(\mu_\Ls^2)\cO_J\sket{\rho^{(0,0)}}
\;.
\end{equation}
This is
\begin{equation}
\sket{\hat\rho^{(0,1)}}
=
-\int_0^{\infty}\!\frac{dk_1^2}{k_1^2}\,
\left(\frac{k_1^2}{Q^2}\right)^{\!\!\epsilon}\,
\left\{
\frac{Q^2}{k_1^2 + Q^2}\,
- \theta(k_1^2 < \mu_\Ls^2)
\right\}
\cC_\LV
\sket{\{\}_0,(0,0)}
\;.
\end{equation}
The first term has a $1/\epsilon$ pole from $k_1^2 \to 0$, but the subtraction from $\cD^{(0,1)}(\mu_\Ls^2)$ eliminates the $k_1^2 \to 0$ singularity. Then $|\hat\rho^{(0,1)})$ is finite at $\epsilon = 0$.

At order $\as^2$, there are three terms in $|\hat \rho)$. Let us look at the contribution from two real emissions:
\begin{equation}
\begin{split}
\sket{\hat\rho^{(2,0)}} ={}& \cO_J \sket{\rho^{(2,0)}} 
- \cD^{(2,0)}(\mu_\Ls^2)\, \cO_J \sket{\rho^{(0,0)}}
\\&
- \cD^{(1,0)}(\mu_\Ls^2)\,\cO_J \sket{\rho^{(1,0)}}
+ \cD^{(1,0)}(\mu_\Ls^2)\,\cD^{(1,0)}(\mu_\Ls^2)\,\cO_J  \sket{\rho^{(0,0)}}
\;.
\end{split}
\end{equation}
This is
\begin{equation}
  \begin{split}
    \sket{\hat\rho^{(2,0)}}=
    \int_0^{\infty}&\frac{dp_2^2}{p_2^2}\,
    \left(\frac{p_2^2}{Q^2}\right)^{\!\!\epsilon}\,
    \int_0^{\infty}\!\frac{dp_1^2}{p_1^2}\,
    \left(\frac{p_1^2}{Q^2}\right)^{\!\!\epsilon}\,
    \\&\times
    \bigg\{
    \frac{p_1^2}{p_1^2 + 2p_2^2}
    \bigg[
    \theta(p_2^2 < Q^2)\,\theta(p_1^2 < Q^2)\,
    \frac{Q^2}{p_1^2 + p_2^2 + Q^2}
    \left(
      1 + \frac{p_1^2 + p_2^2}{Q^2}
    \right)
    \\&\qquad\qquad\qquad
    - \theta(p_2^2 < \mu_\Ls^2)\,\theta(p_1^2 < \mu_\Ls^2)
    \bigg]
    \\&\qquad
    - \theta(p_2^2 < \mu_\Ls^2)\,
    \left[\theta(p_1^2 < Q^2)\,
      \frac{Q^2}{p_1^2 + Q^2}
      \left(
        1 + \frac{p_1^2}{Q^2}
      \right)
      - \theta(p_1^2 < \mu_\Ls^2)\right]
    \bigg\}
    \\&\times
    \cC_\LR^2
    \sket{\{p\}_2,(0,0)}
    \;.
  \end{split}
\end{equation}
Each term here exhibits infrared singularities, but the singularities cancel. Then $(1|\hat\rho^{(2,0)})$ can be evaluated at $\epsilon = 0$. Specifically, for $p_2^2 \to 0$ at fixed $p_1^2$, the first term cancels the third term and the second term cancels the fourth term. For $p_1^2 \to 0$ with fixed $p_2^2$, the first two terms are nonsingular, while the third term cancels the fourth term. When $p_2^2 \to 0$ and $p_1^2 \to 0$, the first term cancels the second term and the third term cancels the fourth term.

The contribution to $|\hat \rho)$ from two virtual emissions is similar. We have
\begin{equation}
\begin{split}
\sket{\hat\rho^{(0,2)}} ={}& \cO_J \sket{\rho^{(0,2)}} 
- \cD^{(0,2)}(\mu_\Ls^2)\,\cO_J  \sket{\rho^{(0,0)}}
\\&
- \cD^{(0,1)}(\mu_\Ls^2)\,\cO_J  \sket{\rho^{(0,1)}}
+ \cD^{(0,1)}(\mu_\Ls^2)\,\cD^{(0,1)}(\mu_\Ls^2)\,\cO_J  \sket{\rho^{(0,0)}}
\;.
\end{split}
\end{equation}
This is
\begin{equation}
  \begin{split}
    \sket{\hat\rho^{(0,2)}}
    =
    \int_0^{\infty}&\frac{dk_2^2}{k_2^2}\,
    \left(\frac{k_2^2}{Q^2}\right)^{\!\!\epsilon}\,
    \int_0^{\infty}\!\frac{dk_1^2}{k_1^2}\,
    \left(\frac{k_1^2}{Q^2}\right)^{\!\!\epsilon}\,
    \\&\times
    \bigg\{
    \frac{k_1^2}{k_1^2 + 2k_2^2}\left[
      \frac{Q^2}{k_1^2 + k_2^2 + Q^2}\,
      - \theta(k_2^2 < \mu_\Ls^2)\,\theta(k_1^2 < \mu_\Ls^2)
    \right]
    \\&\qquad
    - \theta(k_2^2 < \mu_\Ls^2)\,
    \left[\frac{Q^2}{k_1^2 + Q^2}\,
      - \theta(k_1^2 < \mu_\Ls^2)\right]
    \bigg\}
    \\&\times
    \cC_\LV^2
    \sket{\{ \}_0,(0,0)}
    \;.
  \end{split}
\end{equation}
Each term here exhibits $1/\epsilon$ poles, but the poles cancel. The pattern of cancellations is the same as for $|\hat\rho^{(2,0)})$.

The contribution to $|\hat \rho)$ from one real emission and one virtual emission is a little more complicated. We have
\begin{equation}
\begin{split}
\sket{\hat\rho^{(1,1)}} ={}& \cO_J \sket{\rho^{(1,1)}} 
- \cD^{(1,1)}(\mu_\Ls^2)\,\cO_J \sket{\rho^{(0,0)}}
\\&
- \cD^{(1,0)}(\mu_\Ls^2)\,\cO_J  \sket{\rho^{(0,1)}}
- \cD^{(0,1)}(\mu_\Ls^2)\,\cO_J  \sket{\rho^{(1,0)}}
\\&
+  \cD^{(1,0)}(\mu_\Ls^2)\,\cD^{(0,1)}(\mu_\Ls^2)\,\cO_J \sket{\rho^{(0,0)}}
+ \cD^{(0,1)}(\mu_\Ls^2)\,\cD^{(1,0)}(\mu_\Ls^2)\,\cO_J  \sket{\rho^{(0,0)}}
\;.
\end{split}
\end{equation}
We obtain
\begin{equation}
  \begin{split}
    \sket{\hat\rho^{(1,1)}}=
    \int_0^{\infty}&\frac{dp_1^2}{p_1^2}\,
    \left(\frac{p_1^2}{Q^2}\right)^{\!\!\epsilon}\,
    \int_0^{\infty}\!\frac{dk_1^2}{k_1^2}\,
    \left(\frac{k_1^2}{Q^2}\right)^{\!\!\epsilon}\,
    \\&\times
    \bigg\{
    -\theta(p_1^2 < Q^2)\,\frac{Q^2}{k_1^2 + p_1^2 + Q^2}\,
    \left[
      \frac{k_1^2}{k_1^2 + 2p_1^2}
      + \frac{p_1^2}{p_1^2 + 2k_1^2}
    \right]
    \left(1 + \frac{p_1^2}{Q^2}\right)
    \\&\qquad
    +\theta(p_1^2 < \mu_\Ls^2)\,\theta(k_1^2 < \mu_\Ls^2)\,
    \left[
      \frac{k_1^2}{k_1^2 + 2p_1^2}
      + \frac{p_1^2}{p_1^2 + 2k_1^2}
    \right]
    \\&\qquad
    + 
    \theta(p_1^2 < \mu_\Ls^2)\left[\frac{Q^2}{k_1^2 + Q^2}
    - 2\theta(k_1^2 < \mu_\Ls^2)\right]
    \\&\qquad
    + \theta(p_1^2 < Q^2)\,\theta(k_1^2 < \mu_\Ls^2)\,
    \frac{Q^2}{p_1^2 + Q^2}
    \left(1 + \frac{p_1^2}{Q^2}\right)
    \bigg\}
    \\&\times
    \cC_\LR\cC_\LV
    \sket{\{p\}_1,(0,0)}
    \;.
  \end{split}
\end{equation}
Each term here exhibits $1/\epsilon$ poles and singularities but the poles and singularities cancel. Then $|\hat\rho^{(1,1)})$ can be evaluated at $\epsilon = 0$. Specifically, expanding the square brackets gives us seven terms, $T_i$, $i \in \{1,\dots,7\}$. For the singularity in the integrand at $k_1^2 \to 0$ with fixed $p_1^2$, $T_2$ cancels $T_7$ and $T_6$ cancels $T_4 + T_5$.  For the singularity at $p_1^2 \to 0$ with fixed $k_1^2$, $T_1$ cancels $T_5$ and $T_6$ cancels $T_3 + T_7$. For the singularity when $p_1^2\to 0$ and $k_1^2\to 0$, $T_1$ cancels $T_3$, $T_2$ cancels $T_4$, and $T_6$ cancels $T_5$ + $T_7$.

We thus see that $\cD^{-1}(\mu_s^2)$ provides subtraction terms that cancel all of the singularities of $|\rho)$ at order $\as$ and $\as^2$. If we had wanted to construct all of the subtraction terms directly, it would have been somewhat difficult. However, constructing $\cD(\mu_s^2)$ for our toy model was quite simple. Then obtaining the subtraction terms was automatic.

We also note that the perturbative states $|\rho^{(n_\scR,n_\scV)})$ have non-trivial color structures in our toy model. Then the operators $\cD^{(n_\scR,n_\scV)}(\mu_\Ls^2)$ must reflect this color structure. If they do not, the subtractions will not work.

\subsection{The perturbative shower evolution operator}

We are now in a position to construct the order $\as$ and $\as^2$ terms in the perturbative shower generating operator, Eq.~(\ref{eq:Spert}),
\begin{equation}
S_{\rm pert}(\mu_\Ls^2) = \cD^{-1}(\mu_\Ls^2)\,
\mu_\Ls^2\frac{d}{d\mu_\Ls^2}\,\cD(\mu_\Ls^2)
\;.
\end{equation}
The operator $\cS_{\rm pert}$ will have the expansion
\begin{equation}
  \begin{split}
    \cS_{\rm pert}(\mu_\Ls^2) ={}& 
    \as \left(\cS_{\rm pert}^{(1,0)}(\mu_\Ls^2)
      +\cS_{\rm pert}^{(0,1)}(\mu_\Ls^2)\right)
    \\& 
    + \as^2 \left(\cS_{\rm pert}^{(2,0)}(\mu_\Ls^2)
      +\cS_{\rm pert}^{(1,1)}(\mu_\Ls^2)
      +\cS_{\rm pert}^{(0,2)}(\mu_\Ls^2)\right)
    + \cO(\as^3)
    \;,
  \end{split}
\end{equation}
where
\begin{equation}
\begin{split}
\label{eq:cSpert}
\cS_{\rm pert}^{(1,0)}(\mu_\Ls^2) ={}& 
\mu_\Ls^2\frac{d}{d\mu_\Ls^2}\,\cD^{(1,0)}(\mu_\Ls^2)
\;,
\\
\cS_{\rm pert}^{(0,1)}(\mu_\Ls^2) ={}& 
\mu_\Ls^2\frac{d}{d\mu_\Ls^2}\,\cD^{(0,1)}(\mu_\Ls^2)
\;,
\\
\cS_{\rm pert}^{(2,0)}(\mu_\Ls^2) ={}& 
\mu_\Ls^2\frac{d}{d\mu_\Ls^2}\,\cD^{(2,0)}(\mu_\Ls^2)
- \cD^{(1,0)}(\mu_\Ls^2)\,
\mu_\Ls^2\frac{d}{d\mu_\Ls^2}\,\cD^{(1,0)}(\mu_\Ls^2)
\;,
\\
\cS_{\rm pert}^{(1,1)}(\mu_\Ls^2) ={}& 
\mu_\Ls^2\frac{d}{d\mu_\Ls^2}\,\cD^{(1,1)}(\mu_\Ls^2)
- \cD^{(1,0)}(\mu_\Ls^2)
\mu_\Ls^2\frac{d}{d\mu_\Ls^2}\,\cD^{(0,1)}(\mu_\Ls^2)
\\&
- \cD^{(0,1)}(\mu_\Ls^2)\,
\mu_\Ls^2\frac{d}{d\mu_\Ls^2}\,\cD^{(1,0)}(\mu_\Ls^2)
\;,
\\
\cS_{\rm pert}^{(0,2)}(\mu_\Ls^2) ={}& 
\mu_\Ls^2\frac{d}{d\mu_\Ls^2}\,\cD^{(0,2)}(\mu_\Ls^2)
- \cD^{(0,1)}(\mu_\Ls^2)\,
\mu_\Ls^2\frac{d}{d\mu_\Ls^2}\,\cD^{(0,1)}(\mu_\Ls^2)
\;.
\end{split}
\end{equation}
To use this, we first need the derivatives of $\cD^{(n_\scR,n_\scV)}(\mu_\Ls^2)$.

The derivatives of $\cD^{(1,0)}(\mu_\Ls^2)$ and $\cD^{(1,0)}(\mu_\Ls^2)$ are simple
\begin{equation}
  \begin{split}
    \label{eq:dcDmodel}
    \mu_\Ls^2\frac{d}{d\mu_\Ls^2}\,
    \cD^{(1,0)}(\mu_\Ls^2)\sket{\{p\}_m,(c_\scR,c_\scV)} 
    ={}& \left(\frac{\mu^2_\Ls}{Q^2}\right)^{\!\!\epsilon}\,
    \cC_\LR\sket{\{p_1,\dots,p_m, \mu_\Ls\},(c_\scR,c_\scV)}
    \;,
    \\
    \mu_\Ls^2\frac{d}{d\mu_\Ls^2}\,
    \cD^{(0,1)}(\mu_\Ls^2)\sket{\{p\}_m,(c_\scR,c_\scV)} 
    ={}& -
    \left(\frac{\mu^2_\Ls}{Q^2}\right)^{\!\!\epsilon}\,
    \cC_\LV\sket{\{p\}_m,(c_\scR,c_\scV)}
    \;.
  \end{split}
\end{equation}
Here in the state $|\{p_1,\dots,p_m,p_{m+1}\},(c_\scR,c_\scV))$, we have substituted $\mu_\Ls$ for the real number $p_{m+1}$.

The derivative of $\cD^{(2,0)}(\mu_\Ls^2)$ is a little more complicated.
There are two terms. If we rename the integration variable and permute the arguments of the statistical state to match in the two terms, we obtain
\begin{equation}
  \begin{split}
    \label{eq:cD20model}
    \mu_\Ls^2\frac{d}{d\mu_\Ls^2}\,
    \cD^{(2,0)}(\mu_\Ls^2)&\sket{\{p\}_m,(c_\scR,c_\scV)}  
    \\={}&
    \left(\frac{\mu_\Ls^2}{Q^2}\right)^{\!\!\epsilon}\,
    \int_0^{\mu_\Ls^2}\!\frac{dp_{m+1}^2}{p_{m+1}^2}\,
    \left(\frac{p_{m+1}^2}{Q^2}\right)^{\!\!\epsilon}\,
    \left[\frac{p_{m+1}^2}{p_{m+1}^2 + 2\mu_\Ls^2}
      + \frac{\mu_\Ls^2}{\mu_\Ls^2 + 2p_{m+1}^2}\right]
    \\&\qquad\qquad\times
    \cC_\LR^2\sket{\{p_1,\dots,p_m,p_{m+1},\mu_s\},(c_\scR,c_\scV)}
    \;.
  \end{split}
\end{equation}

The derivative of $\cD^{(0,2)}(\mu_\Ls^2)$ is similar to the derivative of $\cD^{(2,0)}(\mu_\Ls^2)$:
\begin{equation}
\begin{split}
\label{eq:cD02model}
\mu_\Ls^2\frac{d}{d\mu_\Ls^2}\,
\cD^{(0,2)}&(\mu_\Ls^2)\sket{\{p\}_m,(c_\scR,c_\scV)} 
\\={}& 
\left(\frac{\mu_\Ls^2}{Q^2}\right)^{\!\!\epsilon}\,
\int_0^{\mu_\Ls^2}\!\frac{dk_1^2}{k_1^2}\,
\left(\frac{k_1^2}{Q^2}\right)^{\!\!\epsilon}\,
\left[\frac{k_1^2}{k_1^2 + 2 \mu_\Ls^2}
+\frac{\mu_\Ls^2}{\mu_\Ls^2 + 2k_1^2}\right]
\cC_\LV^2\sket{\{p\}_m,(c_\scR,c_\scV)}
\;.
\end{split}
\end{equation}
The singularity at $k_1^2 \to 0$ gives us a $1/\epsilon$ pole after integration.

The derivative of $\cD^{(1,1)}(\mu_\Ls^2)$ is more complicated. After combining terms, we obtain
\begin{equation}
  \begin{split}
    \label{eq:cD11model}
    \mu_\Ls^2\frac{d}{d\mu_\Ls^2}\,
    \cD^{(1,1)}(\mu_\Ls^2)&\sket{\{p\}_m,(c_\scR,c_\scV)} 
    \\={}& 
    -\left(\frac{\mu_\Ls^2}{Q^2}\right)^{\!\!\epsilon}
    \int_0^{\mu_\Ls^2}\!\frac{dk_1^2}{k_1^2}
    \left(\frac{k_1^2}{Q^2}\right)^{\!\!\epsilon}
    \left[\frac{k_1^2}{k_1^2 + 2 \mu_\Ls^2}
      +\frac{\mu_\Ls^2}{\mu_\Ls^2 + 2k_1^2}\right]
    \\&\qquad\qquad\quad\times
    \cC_\LR\cC_\LV\sket{\{p_1,\dots,p_m,\mu_\Ls\},(c_\scR,c_\scV)}
    \\&
    -\left(\frac{\mu_\Ls^2}{Q^2}\right)^{\!\!\epsilon}
    \int_0^{\mu_\Ls^2}\!\frac{dp_{m+1}^2}{p_{m+1}^2}
    \left(\frac{p_{m+1}^2}{Q^2}\right)^{\!\!\epsilon}
    \left[\frac{p_{m+1}^2}{p_{m+1}^2 + 2 \mu_\Ls^2}
      +\frac{\mu_\Ls^2}{\mu_\Ls^2 + 2p_{m+1}^2}\right]
    \\&\qquad\qquad\quad\times
    \cC_\LR\cC_\LV\sket{\{p\}_{m+1},(c_\scR,c_\scV)}
    \;.
  \end{split}
\end{equation}
In the first term, there is a pole from the integration region $k_1^2 \to 0$, while in the second term there is a singularity at $p_{m+1}^2 \to 0$.

With the derivatives of $\cD^{(n_\scR,n_\scV)}(\mu_\Ls^2)$ at hand, it is straightforward to use Eq.~(\ref{eq:cSpert}) to construct the shower kernel $\cS_{\rm pert}(\mu_\Ls^2)$. 

The shower kernel at order $\as$ is simple:
\begin{equation}
\begin{split}
\label{eq:cS1model}
\cS_{\rm pert}^{(1,0)}(\mu_\Ls^2)\sket{\{p\}_m,(c_\scR,c_\scV)} 
={}& 
\cC_\LR\sket{\{p_1,\dots,p_m, \mu_\Ls\},(c_\scR,c_\scV)}
\;,
\\
\cS_{\rm pert}^{(0,1)}(\mu_\Ls^2)\sket{\{p\}_m,(c_\scR,c_\scV)} 
={}& -
\cC_\LV\sket{\{p\}_m,(c_\scR,c_\scV)}
\;.
\end{split}
\end{equation}

For the shower kernel for two real emissions, we find
\begin{equation}
  \begin{split}
    \label{eq:cS20model0}
    \cS_{\rm pert}^{(2,0)}(\mu_\Ls^2)&\sket{\{p\}_m,(c_\scR,c_\scV)} 
    \\={}& 
    \left(\frac{\mu_\Ls^2}{Q^2}\right)^{\!\!\epsilon}
    \int_0^{\mu_\Ls^2}\!\frac{dp_{m+1}^2}{p_{m+1}^2}
    \left(\frac{p_{m+1}^2}{Q^2}\right)^{\!\!\epsilon}
    \left[\frac{p_{m+1}^2}{p_{m+1}^2 + 2\mu_\Ls^2}
      + \frac{\mu_\Ls^2}{\mu_\Ls^2 + 2p_{m+1}^2}
      -1\right]
    \\&\qquad\qquad\times
    \cC_\LR^2\sket{\{p_1,\dots,p_m,p_{m+1}, \mu_\Ls\},(c_\scR,c_\scV)}
    \;.
  \end{split}
\end{equation}
The subtraction removes the infrared singularity at $p_{m+1}^2\to 0$, allowing us to set $\epsilon \to 0$. This gives
\begin{equation}
  \begin{split}
    \label{eq:cS20model}
    \cS_{\rm pert}^{(2,0)}(\mu_\Ls^2) \sket{\{p\}_m,(c_\scR,c_\scV)} 
    ={}&
    \int_0^{\mu_\Ls^2}\! dp_{m+1}^2
    \left[\frac{1}{p_{m+1}^2 + 2\mu_\Ls^2}
      - \frac{2}{\mu_\Ls^2 + 2p_{m+1}^2}
    \right]
    \\&\quad\times
    \cC_\LR^2\sket{\{p_1,\dots,p_m,p_{m+1}, \mu_\Ls\},(c_\scR,c_\scV)}
    \;.
  \end{split}
\end{equation}

For the shower kernel for two virtual loops, we find a result that is analogous to what we found above. The subtraction removes the infrared singularity at $k_1^2 \to 0$ so that we can set $\epsilon \to 0$, giving
\begin{equation}
  \begin{split}
    \label{eq:cS02model}
    \cS_{\rm pert}^{(0,2)}(\mu_\Ls^2)\sket{\{p\}_m,(c_\scR,c_\scV)} 
    ={}& 
    \int_0^{\mu_\Ls^2}\!dk_1^2\,
    \left[\frac{1}{k_1^2 + 2 \mu_\Ls^2}
      -\frac{2}{\mu_\Ls^2 + 2k_1^2}
    \right]
    \cC_\LV^2\sket{\{p\}_m,(c_\scR,c_\scV)}
    \;.
  \end{split}
\end{equation}

For the shower kernel with one real emission and one virtual loop, we obtain
\begin{equation}
  \begin{split}
    \label{eq:cS11model0}
    \cS_{\rm pert}^{(1,1)}(\mu_\Ls^2)&\sket{\{p\}_m,(c_\scR,c_\scV)} 
    \\={}& 
    -\left(\frac{\mu_\Ls^2}{Q^2}\right)^{\!\!\epsilon}
    \int_0^{\mu_\Ls^2}\!\frac{dk_1^2}{k_1^2}
    \left(\frac{k_1^2}{Q^2}\right)^{\!\!\epsilon}
    \left[\frac{k_1^2}{k_1^2 + 2 \mu_\Ls^2}
      +\frac{\mu_\Ls^2}{\mu_\Ls^2 + 2k_1^2}
      -1\right]
    \\&\qquad\qquad\quad\times
    \cC_\LR\cC_\LV\sket{\{p_1,\dots,p_m,\mu_\Ls\},(c_\scR,c_\scV)}
    \\&
    -\left(\frac{\mu_\Ls^2}{Q^2}\right)^{\!\!\epsilon}
    \int_0^{\mu_\Ls^2}\!\frac{dp_{m+1}^2}{p_{m+1}^2}
    \left(\frac{p_{m+1}^2}{Q^2}\right)^{\!\!\epsilon}
    \left[\frac{p_{m+1}^2}{p_{m+1}^2 + 2 \mu_\Ls^2}
      +\frac{\mu_\Ls^2}{\mu_\Ls^2 + 2p_{m+1}^2}
      -1\right]
    \\&\qquad\qquad\quad\times
    \cC_\LR\cC_\LV\sket{\{p\}_{m+1},(c_\scR,c_\scV)}
    \;.
  \end{split}
\end{equation}
Again, the subtraction removes the infrared singularities. Then we can set $\epsilon \to 0$, giving
\begin{equation}
  \begin{split}
    \label{eq:cS11model}
    \cS_{\rm pert}^{(1,1)}(\mu_\Ls^2)&\sket{\{p\}_m,(c_\scR,c_\scV)} 
    \\={}& 
    -
    \int_0^{\mu_\Ls^2}\!dk_1^2\,
    \left[\frac{1}{k_1^2 + 2 \mu_\Ls^2}
      -\frac{2}{\mu_\Ls^2 + 2k_1^2}
    \right]
    \cC_\LR\cC_\LV\sket{\{p_1,\dots,p_m,\mu_\Ls\},(c_\scR,c_\scV)}
    \\&
    -\int_0^{\mu_\Ls^2}\!dp_{m+1}^2\,
    \left[\frac{1}{p_{m+1}^2 + 2 \mu_\Ls^2}
      -\frac{ 2}{\mu_\Ls^2 + 2p_{m+1}^2}
    \right]
    \cC_\LR\cC_\LV\sket{\{p\}_{m+1},(c_\scR,c_\scV)}
    \;.
  \end{split}
\end{equation}

Thus Eq.~(\ref{eq:cSpert}) gives us a completely straightforward way to calculate the operators $\cS_{\rm pert}^{(n_\scR,n_\scV)}(\mu_\Ls^2)$. All of these operators at order $\as^1$ and $\as^2$ are infrared finite.

\subsection[The operator $\cV(\mu_\Ls^2)$]
{The operator \boldmath{$\cV(\mu_\Ls^2)$}}

We also construct an operator $\cX(\mu_\Ls^2)$ in Eq.~(\ref{eq:X}). In our toy model, which has no parton distributions, we have
\begin{equation}
\cX(\mu_\Ls^2) = \cD(\mu_\Ls^2)
\;.
\end{equation}
Then we define an operator $\cV(\mu_\Ls^2)$ using Eq.~(\ref{eq:XtoV}),
\begin{equation}
\label{eq:XtoVbis}
\sbra{1} \cX(\mu_\Ls^2) = \sbra{1} \cV(\mu_\Ls^2)
\;.
\end{equation}
In our toy model, at least up to the order that we have defined it, the relation $(1|\cC_\LR = (1|\cC_\LV$ gives us
\begin{equation}
\label{eq:1D}
\sbra{1} \cD(\mu_\Ls^2) = \sbra{1}
\;.
\end{equation}
Thus
\begin{equation}
\cV(\mu_\Ls^2) = 1
\;.
\end{equation}
Then according to the definition Eq.~(\ref{eq:Udef}) (with $\cF = 1$) we have
\begin{equation}
\cU(\mu_2^2,\mu_1^2) = \cU_{\rm pert}(\mu_2^2,\mu_1^2)
\;.
\end{equation}
The full shower operator $\cU$ is the same as $\cU_{\rm pert}$.

\section{Transformation to shower oriented parton distribution functions}
\label{sec:K}

In Sec.~\ref{sec:showerpdfs}, we introduced shower oriented parton distribution functions that are adapted to the choice of the cutoff $\mu_\Ls^2$ used in the shower. The shower oriented parton distributions $f_{a/A}(\eta_\La,\mu^2)$ are related to the five-flavor $\MSbar$ parton distribution functions $f_{a/A}^{\msbar}(\eta_\La,\mu^2)$ by means of a kernel $K$. Following the notation of Sec.~\ref{sec:showerpdfs}, the transformation of the parton distribution for hadron A is
\begin{equation}
\label{eq:fmsbartof}
f_{a/A}^{\msbar}(\eta_\La,\mu^2) = \sum_{a'}\int_0^1\!\frac{dz}{z}\
K_{a a'}^{(\La)}(z,\mu^2,\{p,f\}_m)\,
f_{a'/A}(\eta_\La/z,\mu^2)
\;.
\end{equation}
The kernel $K$ depends on the flavor indices $a'$ and $a$, a momentum fraction variable $z$ and the scale $\mu^2$. We also allow it to depend on the momenta and flavors $\{p,f\}_m$ of the partonic statistical state.  Then $K$ has a perturbative expansion beginning with
\begin{equation}
K^{(\La)}_{a a'}(z,\mu^2,\{p,f\}_m) = 
\delta_{a a'} \delta(1-z)
+ \frac{\as(\mu^2)}{2\pi}\,K^{(\La,1)}_{a a'}(z,\mu^2,\{p,f\}_m)
+ \cO(\as^2)
\;.
\end{equation}

In order to keep the presentation in this paper as simple as possible, we set the masses of all five quarks d,u,s,c,b to zero. In a more complete picture, one uses a variable flavor number scheme. Then, even with $\MSbar$ evolution, we change the renormalization scheme when the scale $\mu$ decreases past the bottom quark mass and then the charm quark mass. Thus the $\beta$ function for $\as$ evolution and the parton evolution kernel change. For the shower oriented parton distributions $f_{a/A}(\eta_\La,\mu^2)$ and $f_{b/B}(\eta_\Lb,\mu^2)$, there is a different dependence on the c- and b-quark masses in the evolution kernels compared to what one uses in the variable flavor number version of $\MSbar$ evolution.

There are various possible versions of $K^{(\La)}_{a a'}(z,\mu^2,\{p,f\}_m)$. The simplest is
\begin{equation}
  \begin{split}
    \label{eq:K1defA}
    K^{(\La,1)}_{a a'}(z,\mu^2,\{p,f\}_m) ={}&
    \delta_{a a'}
    \left[\frac{2 z C_a}{1-z}\,
      \log\!\left(\frac{Q_\LH^2}{(1-z)2p_\La\cdot Q_\LH}\right)\right]_+
    \\&
    +\sum_{\hat a} P^{\rm reg}_{a a'}(z)\,
    \log\!\left(\frac{Q_\LH^2}{(1-z)2p_\La\cdot Q_\LH}\right)
    \\&
    + \delta_{a a'}\, \delta(1-z) \gamma_a\,
    \log\!\left(\frac{Q_\LH^2}{2p_\La\cdot Q_\LH}\right)
    - P^{(\epsilon)}_{a a'}(z)
    \;.
  \end{split}
\end{equation}
The first order $\MSbar$ DGLAP kernel is
\begin{equation}
\label{eq:DGLAPkernel}
P_{a a'}^\msbar(z) = \delta_{a a'}\left[\frac{2 z C_a}{1-z}\right]_+ 
+ P^{\rm reg}_{a a'}(z)
+ \delta_{a a'}\,\delta(1-z)\, \gamma_a
\;.
\end{equation}
Here $C_a = C_\LF$ and $\gamma_a = 3 C_\LF/2$ when $a$ is a quark flavor and $C_\Lg = C_\LA$,  $\gamma_\Lg = 11 C_\LA/6 - 2 T_\LR n_\Lf/3$. The functions $P_{a a'}^{\rm reg}(z)$ and $P_{a a'}^{(\epsilon)}(z)$ are
\begin{equation}
  \begin{aligned} 
    P_{qq}^{\rm reg}(z) &= C_\LF(1-z)\;, &P_{qq}^{(\epsilon)}(z) 
    &= C_\LF(1-z) \;, 
    \\ 
    P_{\Lg \Lg}^{\rm reg}(z) &= 2C_\LA \left(\frac{1-z}{z}
    + z (1-z)\right) \;,
    \quad &P_{\Lg \Lg}^{(\epsilon)}(z) &= 0 \;, 
    \\ 
    P_{q\Lg}^{\rm reg}(z) &= T_\LR\left(1 - 2\,z\,(1-z)\right) \;, 
    &P_{q\Lg}^{(\epsilon)}(z) &= T_\LR 2z(1-z) \;, 
    \\ P_{\Lg q}^{\rm reg}(z) &= C_\LF\left(z + 2\,\frac{1-z}{z}\right)\;, 
    &P_{\Lg q}^{(\epsilon)}(z) &= C_\LF z\;.
  \end{aligned}
\end{equation}
The kernel $K^{(\La,1)}_{a a'}(z,\mu^2,\{p,f\}_m)$ is a distribution in $z$ with a singularity at $z \to 1$. The singularity is represented by the + prescription in the first term of Eq.~(\ref{eq:K1defA}) and by the term proportional to $\delta(1-z)$. The coefficient of $\delta(1-z)$ is associated with how the virtual loop function $D^{(0,1)}(\mu^2)$ is treated. We have here taken a simple choice based on what is in Ref.~\cite{NSThreshold}, but other choices are possible.

The logarithms of $Q_\LH^2 /[(1-z)2p_\La\cdot Q_\LH]$ in Eq.~(\ref{eq:K1defA}) come about as follows. We attempt to calculate $\sbra{1}\cX(\mu^2)$ using the definition (\ref{eq:X}) of $\cX(\mu^2)$. We look at emissions of a parton in the initial state. Call the virtuality associated with this splitting $|k^2| = 2\hat p_\La \cdot \hat p_{m+1}$. We integrate over $|k^2|$ and over the momentum fraction $z$. We use the hardness variable $\Lambda^2$, Eq.~(\ref{eq:Lambdadef}), used in \textsc{Deductor}. This means that there is an upper bound for the integration over $|k^2|$,
\begin{equation}
\label{eq:ksqcut}
|k^2|< \frac{2p_\La\cdot Q_\LH}{Q_\LH^2}\ \mu_\Ls^2
\;.
\end{equation}
There is an infrared divergence coming from the $|k^2| \to 0$ limit of the integration. This divergence is regularized by integrating in $4 - 2\epsilon$ dimensions. Now, dimensional regularization effectively acts as an infrared cutoff on the transverse momentum
\begin{equation}
|k_\LT^2| = (1-z) |k^2|
\;.
\end{equation}
Thus we integrate over $|k_\LT^2|$ with an upper bound 
\begin{equation}
\label{eq:kTcut}
|k_\LT^2| < \frac{2p_\La\cdot Q_\LH}{Q_\LH^2}\,(1-z)\,\mu_\Ls^2
\;.
\end{equation}
The $1/\epsilon$ pole produced by the integration over $|k_\LT^2|$ is removed by the factor $\cZ_\LF(\mu^2)$ in $\cX(\mu^2)$. This leaves us with a $\log(Q_\LH^2/[(1-z) 2p_\La\cdot Q_\LH])$, which multiplies the DGLAP splitting kernel $P_{a\hat a}(z)$. This remaining contribution does not have a $1/\epsilon$ pole. It has a term $\log(\mu^2/\mu_\Ls^2)$, where $\mu_\Ls^2$ is the scale that defines the upper cutoff in the momentum integration and $\mu^2$ is the renormalization scale. We do not see this logarithm because we set these scales equal to each other. The Feynman rules for the splitting functions have some explicit $\epsilon$ dependence, giving a function of the form $f(\epsilon)/\epsilon = f(0)/\epsilon + f'(0) + \cO(\epsilon)$. The term $f'(0)$ gives us the contributions $P_{a a'}^{(\epsilon)}(z)$.

This calculation leaves us with an order $\as$ contribution to $\sbra{1}\cX(\mu^2) = \sbra{1}\cV(\mu^2)$ that we need to eliminate because it does not vanish in the limit $\mu^2 \to 0$. If we did not eliminate this term, we would lose Eq.~(\ref{eq:cVmuf}). The offending contribution can be removed by the factor $\cK(\mu^2)$ in $\cX(\mu^2)$ if we choose the definition (\ref{eq:K1defA}).

We also note that if we use $k_\LT^2$ instead of $\Lambda^2$, Eq.~(\ref{eq:Lambdadef}), as the shower hardness variable, then there is no $\log(1-z)$. Then by redefining $\mu_\Ls^2$ by a factor $Q_\LH^2/[2p_\La\cdot Q_\LH]$, we obtain
\begin{equation}
\begin{split}
\label{eq:K1defB}
K^{(\La,1)}_{a a'}(z,\mu^2,\{p,f\}_m)  ={}&
- P^{(\epsilon)}_{a a'}(z)
\;.
\end{split}
\end{equation}
That is, the shower oriented parton distribution functions are close to the $\MSbar$ parton distribution functions. However, they are not quite equal. That is because for the shower oriented parton distribution functions we are imposing an ultraviolet cutoff with a theta function, while with the $\MSbar$ prescription we subtract a pole $1/\epsilon$.

We now turn to a construction that puts a different cut on parton splitting, leading to a less simple kernel $K^{(\La)}_{a,a'}(z,\mu^2, \{p,f\}_m)$. We retain the ultraviolet cut (\ref{eq:ksqcut}) when $\mu_\Ls^2$ is not too small. However, when $\mu_\Ls^2$ is small, the upper bound Eq.~(\ref{eq:kTcut}) can be very small indeed. We can relax this cut to 
\begin{equation}
\label{eq:kTcutmod}
|k_\LT^2| < \max\!\left[
\frac{2p_\La\cdot Q_\LH}{Q_\LH^2}\,(1-z)\,\mu_\Ls^2,
m_\perp^2\right]
\;,
\end{equation}
where $m_\perp^2$ is, say, $1 \GeV^2$. Then the matching kernel that defines the shower adapted parton distributions is
\begin{equation}
  \begin{split}
    \label{eq:K1defC}
    K^{(\La,1)}_{a,a'}(z,\mu^2,\{p,f\}_m) ={}&
    \delta_{a \hat a}
    \left[\frac{2 z C_a}{1-z}\,
      \log\!\left(
        \min\!\left[
          \frac{Q_\LH^2}{(1-z)2p_\La\cdot Q_\LH},
          \frac{\mu_\Ls^2}{m_\perp^2}
        \right]
      \right)
    \right]_+
    \\&
    +\sum_{a'} P^{\rm reg}_{a a'}(z)\,
    \log\!\left(\min\!\left[
        \frac{Q_\LH^2}{(1-z)2p_\La\cdot Q_\LH},
        \frac{\mu_\Ls^2}{m_\perp^2}
      \right]
    \right)
    \\&
    +
    \delta_{a \hat a}\,
    \delta(1-z)\,\gamma_a
    \log\!\left(
      \min\!\left[
        \frac{Q_\LH^2}{2p_\La\cdot Q_\LH},
        \frac{\mu_\Ls^2}{m_\perp^2}
      \right]
    \right)
    -P^{(\epsilon)}_{a\hat a}(z)
    \;.
  \end{split}
\end{equation}
When $m_\perp^2 \ll \mu_\Ls^2$, this reduces to our previous definition. 
When $\mu_\Ls^2 < m_\perp^2\,Q_\LH^2/2p_\La\cdot Q_\LH$, this becomes, after using Eq.~(\ref{eq:DGLAPkernel}),
\begin{equation}
  \label{eq:K1defD}
  K^{(\La,1)}_{a,a'}(z,\mu^2,\{p,f\}_m) =
  \log\!\left(
    \frac{\mu_\Ls^2}{m_\perp^2}
  \right)
  P^\msbar_{a\hat a}(z)
  -P^{(\epsilon)}_{a\hat a}(z)
  \;.
\end{equation}
If we write Eq.~(\ref{eq:fmsbartof}) to first order as
\begin{equation}
\begin{split}
\label{eq:fmsbartof1}
f_{a/A}(\eta_\La,\mu^2) ={}&
f_{a'/A}^{\msbar}(\eta_\La/z,\mu^2)
\\&
- \frac{\as}{2\pi}
\sum_{a'}\int_0^1\!\frac{dz}{z}\
K_{a,a'}^{(\La,1)}(z,\mu^2,\{p,f\}_m)\,
f_{a'/A}^{\msbar}(\eta_\La/z,m_\perp^2)
\end{split}
\end{equation}
and use Eq.~(\ref{eq:K1defD}), recognizing that the $\MSbar$ kernel generates scale changes in $f_{a/A}^{\msbar}(\eta_\La,\mu^2)$, we find
\begin{equation}
\begin{split}
\label{eq:fmsbartof2}
f_{a/A}(\eta_\La,\mu^2) ={}& 
f_{a/A}^{\msbar}(\eta_\La,\mu^2)
- \left[f_{a/A}^{\msbar}(\eta_\La,\mu^2)
- f_{a/A}^{\msbar}(\eta_\La,m_\perp^2)\right]
\\&
+ \frac{\as}{2\pi}
\sum_{a'}\int_0^1\!\frac{dz}{z}\
P^{(\epsilon)}_{a\hat a}(z)\,
f_{a'/A}^{\msbar}(\eta_\La/z,m_\perp^2)
\\
={}& f_{a/A}^{\msbar}(\eta_\La,m_\perp^2)
+ \frac{\as}{2\pi}
\sum_{a'}\int_0^1\!\frac{dz}{z}\
P^{(\epsilon)}_{a\hat a}(z)\,
f_{a'/A}^{\msbar}(\eta_\La/z,m_\perp^2)
\;.
\end{split}
\end{equation}
That is, with this definition, for small values of the scale $\mu^2$, the shower oriented parton distribution functions approximately equal the $\MSbar$ parton distribution functions at scale $m_\perp^2$. However, at larger scales, the shower oriented parton distribution functions evolve differently from the $\MSbar$ ones. \textsc{Deductor} uses a definition similar to this, except with non-zero c and b quark masses and a corresponding variable flavor number scheme.\footnote{The contribution from $P^{(\epsilon)}_{a\hat a}(z)$ is ignored in \textsc{Deductor}.}

\section{Choosing \boldmath{$Q_\LH$} and \boldmath{$\mu^2$} dynamically}
\label{sec:dynamicalscales}

In Sec.~\ref{sec:scales}, we introduced a vector $Q_\LH$ that is used to set scales and to help define one measure of hardness, $\Lambda^2$, that can be used in the shower. (See Eq.~(\ref{eq:Lambdadef}).) We stated that one can use the intended measurement operator to set $Q_\LH$ globally. There is another possibility: we can use the statistical state $\sket{\rho_\scH}$ that represents the hard scattering, Eq.~(\ref{eq:rhoH}). This state has the expansion
\begin{equation}
\sket{\rho_\scH} = \sum_m \frac{1}{m!} \int\![d\{p\}_m] \sum_{\{f\}_m}\sum_{\{s,s',c,c'\}_m}
\sket{\{p,f,s,s',c,c'\}_m}\sbrax{\{p,f,s,s',c,c'\}_m}\sket{\rho_\scH}
\;.
\end{equation}
Our example process in this paper is Higgs boson production. With this example, the Higgs boson momentum $p_\scH$ is part of $\{p,f,s,s',c,c'\}_m$. We can set $Q_\LH^2 = m_\scH^2$ and set the rapidity of $Q_\LH$  to be the rapidity of $p_\scH$ while letting the transverse part of $Q_\LH$ be zero. Another example would be jet production, for which the Born process is the production of two jets. In this case, we could use whatever infrared safe jet algorithm we like to find the two highest $P_\LT$ jets in $\{p\}_m$. Letting the momenta of these jets be $P_1$ and $P_2$, we could set $Q_\LH^2 = (\vec P_{1,\LT}^2 + \vec P_{2,\LT}^2)/2$, set the rapidity of $Q_\LH$ to the rapidity of $P_1 + P_2$, and set the transverse part of $Q_\LH$ to zero.

This procedure requires some additional definitions since we have defined $\sket{\rho_\scH} = \cD^{-1}(\mu_\scH^2)\sket{\rho(\mu_\scH^2)}$, so that we need $Q_\LH$ and $\mu_\scH^2 = Q_\LH^2$ to create $\sket{\rho_\scH}$, but we cannot use these variables before they have been defined.

Let us see what is needed in the case of jet production, for which the number of final state partons at the Born level is $m = 2$. We content ourselves with what happens with an NLO hard cross section. We have
\begin{equation}
  \begin{split}
    \label{eq:rhoHjet}
    \sket{\rho_\scH} = 
    \left[\frac{\as(\mu^2)}{2\pi}\right]^2
    \bigg\{&
    \sket{\rho^{(2,0)}}
    \\&
    + \frac{\as(\mu^2)}{2\pi}
    \left[\sket{\rho^{(2,1)}(\mu^2)}
      - \cD^{(0,1)}(\mu^2)\sket{\rho^{(2,0)}}
    \right]
    \\&
    + \frac{\as(\mu^2)}{2\pi}
    \left[\sket{\rho^{(3,0)}}
      - \cD^{(1,0)}(\mu^2)\sket{\rho^{(2,0)}}
    \right]
    +\cO(\as^{2})
    \bigg\}
    \;.
  \end{split}
\end{equation}
We need to implement setting $\mu^2$ to $\mu_\scH^2$. In the case of the argument of $\as$, this is a simple replacement. For the Born statistical state $\sket{\rho^{(2,0)}}$, there is nothing further to do. In the terms $\sket{\rho^{(2,1)}(\mu^2)}$ and $\cD^{(0,1)}(\mu^2)\sket{\rho^{(2,0)}(\mu^2)}$, there are cancelling poles. Then from $\cD^{(0,1)}(\mu^2)$ there is a left over $\log{\mu^2}$ from  the cutoff $\mu_\Ls^2$ in the integration over the virtual loop graph. We simply have to set $\mu^2 = \mu_\scH^2$ here. In the term $\sket{\rho^{(3,0)}}$, there is nothing further to do. In the term $\cD^{(1,0)}(\mu^2)\sket{\rho^{(2,0)}}$, the momenta $\{p\}_3$ determine the momenta $\{p\}_{2}$ in $\sket{\rho^{(2,0)}}$ and also the splitting variables $k^2, z, \phi$ in $\cD^{(1,0)}(\mu^2)$. Increasing the ultraviolet cutoff $\mu_\Ls^2$ provides an increased range for $k^2$. As a result, the range in $\{p\}_3$ that is covered increases. Setting $\mu^2 = \mu_\scH^2$, we check whether the point $\{p\}_m$ is inside the allowed range. If it is, we multiply by 1. If it is not, we multiply by zero.

It seems clear that this procedure works beyond NLO, although it becomes more complicated. We work term by term in the expansion of $\sket{\rho_\scH}$. Once we have set $Q_\LH$ and $\mu_\scH^2$, we set $\mu^2 = \mu_\scH^2$ in the argument of $\as$ and in all explicit logs that come from virtual loop integrals. We check whether the point $\{p\}_m$ is generated by a subtraction term with an ultraviolet cutoff $\mu_\scH^2$. If not, we omit this term. Our description here has been algorithmic. To formulate this in terms of operators on the statistical space requires additional notation, which we omit.

\section{Renormalization}
\label{sec:renormalization}

In this paper we use $\MSbar$ renormalization. In particular, this defines $\as$ and the $\MSbar$ parton distributions that we start with. In the parton shower algorithm that we obtain, each element of the calculation is infrared finite and ultraviolet finite in four dimensions. However, part of the cancellation of infrared divergences is tied to the removal of ultraviolet divergences by renormalization. For this reason, the details of the ultraviolet renormalization scheme are significant. In this appendix, we gather the most important formulas that we use, mostly following Ref.~\cite{JCCbook}.

\subsection{Renormalization of the QCD coupling}

In the $\MSbar$ scheme, the renormalization of the coupling is\footnote{We find this definition useful for our purposes. Ref.~\cite{JCCbook} uses a different strong coupling, $\tilde \alpha_\Ls$, with $\as = S_\epsilon\tilde \alpha_\Ls$.}
\begin{equation}
  \as^{\bare}  S_\epsilon = Z_\alpha(\mur)\,\mu^{2\epsilon}\, \as(\mur)\;\;,
\end{equation}
where $\mu$ is the renormalization scale, $\as^{\bare}$ has mass dimension $2\epsilon$, and
\begin{equation}
  S_\epsilon = \frac{(4\pi)^\epsilon}{\Gamma(1-\epsilon)}
  \;.
\end{equation}
The renormalization constant of the strong coupling is given as a sum,
\begin{equation} 
  Z_\alpha(\mur) 
  = 1 
  + \sum_{n=1}^\infty \left[\frac{\as(\mur)}{2\pi}\right]^n
  \sum_{k=1}^n\frac{Z_{\alpha}^{[n,k]}}{\epsilon^k}
  \;\;.
\end{equation}
The scale independent coefficients of the singularities, $Z_{\alpha}^{[n,k]}$, can be given in terms of the expansion parameters of the $\beta(\as)$ function by a recursion relation:
\begin{equation} 
  Z_{\alpha}^{[n,k+1]} = -\frac1n\sum_{l=k}^{n-1}(l+1)\beta_{n-l}\, Z_{\alpha}^{[l,k]}\;\;, \qquad
  Z_{\alpha}^{[n,1]}  = -\frac1n\beta_n\;\;.
\end{equation}
The running coupling obeys the evolution equation
\begin{equation} 
  \mur\frac{d\as(\mur)}{d\mur} 
  = - \as(\mur) \,\big(\epsilon + \beta(\as) \big) 
  = -\as(\mur) \left(
    \epsilon + \sum_{n=1}^\infty\left[\frac{\as(\mur)}{2\pi}\right]^n\beta_{n} 
  \right)
  \;,
\end{equation}
where the first two $\beta_i$ coefficients are
\begin{equation} 
  \beta_1 = \frac{11C_A-4T_R n_\Lf}6\;\;,\qquad 
  \beta_2 = \frac{17 C_A^2 -10C_A T_R n_\Lf - 6 C_F T_R n_\Lf}{6}\;\;.
\end{equation}

\subsection{Renormalization of the parton distribution functions}

The $\MSbar$ parton distribution functions enter the cross section formula (\ref{eq:NkLOsigma}) in the form
\begin{equation}
  \begin{split}
    [\cF_{\msbarsub}(\mu^2)\circ &\cZ_F(\mu^2)]\sket{\{p,f,s,s',c,c'\}_m}
    \\
    ={}& \sum_{a',b'} \int_0^1\!\frac{dz_\La}{z_\La}
    \int_0^1\!\frac{dz_\Lb}{z_\Lb}\ 
    \frac{f_{a'/A}^\msbar(\eta_\La/z_\La,\mu^2)\,
    f_{b'/B}^\msbar(\eta_\Lb/z_\Lb,\mu^2)}
    {n_\Lc(a) n_\Ls(a)\, n_\Lc(b) n_\Ls(b)\ 4 p_\La\cdot p_\Lb}
    \\&\qquad\quad\times
    Z_{F}(a,a';z_\La,\as(\mu^2))\, Z_{F}(b,b';z_\Lb,\as(\mu^2))\,
    \sket{\{p,f,s,s',c,c'\}_m}
    \;.
  \end{split}
\end{equation}
The renormalization factor $\cZ_F$ is a product, so that each of the two parton distributions is transformed separately. The renormalization kernel $Z_F$ relates the renormalized parton distribution to the bare parton distribution:
\begin{equation}
  \label{eq:PDFrenorm}
  f_{a/A}^\bare(\eta) = 
  \sum_{a'} \int_0^1\!\frac{dz}{z}\,Z_{F}(a,a';z,\as(\mu^2))\, 
  f_{a'/A}^\msbar(\eta/z,\mu^2)
  \;.
\end{equation}
The kernel has a perturbative expansion
\begin{equation}
  \begin{split}
    Z_F(a,a'; z,\as(\mur))
    ={}&  \delta_{a,a'}\,\delta(1-z)
    + \sum_{n=1}^\infty \left[\frac{\as(\mur)}{2\pi}\right]^n 
    \sum_{k=1}^n\frac{Z^{[n,k]}_{a,a'}(z)}{\epsilon^k} 
    \;\;.
  \end{split}
\end{equation}

It follows from the requirement that $f^{\rm bare}_{a/A}(\eta)$ is independent of $\mur$ and $f^\msbar_{a/A}(\eta,\mu^2)$ has no poles that the renormalized parton distribution function obeys the DGLAP evolution equation, 
\begin{equation}
  \mur\frac{d f^\msbar_{a/A}(\eta,\mu^2)}{d\mur} =
  \int_0^1\frac{dz}{z}\, P_{a,a'}(z, \mur)\,f^\msbar_{a'/A}(\eta/z,\mu^2) 
  \;,
\end{equation}
where
\begin{equation}
P_{a,a'}(z, \mur) =  \sum_{n=1}^\infty \left[\frac{\as(\mur)}{2\pi}\right]^n P_{a,a'}^{(n)}(z)
\end{equation}
with
\begin{equation}
P_{a,a'}^{(n)}(z) = n\,Z^{[n,1]}_{a,a'}(z)
\;.
\end{equation}
The coefficients of $1/\epsilon$ to higher powers are then determined by the recursion relation
\begin{equation}
  Z_{a,a'}^{[n,k+1]}(z) = \frac1n \sum_{l=k}^{n-1}
  \int_z^1 \frac{dx}{x}\,\sum_c
  Z_{a,c}^{[l,k]}(z/x)
  \left[P^{(n-l)}_{c,a'}(x) - \delta_{c,a'}\delta(1-x)\, l\,\beta_{n-l}\right]
  \;.
\end{equation}

We can also write the evolution kernel as
\begin{equation}
  \begin{split}
  P_{a,a'}(z, \mur) = {}& 
  - \int_z^1 \frac{dx}{x}\,\sum_c
  Z_F^{-1}(a,c; z/x, \as(\mur))\
  \mur \frac{dZ_F(c,a'; z, \as(\mur))}{d\mur}
  \;.
 \end{split}
\end{equation}
The factors contain poles $1/\epsilon^k$, but the poles cancel.

\subsection{Renormalization scale dependence}

The physical states $\sket{\rho(\mur)}$ defined in Eq.~(\ref{eq:rhoexpansion})  represent the quantum density operator of the partonic scattering. It is constructed from amplitudes $\ket{M(\{p,f\}_m)}$ and conjugate amplitudes $\bra{M(\{p,f\}_m)}$. We include the proper Lehmann-Symanzik-Zimmermann (LSZ) factors for the incoming and outgoing partons so that the amplitudes are S-matrix elements in the renormalized theory. With massless partons and dimensional regularization in Feynman gauge, this amounts to multiplying by field strength renormalization factors $Z_\psi^{-1/2}$ or $Z_A^{-1/2}$ for each external leg of an amputated graph. Then $\sket{\rho(\mur)}$ is independent of the renormalization scale if calculated at all perturbative orders. When we calculate it up to order $\as^k$ as in Eq.~(\ref{eq:rhoexpansion}), we have
\begin{equation}
  \mur\frac{d}{d\mur} \sket{\rho(\mur)} = \cO(\as^{k+1})
  \;.
\end{equation}

Let us consider next the infrared sensitive operator $\cD(\mur)$. This operator depends on two scales, the renormalization scale $\mur$ and shower scale $\mus$. We have set these scales to equal each other, but in this appendix we highlight their separate roles. Thus we write $\cD(\mur, \mus)$ with two arguments and let the operator with only one argument denote
\begin{equation}
  \cD(\mur) = \cD(\mur, \mur)
  \;.
\end{equation}
For any basis state $\sket{\{p,f,s,s',c,c'\}_m}$, the state $\cD(\mur,\mus)\sket{\{p,f,s,s',c,c'\}_m}$ is to be defined so as to have the properties of a physical statistical state $\sket{\rho(\mur)}$. It is constructed in the renormalized theory and has the proper field strength renormalization factors for its external partons so that $\cD(\mur,\mus)$ is independent of the renormalization scale up to the order that we calculate. That is, if $\cD(\mur,\mus)$ is calculated up to order $\as^k$, we have
\begin{equation}
  \label{eq:deriv-D-renorm}
  \mur\frac{\partial}{\partial\mur} \cD(\mur,\mus) = \cO(\as^{k+1})
  \;.
\end{equation}

In order to calculate the parton shower generators $\cS_{\rm pert}(\mu^2)$ and $\cS(\mu^2)$, we need the total derivative of $\cD(\mu^2)$ with respect to the common scale. Since $\cD(\mur,\mus)$ is independent of $\mur$, this is actually the derivative with respect to $\mus$:
\begin{equation}
\begin{split}
\label{eq:cDtotalderiv}
    \frac{d}{d\mur}  \cD(\mur)
    = {}& \left. \frac{\partial}{\partial\mur} \cD(\mur,\mus)\right|_{\mus=\mur}
    +   \left. \frac{\partial}{\partial\mus} \cD(\mur,\mus)\right|_{\mus=\mur}
    \\
    ={}&
    \left. \frac{\partial}{\partial\mus} \cD(\mur,\mus)\right|_{\mus=\mur}
    +\cO(\as^{k+1})
    \;.
\end{split}
\end{equation}
The perturbative expansion of this is
\begin{equation}
  \mur\frac{d}{d\mur}  \cD(\mur) = \sum_{n=1}^k \left[\frac{\as(\mur)}{2\pi}\right]^n
  \left.\mur\frac{\partial}{\partial\mus} \cD^{(n)}(\mur,\mus)\right|_{\mus=\mur}
  + \cO(\as^{k+1})
  \;.
\end{equation}
Using this, the generators of the perturbative shower operator can be obtained from Eqs.~(\ref{eq:Spert}) and (\ref{eq:Dinvexpansion}) as
\begin{equation}
 \frac{1}{\mur}\, S^{(n)}_{\rm pert}(\mur)= \left[\frac{\partial}{\partial\mus} \cD^{(n)}(\mur,\mus)
    - \sum_{k=1}^{n-1}\widetilde \cD^{(k)}(\mur)\frac{\partial}{\partial\mus} \cD^{(n-k)}(\mur,\mus)
  \right]_{\mus=\mur}
  \;.
\end{equation}
This simplifies the shower generators. The operators $\cD^{(n)}(\mur,\mus)$ always contains a theta function like $\theta(\Lambda^2 < \mus)$ or the equivalent constraint for the loop contributions. The partial derivative turns one of these theta functions into a Dirac delta function.

The inclusive infrared finite operator $\cV(\mur)$ is defined by the condition (\ref{eq:XtoV}). This operator is derived from the infrared sensitive operator $\cD$ and, just like $\cD$, depends on both the renormalization and shower scales. Thus we write $\cV(\mur, \mus)$ with two arguments and let the operator with only one argument denote
\begin{equation}
  \cV(\mur) = \cV(\mur, \mur)
  \;.
\end{equation}
Using Eqs.~(\ref{eq:X}), (\ref{eq:XtoV}), we have
\begin{equation}
  \label{eq:1cVcF}
  \sbra{1}\cV(\mur, \mus) \cF(\mur)
  =
  \sbra{1}\!\left[\cF(\mu^2)\circ \cK(\mu^2)\circ\cZ_F(\mu^2)\right]
  \cD(\mur, \mus)
  \;.
\end{equation}
Recall that $\left[\cF(\mu^2)\circ \cK(\mu^2)\circ\cZ_F(\mu^2) \right]$ is independent of $\mu^2$. We have just seen that $\cD(\mu^2\!, \mus)$ is independent of $\mur$ at fixed $\mus$. Recall from Sec.~\ref{sec:cVdef} that we fix the color and spin content of $\cV(\mur, \mus)$ to make Eq.~(\ref{eq:1cVcF}) work for $\sbra{1}\cV(\mur, \mus)$. Once we have done that for one choice of $\mur$ at a given $\mus$, we can keep the same operator $\cV(\mur, \mus)$ for other choices of $\mur$ and Eq.~(\ref{eq:1cVcF}) will continue to hold. That is, we can define $\cV(\mur, \mus)\cF(\mu^2)$  so that it is independent of $\mur$ up to whatever order we calculate:
\begin{equation}
\label{eq:diffV-renorm}
  \frac{\partial}{\partial\mur} \left[\cV(\mur, \mus) \cF(\mur)\right] = 
   0  + \cO(\as^{k+1})
   \;.
\end{equation}
This gives us
\begin{equation}
  \cV(\mur, \mus)^{-1}\frac{\partial}{\partial\mur} \cV(\mur, \mus) = 
   -\left[\frac{d}{d\mur} \cF(\mur)\right]\cF^{-1}(\mur)
    + \cO(\as^{k+1})
   \;.
\end{equation}

The generator $\cS_\cV(\mu^2)$ of $\cU_\cV(\mu_2^2,\mu_1^2)$ is defined in Eq.~(\ref{eq:SVdef}) as
\begin{equation}
\frac{1}{\mur}\,\cS_\cV(\mu^2) =
\cV^{-1}(\mu^2)\left[\frac{\partial}{\partial\mur}\cV(\mur, \mus)
+ \frac{\partial}{\partial\mus}\cV(\mur, \mus)\right]_{\mus = \mur}
\;.
\end{equation}
This gives us
\begin{equation}
\label{eq:SVdefencore}
  \frac{1}{\mur}\,\cS_\cV(\mur) = \cV^{-1}(\mur)\left.\frac{\partial}{\partial\mus}\cV(\mur,\mus)\right|_{\mus=\mur}
  - \frac{d\cF(\mur)}{d\mur}\cF^{-1}(\mur)
  \;.
\end{equation}
Here the first term represents the evolution of the perturbative part of $\cV$ and the second term gives the evolution of the parton distribution functions. This generalizes Eq.~(\ref{eq:SV1storder}) for the first order contribution to $\cS_\cV$.

Finally the generator of the probability preserving shower evolution operator is given in Eqs.~(\ref{eq:Sdef3}) and (\ref{eq:Spert}). We can simplify this by using Eqs.~(\ref{eq:cDtotalderiv}) and (\ref{eq:SVdefencore})
\begin{equation}
  \begin{split}
    \label{eq:Ssimplified}
    \frac{1}{\mur}\cS(\mur)
    ={}&
    \left.\cV(\mur)\cF(\mur) \cD^{-1}(\mur)\, \frac{\partial\cD(\mur,\mus)}{\partial\mus}
      \,\cF^{-1}(\mur)\, \cV^{-1}(\mur)\right|_{\mus=\mur}
    \\
    &-\left.\frac{\partial\cV(\mur,\mus)}{\partial\mus}\,\cV^{-1}(\mur)\right|_{\mus=\mur}
    \;.
  \end{split}
\end{equation}
%


\end{document}